 \documentclass{elsart}

\usepackage{graphicx}

\begin{document}


  
\begin{frontmatter}
   \title{Failure of classical traffic flow theories: \protect\newline Stochastic   highway capacity  and  automatic driving  }
 
\author{Boris S. Kerner $^1$}

 \address{$^1$
Physics of Transport and Traffic, University Duisburg-Essen,
47048 Duisburg, Germany}


\maketitle

\begin{abstract}
In a mini-review  [Physica A {\bf 392} (2013) 5261--5282] it has been shown that classical traffic flow theories and models failed to
explain empirical traffic breakdown --   a phase transition from metastable free flow   to 
synchronized flow   at highway bottlenecks.   The main objective of
  this mini-review is to study the consequence of this failure of classical traffic-flow theories
  for an analysis of empirical stochastic highway capacity
  as well as for  the    effect of automatic driving vehicles
	and cooperative driving on traffic flow. To reach this goal,
  we show a deep connection between the understanding of empirical stochastic highway capacity
and  a reliable analysis of  automatic driving vehicles in traffic flow.   
With the use of simulations in the framework of three-phase traffic theory,
a probabilistic analysis of the effect of automatic driving vehicles  
 on a mixture
	traffic flow consisting of a random distribution
	of automatic driving and manual driving  vehicles has been made.
We have found that the parameters of
     automatic driving vehicles   
	   can   either decrease or increase
   the probability of traffic breakdown. The increase in
   the probability of traffic breakdown, i.e.,
	the deterioration of the performance of  the
	traffic system can occur already at a small percentage (about 5$\%$) of 
		  automatic driving   vehicles.
		The increase in
   the probability of traffic breakdown through automatic driving   vehicles  can be realized,
	even if any platoon of
	 automatic driving vehicles satisfies condition for string stability.
    \end{abstract}
\end{frontmatter}
\tableofcontents
\newpage

 \section{Introduction. The reason for paradigm shift in traffic and
transportation science  \label{Int}}
 
 A current effort   of many car-developing companies is devoted to the development of
 automatic driving vehicles\footnote{Automatic driving is also called automated driving. Respectively, automatic driving vehicles are
  also called automated driving vehicles.}. It is assumed that  the future
  vehicular traffic   consisting  of human driving and automatic driving vehicles
  should considerably enhance   highway capacity. 
	
	Highway capacity  is limited by traffic breakdown,
i.e., a transition from free flow to 
  congested 
  traffic~\cite{Greenshields1935,May,Manual2000,Manual2010,Hall1986A,Hall1987A,Hall1991A,Hall1992A10,Elefteriadou1995A,Persaud1998B,Lorenz2000A10,Brilon310,Brilon210,Brilon,BrilonISTTT2009,ach_Elefteriadou2014A,ach_ElefteriadouBook2014}. 
  Traffic breakdown with resulting traffic congestion occurs usually at
a road 
bottleneck (see, e.g.,~\cite{Greenshields1935,May,Manual2000,Manual2010,Hall1986A,Hall1987A,Hall1991A,Hall1992A10,Elefteriadou1995A,Persaud1998B,Lorenz2000A10,Brilon310,Brilon210,Brilon,BrilonISTTT2009},

~\cite{ach_Elefteriadou2014A,ach_ElefteriadouBook2014,Ceder1975A10,Ceder197610,Hall1988B10,Hall1994A10,Hall1993A10,May1964A10,May1963A10,Drake1967A10,Cassidy1989A10,Chin1991A10,Schoen1995A10,Persaud1986A10,Persaud1989A10,Persaud1990A10,Persaud1988A10,Persaud1991A10,Persaud2001A10,Ringert1993A10},

~\cite{Banks1995A10,Banks1989A10,Banks1990B10,Banks1990A10,Banks1991A10,Banks1991B10,Banks1993A10,Banks1999A10,Banks2002A10,Hsu1993A10,Easa1980A10,Edie1961A10,Edie1965A10,Edie195810,Edie1960A10,Edie1980A10,Wemple1991A10,Urbanik199110},

~\cite{Treiterer1967A10,Treiterer197510,Treiterer1974First,Treiterer1966A10,Koshi10,Koshi1984A10,Koshi2003A10,NeubertLee10,Neubert10,Neubert2000A10,Newman1963A10,Nakayama_2009First,Athol1965A10,Athol1965B10,Hess1963A10,Hess1965A10,Greenburg1960A10},

~\cite{Dudek1973A10,Dudek1982A10,Forbes1990A10,Forbes1967A10,Forbes1968A10,Gazis1969A10,Gazis1962First,Jones1962A10,Drew1965B10,Drew1965A10,Daou1964A10,Hillegas1974A10,Haynes1965A10,Buckley1968A10,Mika1969A10,Courage1969A10,Jones1970A10},

~\cite{Miller1970A10,Lam1970A10,Gafarian1971A10,Munjal1973A10,Munjal1971B10,Munjal1971First,Munjal1971C10,Munoz2001A10,Pahl1971A10,Heyes1972A10,Martin1973A10,Breiman1973A10,Breiman1977A10,Breiman1977B10,Dendrinos1978A10,Tolle1974A10,Wasielewski1981A10,Branston1976A10,Linzer1979A10,Kohler1979A10,McDermott198010,Dudash1983A10},

~\cite{Hurdle1983A10,Hurdle1986A10,Mahalel1983A10,Westland1998A10,Owens1988A10,Bennet1995A10,Iwasaki1991A10,Luttinen1992A10,Bovy10,Kockelman1998A10,VanGoeverden1998A10,Windover1998A10,Yukawa2003A10,Cassidy2001AA10,Cassidy2006First,Bertini1998A10,Bertini1999B10,Cassidy1997A10,MD200210,Smilowitz1999First,Rehborn2011},

~\cite{Nishinari2001A10,Nishiari2002A10,Hoogendoorn1998A10,TreiberWilson2011A,vanLint2009A,ZielkeBertini2008A,BertiniHansen2005A,BertiniLindgren2006A,Koller2015A,Rehborn2014A,KR1996A10_Int,KR1996B10_Int,KR199710_Int,KR1998A10_Int,KR1998B10_Int,KernerTEC2013A,KernerTGF2015A,KRPKl2011A,KernerRSch201310_Int,Kerner1977Pa,ASDA_FOTO_A,FranHoll2015}
 and references in reviews,
books, and conference 
proceedings

\cite{Haight1963A,Drew10,Mannering1998First},
~\cite{Prigogine,Da,Pa1983,Cremer,Leutzbach,GartnerEd,Wh2,Wid,New,Brackstone,Gazis,Wolf10,Chowdhury,Hel9710,Helbing,Nagatani},

~\cite{Nagel,Mahnke,Bellomo,Maerivoet,Piccoli,Hesham10,Rakha2009First,Sch,TreiberD,Treiber,Lesort10,Ceder10,Taylor10},

~\cite{ISTTT2005,ISTTT2007,ISTTT2009,ISTTT2011,SW110,SW210,SW310,SW410,SW510,SW710,SW910,Gartner198710,RK2009,Kerner2009B,Kerner2008A,KernerBook,KernerBook2,Kerner2008D,Kerner2011CD,MiniReview,Kerner_Review2}).
Road bottlenecks are 
  caused, for example,
  by  road works, on- and off-ramps, road gradients, 
  reduction of lane number (see, e.g.,~\cite{May,Manual2000,Manual2010}), slow moving vehicles (called $\lq\lq$moving 
  bottlenecks")~\cite{Gazis1992,Newell1993,Newell1988,Daganzo2002,Lebacque1998,Leclercq,Daganzo2003,Rakha2014A,Rakha2014B,KKl2010A}.
Therefore, to understand
the nature of highway capacity of real traffic,   empirical features of traffic breakdown at a bottleneck should be known.
	
	\subsection{Achievements of empirical study of traffic breakdown \label{Ach_S}}
 
\begin{figure} 
  \begin{center}
\includegraphics[scale=.6]{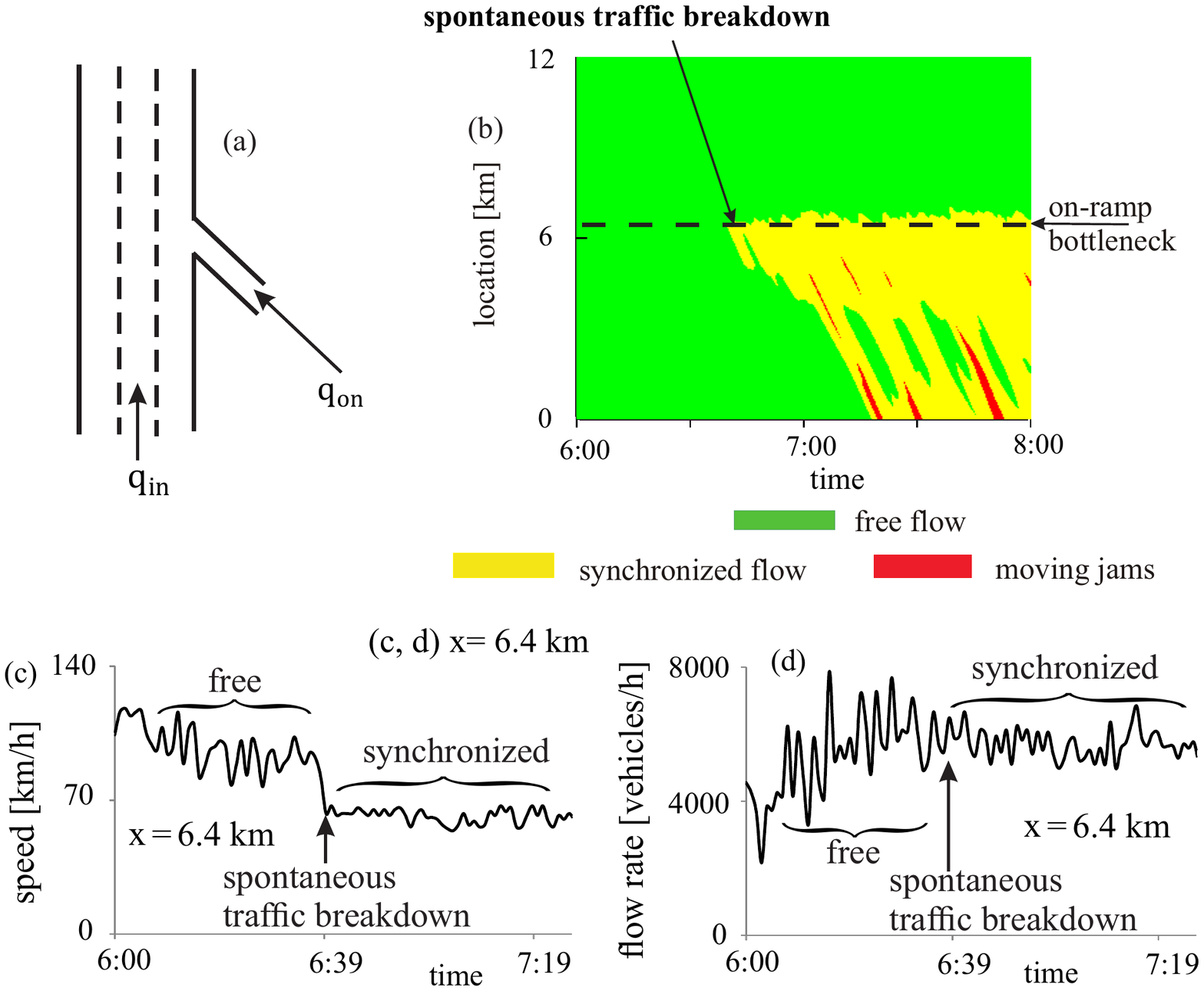}
 \end{center}
\caption{Empirical features of spontaneous  traffic breakdown (spontaneous F$\rightarrow$S transition)
   at on-ramp 
bottleneck. Real field traffic
 data measured on three-lane freeway A5-South in Germany on April 15, 1996:
 (a) Schema of three-lane freeway with on-ramp bottleneck.
(b) Speed  data  measured with road detectors installed 
 along    road section;   data is
presented in space and time with   averaging   method described in Sec.~C.2 of~\cite{KernerRSch201310_Int}.
(c, d) 1-min average data for speed (c)
  and flow rate (d) as time functions  measured at location $x=$ 6.4 km.
  Empirical results
   are qualitatively the same as those found from studies of real field traffic data measured 
 in different countries~\cite{May,Manual2000,Manual2010,Hall1986A,Hall1987A,Hall1991A,Hall1992A10,Elefteriadou1995A,Persaud1998B,Lorenz2000A10,Brilon310,Brilon210,Brilon,BrilonISTTT2009,ach_Elefteriadou2014A,ach_ElefteriadouBook2014}. Free -- free flow, synchronized -- synchronized flow.
}
\label{OnRampSp150496}
\end{figure}

  Beginning from the classical work by Greenshields~\cite{Greenshields1935},
a great effort has been made to understand the empirical features of 
traffic breakdown (see, 
e.g.,~\cite{May,Manual2000,Manual2010,Hall1986A,Hall1987A,Hall1991A,Hall1992A10,Elefteriadou1995A,Persaud1998B},
~\cite{Lorenz2000A10,Brilon310,Brilon210,Brilon,BrilonISTTT2009,ach_Elefteriadou2014A,ach_ElefteriadouBook2014}).
Traffic breakdown at a highway bottleneck is a 
local phase transition from free flow (F) to congested traffic 
whose downstream front is usually fixed at the bottleneck
 location (Fig.~\ref{OnRampSp150496} (a, b))~\cite{May,Manual2000,Manual2010,Hall1986A,Hall1987A,Hall1991A,Hall1992A10,Elefteriadou1995A,Persaud1998B,Lorenz2000A10,Brilon310,Brilon210,Brilon,BrilonISTTT2009,ach_Elefteriadou2014A,ach_ElefteriadouBook2014}.
In three-phase traffic theory, such congested traffic is called synchronized flow (S)~\cite{KernerBook,KernerBook2}. In other words,
using the terminology of the three-phase traffic theory, traffic breakdown is a transition from free flow to synchronized flow (called F$\rightarrow$S transition)~\cite{KernerBook,KernerBook2}.
However, it should be emphasized that
  as long as features of synchronized flow are not discussed (this discussion 
	will be done in Sec.~\ref{Nuc_Nature_S}), the term {\it synchronized flow}
is nothing more as only the definition of congested traffic whose downstream front is fixed at the bottleneck.

During traffic breakdown vehicle speed sharply decreases (Fig.~\ref{OnRampSp150496} (c)).  For this reason, 
traffic breakdown is also called {\it speed drop} or 
     {\it speed breakdown}~\cite{May,Manual2000,Manual2010,Hall1986A,Hall1987A,Hall1991A,Hall1992A10,Elefteriadou1995A,Persaud1998B,Lorenz2000A10},
		~\cite{Brilon310,Brilon210,Brilon,BrilonISTTT2009,ach_Elefteriadou2014A,ach_ElefteriadouBook2014}. In contrast, 
 after traffic breakdown has occurred the flow rate can remain 
as large as in an initial free 
flow (Fig.~\ref{OnRampSp150496} (d))\footnote{After traffic breakdown at the bottleneck has occurred, a congested pattern emerges and further develops upstream of the bottleneck.
Empirical features of the congested pattern development can be found in the book~\cite{KernerBook}. 
However, it should be emphasized that the above statement that the flow rate in free flow
 downstream of the bottleneck 
after the breakdown has occurred
 can remain 
as large as in an initial free flow is not often satisfied, when  
  due to the so-called pinch effect in synchronized flow upstream of the bottleneck, wide moving jams emerge   in the synchronized flow.
In this case, the congested pattern can exhibit a very complex spatiotemporal structure consisting of the synchronized flow and wide moving jam traffic phases of congested traffic.
The maximum  flow rate in the outflow from a wide moving  jam is considerably smaller than the maximum possible flow rate in synchronized flow~\cite{KernerBook}.
This is one of  the reasons why the flow rate in the outflow of a congested pattern at the bottleneck (called often $\lq\lq$discharge flow rate"), as well-known from
 empirical observations (e.g.,~\cite{May}),  
can become considerably smaller than the flow rate in free flow before the breakdown has occurred.
However, a consideration of the physics of the development of   congested patterns and their empirical features are out of scope of this mini-review.}~\cite{Hall1986A,Hall1987A,Hall1991A,Hall1992A10,Elefteriadou1995A,Persaud1998B,Lorenz2000A10,ach_Elefteriadou2014A,ach_ElefteriadouBook2014}. 
   In the review article,
    the flow rate in free flow  
  at the bottleneck  is denoted by $q=q_{\rm sum}$.
	
	In 1995, Elefteriadou {\it et al.}~\cite{Elefteriadou1995A,ach_Elefteriadou2014A,ach_ElefteriadouBook2014}  found that traffic breakdown at a highway bottleneck has a 
stochastic (probabilistic) 
behavior. This
 means the following: At a given flow rate $q_{\rm sum}$ in free flow at
  the bottleneck traffic breakdown can occur but it should not necessarily occur. 
 Thus   on one day traffic breakdown occurs, however, on another day at the same flow rate $q_{\rm sum}$  traffic breakdown is not observed.
Studying the probability for the probabilistic breakdown phenomenon
at a freeway bottleneck, in 1998  Persaud {\it et al.}~\cite{Persaud1998B}   discovered that  
the probability $P^{\rm (B)}(q_{\rm sum})$ of traffic breakdown
is an increasing function of the flow rate $q_{\rm sum}$ in free flow  
  at the bottleneck. 
The empirical result  of Persaud {\it et al.}~\cite{Persaud1998B}  has also been found  for freeways in the USA
by  Lorenz and Elefteriadou~\cite{Lorenz2000A10} 
as well as for German freeways by Brilon
{\it et al.}~\cite{Brilon310,Brilon210,Brilon,BrilonISTTT2009}.

  Traffic parameters, like weather, percentage of long vehicles in traffic flow, shares of aggressive and timid drivers  are stochastic time-functions. Thus it is generally assumed that  the stochastic nature of real traffic breakdown might be explained by classical traffic flow theories,  in which stochastic traffic parameters should be taken into account (see, e.g.~\cite{Hall1986A,Hall1987A,Hall1991A,Hall1992A10,Elefteriadou1995A,Persaud1998B,Lorenz2000A10,ach_Elefteriadou2014A,ach_ElefteriadouBook2014,Treiber} and references there). 
  \begin{itemize}
\item In contrast with this general accepted 
assumption~\cite{Hall1986A,Hall1987A,Hall1991A,Hall1992A10,Elefteriadou1995A,Persaud1998B,Lorenz2000A10,ach_Elefteriadou2014A,ach_ElefteriadouBook2014,Treiber},
in Sec.~\ref{Nuc_Nature_S} we will explain that the sole  knowledge of the above-mentioned
   features of empirical   traffic breakdown at highway bottlenecks and highway capacity
revealed and reviewed in~\cite{Greenshields1935,May,Manual2000,Manual2010,Hall1986A,Hall1987A,Hall1991A,Hall1992A10,Elefteriadou1995A},
~\cite{Persaud1998B,Lorenz2000A10,Brilon310,Brilon210,Brilon,BrilonISTTT2009,ach_Elefteriadou2014A,ach_ElefteriadouBook2014}
  is  {\it not} sufficient to disclose the physical nature 
  of  traffic   breakdown and associated stochastic highway capacity. Indeed, we will find that empirical stochastic highway capacity exhibits the nucleation nature that contradicts basic  results of  classical traffic flow theories. 
	In particular, in Sec.~\ref{versus_Nature_S} we will show that the
	classical understanding of stochastic highway capacity that is generally 
	accepted~\cite{Hall1986A,Hall1987A,Hall1991A,Hall1992A10,Elefteriadou1995A,Persaud1998B,Lorenz2000A10,ach_Elefteriadou2014A,ach_ElefteriadouBook2014} is invalid for real traffic.
 \end{itemize} 

\subsection{Basic   assumption of three-phase traffic theory   \label{Basic_P_Sp_S}} 

As emphasized 	in Sec.~\ref{Ach_S}, real traffic breakdown at a road bottleneck is an F$\rightarrow$S transition.   
 To explain  features of empirical   traffic breakdown at highway 
bottlenecks, in  three-phase traffic theory introduced by the author in 1996-2002 (three-phase theory, for short) 
has been assumed that traffic breakdown is the F$\rightarrow$S transition at the bottleneck
that occurs in  {\it metastable  free 
flow}~\cite{KernerBook,KernerBook2,Kerner1998E,Kerner1998C,Kerner1999B,Kerner1999D,Kerner2000D,Kerner2000A,Kerner2000B,Kerner2001A,Kerner2001B,Kerner2002A,Kerner2002B,Kerner2002C,Kerner2002D,Kerner2003A,Kerner2004A}.
Thus in the three-phase theory the term {\it traffic breakdown}
is a synonym of the term  {\it F$\rightarrow$S transition} occurring in metastable free flow at the bottleneck.

The term $\lq\lq$metastable free flow with respect to 
 the F$\rightarrow$S transition at a bottleneck" means that 
a small enough disturbance (speed, density, and/or flow rate)
 in free flow at the bottleneck decays. Therefore, in this case free flow persists
 at the bottleneck over time. However, when a   disturbance of a large enough amplitude
 appears in free flow in a neighborhood of the bottleneck, 
an F$\rightarrow$S transition occurs at the bottleneck. 
In accordance with  other metastable systems of natural 
science~\cite{Haken1977,Gardiner,NP,Scholl1987,Vasilev}, 
~\cite{Michailov1,Michailov2,KO,Chandrasekhar1961,NIE95,Pound,Sanz}
 such a  local disturbance in free traffic flow can be called
  a {\it nucleus} for traffic breakdown (F$\rightarrow$S transition) at a bottleneck. 
	 \begin{itemize}
\item A  nucleus  for traffic breakdown (F$\rightarrow$S transition) at a bottleneck is
a time-limited local disturbance in free   flow that 
does lead to traffic breakdown at the bottleneck. 
 \end{itemize}
This means that traffic breakdown at the bottleneck exhibits the nucleation nature: 
If the nucleus for traffic breakdown occurs in free flow at the bottleneck, 
traffic breakdown does occurs. In contrast, as long as no nucleus appears, 
no traffic breakdown occurs in a metastable state of free flow.  
  \begin{itemize}
\item The basic assumption of the three-phase theory is that traffic breakdown
at a bottleneck is the F$\rightarrow$S transition  that exhibits {\it the nucleation nature}. 
 \end{itemize}

The   basic assumption of the three-phase theory can mathematically be formulated  
as follows~\cite{KernerBook,KernerBook2,Kerner1998E,Kerner1998C,Kerner1999B,Kerner1999D,Kerner2000D,Kerner2000A},
~\cite{Kerner2001B,Kerner2002A,Kerner2002B,Kerner2002C,Kerner2002D,Kerner2003A,Kerner2004A}:
\begin{equation}
P^{\rm (B)}(q_{\rm sum})=P^{\rm (B)}_{\rm nucleus}(q_{\rm sum}),
\label{C1}
\end{equation}
where $P^{\rm (B)}(q_{\rm sum})$ is a
 flow rate dependence of the probability that during a
 given time interval $T_{\rm ob}$ traffic breakdown (F$\rightarrow$S transition)
occurs in free flow at the bottleneck, 
  $P^{\rm (B)}_{\rm nucleus}(q_{\rm sum})$   is the flow-rate dependence of the probability that during the time interval $T_{\rm ob}$   a nucleus for traffic breakdown occurs spontaneously in this free flow at the bottleneck.
A mathematical nucleation theory of traffic breakdown can be found 
 in~\cite{KKl2009B,KKl2006B,KKl2006C}.

   \subsection{Empirical proof of nucleation nature  of traffic breakdown  \label{Nuc_Nature_S}}
	
\begin{figure} 
  \begin{center}
\includegraphics[scale=.6]{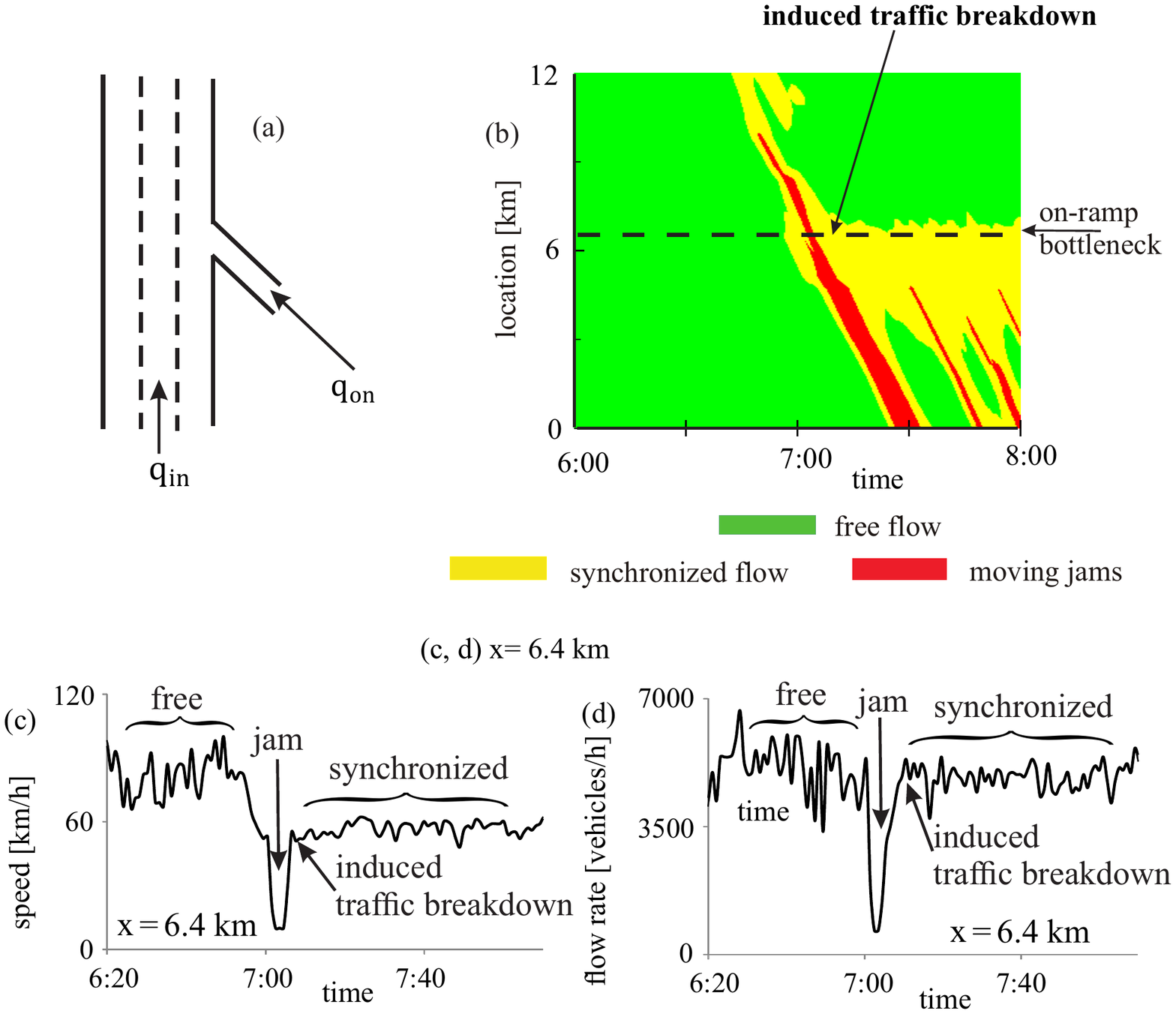}
 \end{center}
\caption{Empirical features of   induced traffic breakdown (induced F$\rightarrow$S transition)  at on-ramp 
bottleneck. Real field traffic
 data measured on three-lane freeway A5-South in Germany   
on March 22, 2001: (a) Schema of three-lane freeway with on-ramp bottleneck
that is the same as that in Fig.~\ref{OnRampSp150496} (a).
(b) Speed  data  measured with road detectors installed 
 along    road section;   data is
presented in space and time with   averaging   method described in Sec.~C.2 of~\cite{KernerRSch201310_Int}.
(c, d) 1-min average data for speed (c)
  and flow rate (d) as time functions  measured at location $x=$ 6.4 km.
	Free -- free flow, synchronized -- synchronized flow, jam -- wide moving jam.
}
\label{OnRampInd220301}
\end{figure}

	In real traffic flow, there always different drivers and
 vehicles. Therefore, to perform a clear empirical
   proof of the nucleation nature of traffic breakdown that is independent of  differences in vehicle and driver characteristics in free flow,
   we   distinguish between {\it empirical spontaneous} traffic breakdown 
 (Fig.~\ref{OnRampSp150496}) and 
 {\it empirical induced} traffic breakdown 
 (Fig.~\ref{OnRampInd220301})~\cite{KernerBook,KernerBook2,MiniReview,Kerner_Review2}:
\begin{enumerate}
\item [1.]	Empirical spontaneous traffic breakdown is defined as follows. If before traffic breakdown occurs at the bottleneck, there is free flow at the bottleneck as well as upstream and downstream in a neighborhood of the bottleneck, then traffic breakdown 
at the bottleneck is called spontaneous traffic breakdown (Fig.~\ref{OnRampSp150496}). 
\item [2.]	Empirical induced traffic breakdown at the bottleneck is traffic breakdown induced by the propagation of a spatiotemporal congested traffic pattern. This congested pattern has occurred earlier than the time instant of traffic breakdown at the bottleneck and at a different road location (for example at a downstream bottleneck) than the bottleneck location 
(Fig.~\ref{OnRampInd220301})\footnote{Here, the following question arises:
Why and when can traffic congestion occurring due to moving jam propagation to a bottleneck location be considered $\lq\lq$induced traffic breakdown"?
  Many researches consider
upstream propagation of   traffic congestion occurring initially at a downstream bottleneck that forces
congested traffic at an upstream   bottleneck as the {\it spillover} effect, not traffic breakdown.
 Indeed, when a wide moving jam
shown in Fig.~\ref{OnRampInd220301} (b) reaches  the bottleneck, 
the jam can  be considered  {\it spillover}: The jam forces
congested traffic at the bottleneck.
However, due to the upstream jam propagation,
the jam can   be considered as
 spillover {\it only}  during a short time interval:
When  the jam is far away upstream of the bottleneck, the jam does not force congested traffic
  at the bottleneck any more. 
  However,  we do not use the  term {\it spillover}. The reason for this has been explained 
	in~\cite{Waves}:
There can be  several  {\it qualitatively different} empirical spillover effects
      that should be considered separately each other. Only some of these different
	spillover effects as that shown in Fig.~\ref{OnRampInd220301} (b) can 
	be considered $\lq\lq$induced traffic breakdown". 
}. 
\end{enumerate}

 In contrast with Fig.~\ref{OnRampSp150496} (b--d) in which synchronized flow  has spontaneously
emerged at the on-ramp bottleneck,  in Fig.~\ref{OnRampInd220301} (b) synchronized flow  
has been induced  at the on-ramp bottleneck
due to
the propagation of a wide moving jam through  the bottleneck. 
In both cases (Figs.~\ref{OnRampSp150496} (c, d) and~\ref{OnRampInd220301} (c, d)),
synchronized flow resulting from the
breakdown at the bottleneck is self-maintained under free flow conditions downstream of the bottleneck.
Empirical  features of  synchronized flow resulting from the induced breakdown
(at  $t>$ 7:07 in Fig.~\ref{OnRampInd220301} (b--d)) 
are qualitatively identical to those found in synchronized flow resulting from empirical spontaneous traffic
breakdown (Fig.~\ref{OnRampSp150496} (b--d)). 	This means that
after the breakdown has occurred, characteristics of
synchronized flow that has been formed at the bottleneck do not depend on whether synchronized flow has occurred
due to   empirical spontaneous breakdown
(Fig.~\ref{OnRampSp150496} (c, d)) or due to   empirical induced  breakdown
 (Fig.~\ref{OnRampInd220301} (c, d)).  
	In particular, as in the case of empirical
spontaneous  breakdown, in the case of empirical
induced traffic breakdown the flow rate in synchronized flow resulting from the breakdown
 can be as high as the flow rate in free flow just before the breakdown has occurred (location $x=$ 6.4 km  in Figs.~\ref{OnRampSp150496} (d) and~\ref{OnRampInd220301} (d));
 this is in contrast with the wide moving jam within which the flow 
rate is very small~\cite{KernerBook,KernerBook2} 

In~\cite{KernerBook,KernerBook2,Waves},
it has been shown that in
real field traffic data
there can be found many different scenarios of empirical spontaneous and induced traffic breakdowns. All these scenarios show qualitatively the same nucleation nature of traffic breakdown.
	\begin{itemize}
\item
The  terms  {\it empirical spontaneous} traffic breakdown and {\it empirical induced}
 traffic breakdown  at a bottleneck
distinguish     different {\it  sources}  of  
    a  nucleus that occurrence leads to   traffic 
		breakdown~\cite{KernerBook,KernerBook2,Kerner1998E,Kerner1998C,Kerner1999B,Kerner1999D,Kerner2000D,Kerner2000A,Kerner2000B},
		
		\cite{Kerner2001A,Kerner2001B,Kerner2002A,Kerner2002B,Kerner2002C,Kerner2002D,Kerner2003A,Kerner2004A,Waves}:
\item
As found in~\cite{Waves},  the source of   empirical spontaneous breakdown is usually
  one of the waves in free flow that reaches a permanent speed disturbance localized at a highway 
	bottleneck\footnote{A wave acts as a nucleus for traffic breakdown only when the wave reaches the location of the permanent local speed
disturbance  in free flow at  the bottleneck~\cite{Waves}.  An explanation of this empirical results is as follows.
A decrease in the free flow speed
 within   the   permanent local speed disturbance becomes larger, when the wave 
 reaches the effective bottleneck location.
 This is because within the wave the flow rate is larger and the speed is smaller than outside the wave.
For this reason, the location of the permanent   disturbance determines 
 the effective location of the bottleneck at which traffic breakdown occurs.
In~\cite{Waves} has been found that 
 the physics of the occurrence of empirical nuclei for empirical traffic breakdown at highway bottlenecks
 is explained
by an interaction of a wave in free flow with a permanent   speed disturbance localized at 
 the effective location of the bottleneck. This is independent on whether there are trucks in traffic flow or not.}.
	 The source of   empirical induced breakdown is a localized moving congested pattern
	that reaches the location of the bottleneck (in Fig.~\ref{OnRampInd220301} (b), this localized pattern
  is
a wide moving jam).
\item The empirical evidence of induced F$\rightarrow$S transition is
the empirical proof of the metastability of free flow with respect to traffic breakdown (F$\rightarrow$S transition)  (Fig.~\ref{OnRampInd220301} (b--d)). This empirical
 proof  is independent
   on the degree of the heterogeneity of real vehicular traffic. 
 \end{itemize}
Indeed, the empirical evidence of induced traffic breakdown is the empirical proof that at a given flow rate at a bottleneck there can be one of two different traffic states at the bottleneck: (i) A traffic state related to free flow and (ii) a congested traffic state labeled as synchronized flow in Fig.~\ref{OnRampInd220301}(b). Due to the upstream propagation of a localized congested pattern, a transition from the state of free flow to the state of synchronized flow, i.e., traffic breakdown is induced. 
A more detailed discussion of the empirical proof of the nucleation nature of
real traffic breakdown can be found in~\cite{Waves}.

We can make the following conclusions:
\begin{itemize}
\item  The nucleation nature of real traffic breakdown at road bottlenecks is the fundamental empirical result that changes basically the theoretical fundamentals of transportation science.
\item For this reason, the empirical metastability of free flow with respect to the F$\rightarrow$S transition (traffic breakdown) at
a highway bottleneck
can be considered the  empirical fundament  of transportation science.
 \end{itemize}
 Therefore, rather than
  features of traffic congested patterns resulting from traffic breakdown, in this min-review   we analyze
the impact of the nucleation nature of real traffic breakdown on stochastic highway capacity and on characteristics  
of intelligent transportation systems (ITS).

\subsection{Failure of applications of classical traffic and transportation theories
for analysis of intelligent transportation systems (ITS) and traffic network optimization \label{Crit_Cla_S}}

      Generally accepted classical   traffic and transportation
  theories  have had a great impact on the understanding of many empirical traffic phenomena. 
  In particular, the Lighthill-Whitham-Richards (LWR) model~\cite{LW10,Richards10} and associated kinetic macroscopic
 traffic flow models as well as in traffic flow models of
 the General Motors (GM) model class, diverse driver behavioral characteristics
 related to real traffic have been discovered and 
incorporated~\cite{Haight1963A,Prigogine,Da,Pa1983,Cremer,Leutzbach,GartnerEd},

\cite{Wh2,Wid,New,Brackstone,Gazis,Chowdhury,Helbing,Nagatani},

\cite{Nagel,Mahnke,Bellomo,Maerivoet,Piccoli,Sch,TreiberD,Treiber}.

  However, as explained in~\cite{MiniReview,Kerner_Review2}, the
  classical   traffic and transportation
  theories
  have nevertheless failed by 
their applications in the real 
world. 
Even several decades of a very intensive effort to improve and validate 
  network optimization and control   models based on the classical traffic and transportation theories
   had have no success. Indeed,   there can be found   no  examples 
where on-line implementations of the network optimization models   based on these classical traffic and transportation theories 
could reduce congestion   in real traffic and transportation networks.

This failure of classical traffic and transportation theories is  explained 
 as 
 follows~\cite{MiniReview,Kerner_Review2}:
 \begin{itemize}
\item The LWR model~\cite{LW10,Richards10}   cannot show the nucleation nature of traffic breakdown at highway bottlenecks. For this reason, the LWR theory as well as  further theoretical
approaches based on
 the LWR theory, like
Daganzo's cell transmission model (CTM)~\cite{Daganzo1994,Daganzo1995Cell}, 
  N-curves~\cite{Daganzo2002,MD200210,Lin1997}, 
 and a macroscopic fundamental diagram (MFD)~\cite{Daganzo2007A,Daganzo2008A}
(see also  references in recent publications~\cite{LeclercqMFD2014,HollandMFD2015A,RakhaMFD2015A,XiongMFD2015A,SchweizMFD2015A})
are inconsistent with 
the nucleation nature of real traffic breakdown at road bottlenecks.
Applications of these approaches for an analysis    of the effect of
ITS on traffic flow, which are widely used by  many researchers,
  do lead
to invalid (and sometimes incorrect) conclusions about the ITS performance in real traffic.
\item In traffic flow models belonging to the  GM model class
introduced by Herman, Gazis,  Montroll, Potts,   Rothery, 
and Chandler~\cite{GH195910,Gazis1961A10,GH10,Chandler}, traffic breakdown is associated with the classical traffic flow instability
caused by a time  delay in vehicle deceleration due to driver reaction time. As has been firstly shown by Kerner and Konh{\"a}user~\cite{KK1993,KK1994},
  the classical traffic flow instability revealed by Herman, Gazis,  Montroll, Potts,   Rothery, 
and Chandler~\cite{GH195910,Gazis1961A10,GH10,Chandler}
   leads to a phase transition from free flow (F) to a moving jam(s) (J) 
	(called F$\rightarrow$J transition). The classical traffic flow instability 
has been incorporated in a huge number of traffic flow models; examples are the well-known 
optimal velocity (OV) model by  
Newell~\cite{New6110,Newell1963A,Newell1981}, a stochastic version of Newell's model~\cite{Newell_Stoch}, 
the Nagel-Schreckenberg (NaSch) cellular automaton (CA) model~\cite{NS10,B199810}, 
Gipps model~\cite{Gipps10,Gipps198610}, a stochastic model by Krau{\ss} {\it et al.}~\cite{Kra10,Kra_PhD10}, 
Payne's macroscopic model~\cite{ach_Pay197110,ach_Pay197910},
Whitham's model~\cite{Whitham10}, Wiedemann's model~\cite{Wid}, 
 the OV model
by Bando {\it et al.}~\cite{B1994A10,B1995A10,B1995B10}, Treiber's intelligent 
driver model~\cite{Helbing200010},  
 the Aw-Rascle macroscopic model~\cite{ach_Aw200010}, 
a full velocity difference OV model    by Jiang {\it et al.}~\cite{ach_Jiang2001A10},
a lattice model by Nagatani~\cite{fail_Nagatani1998A,fail_Nagatani1999A},
and a huge number of other traffic flow models. There is a huge number of other
traffic flow models belonging
   to the GM model; some of the models of the GM model class 
as well as results of their analysis can be found, for example,  in

\cite{Rakha2014A,Rakha2014B,Nagel1998A,Kuhne1984A,Kuehne1991,KKR1995A,KKS1995,KKS1996A,KKlK1997AA,HerKer1998,HermanAr198410,Tang10,Nishinari2001BA10,Sugiyama1997A10,Shvetsov199910,Tadaki2002AA10,Tomer200010,Tomer2002A10},

\cite{KestingPhil2010,Wilson2001A10,Wilson2008A,ChamberlayneHesham2012A,Berg2000A10,Berg2001A10,Brockfeld200310,Brockfeld200410,Fritzsche10,Hilliges199510,Hoogendoorn1999A10,Kurze1995A10,Lee199810,Lee1999A10,Lee2000A10,Lee2000B10,OssenHoogendoorn2011A,SchadschneiderSch1993A,SchadschneiderSch1995A,RickertNagel1996A,EsserSchreckenberg1997A,SchreckenbergUsadel1998A,SchreckenbergSanten1998B},

\cite{KnospeSchreckenberg1999A,KaumannSchreckenberg1999B,SchreckenbergBarlovic2001A,WahleSchreckenberg2002A,HuisingaSchreckenberg2002A,KnospeSchreckenberg2004C,KnospeSchreckenberg2004D,Hoogendoorn2010E,ach_Jiang2002C10,ach_Jiang2001C10,ach_Jiang2002A10,Sipahi2010E,Hino2013A,Sugiyama2013A,PengPhysA_2013A,Ngoduy2009A,GaoLi2007A,Neto2011A,TangLi2011A,SunPeng2011A,LiWang2011A,GeMeng2011A},

\cite{LvSong2011A,JinWang2011A,TianYuan2012A,Vasic2012A,ZhangLi2012A,ZhuZhang2012A,ZhangWu2012A,NaitoNagatani2012A,YangJin2013A,Peng2015AA,Peng2016BB,fail_Nagatani1999B,fail_Nagatani1999C,fail_Peng2013A,fail_Peng2011A,fail_Peng2011B,fail_Peng2011C,fail_Peng2012A,fail_Peng2012B,fail_Peng2015AB,fail_Peng2010A},

\cite{fail_Zhang2014A,fail_Zhang2015B,fail_Zhang2015AA,fail_ZhuZhang2012A,fail_GeCheng2008A,fail_Gupta2013A,fail_Gupta2014A,fail_Gupta2015A,fail_WangGao2014A,fail_WangGao2013A,fail_CaoShi2015A}

and reviews~\cite{Cremer,Leutzbach,GartnerEd,Brackstone,Gazis,Chowdhury,Helbing,Nagatani,Nagel,Maerivoet,Hesham10,Sch,TreiberD,Treiber,Saifuzzaman2015A}). 
An F$\rightarrow$J transition exhibits a nucleation nature; however,
 the nucleation nature of the F$\rightarrow$J transition contradicts to the nucleation nature of empirical traffic breakdown: Rather than an F$\rightarrow$J transition, real  
traffic breakdown is the F$\rightarrow$S transition. 
\item	Classical
models for dynamic traffic assignment, control and optimization of traffic and transportation networks, for example, which are
based on 
Wardrop's user equilibrium (UE) and system optimum (SO) principles~\cite{Wardrop}
(see, e.g.,~\cite{SchweizMFD2015A,Merchant1978A,Merchant1978B,Bell1992A10,Bell1995A10,Bell2000A10,Bell2002A10,Bell1997B10,Ceylana200510,Yang199710,Yang1998A10,Yang2000A10,Yang199810,Bell1991B10,Bell198310,Bell199110,Bell1995B10,Heydecker2005A,Daganzo1998A10},

~\cite{GonzalesDaganzo2012A,Mahmassani_87First,Mahmassani_Herman,Mahmassani_Peeta1993,PeetaMahmassani_1995B,PeetaMahmassani_1995C,Mahmassani_Kotzinos1997A,Mahmassani_Abdelfatah1999A,Mahmassani_Chiu2002A,Mahmassani_Huynh2002A,Mahmassani_Zhou2008A,Mahmassani_Zhang2008A},

\cite{Friesz1993A,Friesz1989A,Friesz1988A,Friesz1987A,Friesz1994A,Friesz1992A,Friesz1985A,Friesz1983B,Friesz1990B,Friesz2013A,Friesz2013B,Nesterov2003A,DoanUkkusuri2015A,MosheDTA2015A,YangJiangDTA2014A,Amirgholy2016A,XuKitthamkesorn2015A,AhipasMeskarian2015A,LiuWang2015A,LamLoWong2014A,ShenZhang2014A,CareyHumphreys2014A}

and references in
reviews~\cite{Sheffi1984,Bell1997A,Ran_96First,Mahmassani2001A,Rakha_Int1}),   failed
due to the metastability of empirical free flow 
 with respect to  an F$\rightarrow$S transition  at a network bottleneck. This is explained as 
follows~\cite{MiniReview}. The   objective of these and other classical approaches
 to dynamic traffic assignment, control, and
 optimization of a traffic network is the minimization of   travel times (and/or other  $\lq\lq$travel costs")
 in the network. However,
		this leads to a  considerable increase in the probability of traffic breakdown
		(F$\rightarrow$S transition)
		on some of the network links~\cite{MiniReview,BM}. The increase in the breakdown probability results in
		  the   deterioration of the performance of the
		traffic system. 
\end{itemize} 
Thus the classical traffic and transportation theories are not consistent
with the nucleation nature of empirical   traffic breakdown   at a highway 
bottleneck. This is due to the fact that the nucleation nature of empirical   traffic breakdown have been understood only during last 20 years. In contrast, the classical theoretical
works, in particular, made by
Wardrop~\cite{Wardrop}, Lighthill, Whitham, and Richards~\cite{LW10,Richards10},
Herman, Gazis,  Montroll, Potts,   Rothery, 
and Chandler~\cite{GH195910,Gazis1961A10,GH10,Chandler}, Newell~\cite{New6110,Newell1963A},
Kometani and Sasaki~\cite{KS,KS1959A,KS1961A}, Prigogine~\cite{Prigogine1961},  
Reuschel~\cite{Reuschel1950},
and Pipes~\cite{Pipes1953A}
 that are the  basic for
 the generally accepted fundamentals and methodologies 
of traffic and transportation theory have been introduced in the 
 1950s--1960s. 
These and other scientists whose ideas led to the classical fundamentals and methodologies of traffic and transportation theory 
 could not know the nucleation nature of  real traffic breakdown at road bottlenecks.

  Because
none of the classical traffic flow models can show the
F$\rightarrow$S transition in metastable free flow at the bottleneck, 
as already emphasized in~\cite{MiniReview},
the application of these classical models for an analysis    of the effect of
ITS
on traffic flow, which is generally accepted by traffic and transportation researchers,
  do lead
to invalid (and sometimes incorrect) conclusions about the ITS performance in real traffic. This criticism
is related to all ITS that affect traffic flow, for example,
on-ramp metering
(see, e.g.,~\cite{Arnold1998A,Pa199010,Pa199110,Pa199710,Pa2007B10,Pa200710,PaOn2015A,CarlsonPaOn2014A}),
variable speed limit control  (see, 
e.g.,~\cite{CarlsonPaOn2014A,Abdel-Aty2006A,AllabyH2006A,CarlsonPa2010A,CarlsonPa2010B,CarlsonPa2011A,CarlsonPa2013A,CastroMonzon2013A},

~\cite{ChenHegyi2014A,HegyiDeSchutter2005A,KhondakerKat2015A,KhondakerKat2015B}),
dynamic traffic assignment

 (see, e.g.,~\cite{SchweizMFD2015A,Merchant1978A,Merchant1978B,Bell1992A10,Bell1995A10,Bell2000A10,Bell2002A10,Bell1997B10,Ceylana200510,Yang199710,Yang1998A10,Yang2000A10,Yang199810,Bell1991B10,Bell198310,Bell199110,Bell1995B10,Heydecker2005A,Daganzo1998A10,GonzalesDaganzo2012A},

\cite{Mahmassani_87First,Mahmassani_Herman,Mahmassani_Peeta1993,PeetaMahmassani_1995B,PeetaMahmassani_1995C,Mahmassani_Kotzinos1997A,Mahmassani_Abdelfatah1999A,Mahmassani_Chiu2002A,Mahmassani_Huynh2002A,Mahmassani_Zhou2008A,Mahmassani_Zhang2008A},

\cite{Friesz1993A,Friesz1989A,Friesz1988A,Friesz1987A,Friesz1994A,Friesz1992A,Friesz1985A,Friesz1983B,Friesz1990B,Friesz2013A,Friesz2013B,Nesterov2003A,DoanUkkusuri2015A,MosheDTA2015A,YangJiangDTA2014A,Amirgholy2016A,XuKitthamkesorn2015A,AhipasMeskarian2015A,LiuWang2015A,LamLoWong2014A,ShenZhang2014A,CareyHumphreys2014A}

 and references in
reviews~\cite{Sheffi1984,Bell1997A,Ran_96First,Mahmassani2001A,Rakha_Int1})
and many other ITS-applications
 (e.g.,~\cite{AkamatzuISSTT2007,SzetoISSTT2007,ChowISSTT2007,Bell2009AISSTT,QianZhang2012B,ChenZhou2011A,ZhongSumalee2011A,Iryo2011A,SunCline2011A,NakayamaAISSTT,MounceAISSTT,KalafatosAISSTT,Acierno2012A,Cipriani2012A,WangMeng2013A,YangWang2012A},

\cite{JinWL2012A,QianZhang2012A,DoanUkkusuri2012A,XieW2012A,Luathep2011A,Keyvan-Ekbatani2012A,Haddad2012A,Keyvan-Ekbatani2013A,NieY2012A,ZhangMahmassani2013A,ChungYao2012A,KurzhanskiyPhil2010,Smith2013A,Maruyama2012A,Asakura2012A,Szeto2012A,Varia2013A,Kumar_2012A,YangJin2015A}).

Unfortunately,    this critical conclusion is also related to 
most studies of the effect of adaptive 
	cruise control (ACC)  and other vehicle systems on traffic flow, in particular,
considered and/or reviewed in
  Ref.~\cite{TreiberD,Treiber,KestingPhil2010,Shrivastava2002A,Zhou2005A,Kesting2008A,Ngoduy2012A,Ngoduy2013A,Shladover2012A,Shladover2002A,Suzuki2003A},
	
	\cite{Lin2009A,Martinez2007A,Papageorgiou2015A,Papageorgiou2015B,Wagner2015A,Friedrich2015A,Papageorgiou2015C,Papageorgiou2015DA,Benz2016A,LevinACC2015A}. In other words, because the classical generally accepted traffic flow models cannot show 
the empirical features of metastable free flow at highway bottlenecks, the application of these models and associated simulation tools for a study of the effect of automatic driving vehicles on   traffic flow
leads to incorrect conclusions. For this reason, such simulations (see, for 
example,
\cite{TreiberD,Treiber,KestingPhil2010,Shrivastava2002A,Zhou2005A,Kesting2008A,Ngoduy2012A,Ngoduy2013A,Shladover2012A,Papageorgiou2015A,Papageorgiou2015B,Papageorgiou2015C,Papageorgiou2015DA,Benz2016A,LevinACC2015A})
  cannot also be used for the development of reliable systems for   automatic driving vehicles. 
  This criticism is also related to the use of    well-known simulation tools based on the classical traffic flow theories 
like simulation tools VISSIM (Wiedemann model~\cite{Wid}) and SUMO (Krau{\ss} model~\cite{Kra10})
  (see, e.g.,~\cite{Wagner2015A,Friedrich2015A,Benz2016A}).

		\subsection{Breakdown minimization  (BM) principle
		for optimization of traffic and transportation networks \label{BM_Cla_S}}
    
		The  minimization of $\lq\lq$travel costs" in traffic and transportation networks, which
		is performed with classical
models for dynamic traffic assignment, control, and optimization in the 
networks (Sec.~\ref{Crit_Cla_S}),
		ignores the metastability of empirical free flow  
  with respect to
an F$\rightarrow$S transition at a bottleneck~\cite{MiniReview}. This
		can lead to  the   deterioration of the performance of the
		traffic system. For this reason,  in 2011 
		the author  introduced
    a breakdown minimization principle (BM principle) 
		for the optimization of traffic and transportation networks~\cite{BM}. 
		The   
  basis assumption used in the formulation of   the BM principle
	is the metastability of empirical free flow  
  with respect to
an F$\rightarrow$S transition at a bottleneck.
    The BM principle can be formulated as follows~\cite{BM}:
\begin{itemize}
\item 
 The BM principle states that the optimum of a traffic network with $N$
    bottlenecks is reached,
  when dynamic traffic optimization and/or control are performed in the network 
  in such a way that the probability for   occurrence of either induced or spontaneous 
	traffic breakdown 
  in at least one of the network bottlenecks during a given observation time reaches
   the minimum possible value. 
   \item The BM principle is equivalent to the maximization of the probability that 
	either induced or spontaneous traffic breakdown occurs at none of the network bottlenecks.
	 \end{itemize}
	A detailed consideration of the BM principle is a special subject that is out of scope of this mini-review.

	\subsection{Infinite number of stochastic highway capacities in three-phase theory \label{Infin_Int_S}} 

The empirical nucleation nature of real traffic breakdown
(F$\rightarrow$S transition) at road bottlenecks leads to 
the assumption of the three-phase  theory that at any time instant there are the infinite number of highway capacities~\cite{KernerBook,KernerBook2,MiniReview,Kerner2004A}. Indeed,
 in accordance with empirical results of Sec.~\ref{Nuc_Nature_S},
there should be
 a  range of the flow rate $q_{\rm sum}$ in free flow within which traffic breakdown can be induced
in free flow at a bottleneck.
Therefore,
within this flow rate range free flow is in a metastable state with respect to an F$\rightarrow$S transition. Empirical observations show that
this range of the flow rate is {\it limited}: When the flow rate $q_{\rm sum}$ in free flow
at a bottleneck  is smaller than some   minimum highway capacity   $C_{\rm min}$ no traffic breakdown can be induced at a bottleneck. On contrary, when the flow rate $q_{\rm sum}$ in free flow is larger than some   maximum highway capacity  $C_{\rm max}$, traffic breakdown should occur with probability $P^{\rm (B)}=1$. 

For these reasons, in the three-phase  theory it is assumed that
the metastability of free flow  
  with respect to
an F$\rightarrow$S transition at a 
bottleneck is realized
under the following conditions~\cite{KernerBook,KernerBook2,MiniReview,Kerner2004A}: 
\begin{equation}
C_{\rm min}\leq q_{\rm sum} < C_{\rm max}.
\label{meta_F}
\end{equation} 
  It is assumed in the three-phase  theory that when the flow rate $q_{\rm sum}$ satisfies conditions (\ref{meta_F}), traffic breakdown can be induced
at the bottleneck.
This explains why in three-phase traffic theory highway capacity of free flow at a bottleneck is defined through the empirical evidence of empirical 
induced traffic breakdown as follows:
\begin{itemize}
\item
{\it At any time instant}, there are the infinite
number of the flow rates $q_{\rm sum}$ in free flow at a bottleneck at which traffic breakdown can be induced at the bottleneck. These
 flow rates are the infinite number of the capacities of free flow at the bottleneck. The range of these capacities of free flow
at the bottleneck is limited by the minimum highway capacity $C_{\rm min}$ and the maximum highway capacity $C_{\rm max}$.
\end{itemize} 
Recently, the theoretical conclusion that at any time instant there are the infinite number of road capacities  have been generalized for a city bottleneck due to traffic 
signal~\cite{KernerCity2011B,Kerner_EPL,Kerner_2014}.

\subsection{About traffic flow models and some ITS-developments in the framework of three-phase theory}

The three-phase  theory is a qualitative theory that consists of several 
hypotheses~\cite{Kerner1998E,Kerner1998C,Kerner1999B,Kerner1999D,Kerner2000D,Kerner2000A,Kerner2000B,Kerner2001A},
\cite{Kerner2001B,Kerner2002A,Kerner2002B,Kerner2002C,Kerner2002D}. Some of these hypotheses
have been discussed  in Secs.~\ref{Basic_P_Sp_S} and~\ref{Infin_Int_S}. We can expect that  
a diverse variety of different mathematical approaches and models can be
 developed in the framework of the three-phase theory.
 
The Kerner-Klenov model introduced in 
   2002~\cite{KKl} was the
	first mathematical traffic flow model in the framework of the three-phase traffic 
  theory that
   can show and explain
    traffic breakdown  by the F$\rightarrow$S transition in the metastable free flow at the bottleneck as
    found in real field traffic data. Some months later,
		Kerner, Klenov, and Wolf   developed a CA model
		in the framework of the three-phase  
  theory (KKW CA model)~\cite{KKW}. Based on the
	KKW CA model,   the KKS (Kerner-Klenov-Schreckenberg) CA model~\cite{KKS2011}  
	and the KKSW (Kerner-Klenov-Schreckenberg-Wolf) CA model~\cite{KKHS2013,KKS2014A} have been developed for a more detailed 
	description of empirical features of real traffic.
	
	The Kerner-Klenov stochastic three-phase traffic flow model  
has   further been developed for different 
applications in~\cite{KKl2010A,Kerner_EPL,Kerner_2014,KKl2003A,KKl2009A,KKl2004A,Kerner_Hyp,Heavy,KKl2006AA,KernerH2006A,KernerH2006B,KernerH2006C},

\cite{KernerCon2005A,KernerCon2005B,KernerCon2005C,KernerCon2005D,KernerLim2007A,Brakemeier2007A,Brakemeier2009A,KernerCOM2015A,Kerner2003E,Kerner2003G,Kerner2015B}, 

in particular to simulate  
on-ramp metering~\cite{KernerCon2005A,KernerCon2005B,KernerCon2005C,KernerCon2005D}, 
speed limit control~\cite{KernerLim2007A}, 
traffic assignment~\cite{BM}, traffic at heavy bottlenecks~\cite{Heavy}  
and on moving bottlenecks~\cite{KKl2010A}, features of heterogeneous traffic flow consisting of different vehicles and drivers~\cite{KKl2004A}, 
jam warning methods~\cite{KRPKl2011A,KernerRSch201310_Int},	vehicle-to-vehicle (V2V) 
	communication~\cite{Brakemeier2007A,Brakemeier2009A,KernerCOM2015A}, 
	the ACC performance~\cite{Kerner2003E,Kerner2003G}, 
 traffic breakdown at signals in city traffic~\cite{KernerCity2011B,Kerner_EPL,Kerner_2014,KernerCity2014BC},
		over-saturated city traffic~\cite{KernerCity2014B}, 
		vehicle fuel consumption in traffic networks~\cite{HermannsCity2015,HemmerleCity2015} based on a cumulative vehicle acceleration~\cite{KernerCity2014D}.

		Over time several scientific groups have
used  hypotheses of the three-phase theory and 
developed new   models and new results in the framework of the three-phase  
 theory (e.g.,~\cite{Davis,Davis2004B9,Davis2006,Davis2008A9,Davis2014C,Lee_Sch2004A,Jiang2004A,Gao2007,Davis2006b,Davis2006d,Davis2006e,Davis2010,Davis2011},

\cite{Jiang2007A,Jiang2005A,Jiang2005B,Jiang2007C,Pott2007A,Li,Wu2008,Laval2007A8,Hoogendoorn20088,Wu2009,Jia2009,Tian2009,He2009,Jin2010,Klenov,Klenov2,Kokubo,LeeKim2011,Jin2011},

\cite{Neto2011,Zhang2011,Wei-Hsun2011IEEEA,Lee2011A,Tian2012,Kimathi2012B,Wang2012A,Tian2012B,Qiu2013,YangLu2013A,KnorrSch2013A,XiangZhengTao2013A},

\cite{Mendez2013A,Rui2014A,Hausken2015A,Tian2015A,Rui2015A,Rui2015B,Rui2015C,Rui2015D,Xu2015A,Davis2015A}). 

In particular, new models
 in the framework of  the three-phase theory have been introduced in the works
by Jiang, Wu, Gao, {\it et al.}~\cite{Jiang2004A,Gao2007}, Davis~\cite{Davis,Davis2006},
  Lee, Kim, Schreckenberg,  {\it et al.}~\cite{Lee_Sch2004A},  
Schreckenberg, Schadschneider,	Knorr, {\it et al.}~\cite{Pott2007A,KnorrSch2013A}, 
as well as  Tian, Treiber, Jia, Ma, Jiang,
{\it et al.}~\cite{Tian2015A,Rui2015A,Rui2015B}.  

Through the  use of traffic models  
	in the framework of the three-phase   theory, Davis has derived a number of novel results related to ITS applications, in particular, for
	cooperative vehicle control to avoid synchronized flow at 
	bottlenecks~\cite{Davis2006b},  for
	wirelessly connected ACC-vehicles~\cite{Davis2015A}, for predicting travel time to limit congestion~\cite{Davis2010}, 
	for realizing Wardrop equilibria with real-time traffic information~\cite{Davis2006e},
	for traffic control at highway bottlenecks~\cite{Davis2006d},
	and for on-ramp metering near the transition to the synchronous flow phase~\cite{Davis2006}.
	Davis was also one of the first who studied 
		the effect of ACC-vehicles on traffic flow with 
	a three-phase traffic flow model~\cite{Davis2004B9}.

	Recently, Jiang,  {\it et al.}~\cite{Rui2014A,Rui2015D}
	have performed    traffic experiments on an open section of a road
	that have revealed  new features of
	growing   disturbances of speed reduction
	in synchronized flow leading to S$\rightarrow$J transitions; additionally,
	Jiang's  microscopic experimental results have confirmed the hypothesis of  
		the three-phase theory about    two-dimensional (2D) states of traffic 
		flow in the flow--density plane 
		(or in the space-gap--speed plane)~\cite{Kerner1998E,Kerner1998C,Kerner1999B,Kerner1999D,Kerner2000D,Kerner2000A,Kerner2000B},
		\cite{Kerner2001A,Kerner2001B,Kerner2002A,Kerner2002B,Kerner2002C,Kerner2002D}\footnote{A more detailed consideration of the hypothesis of the three-phase theory 
		about  2D-states of traffic 
		flow~\cite{Kerner1998E,Kerner1998C,Kerner1999B,Kerner1999D,Kerner2000D,Kerner2000A,Kerner2000B},
		\cite{Kerner2001A,Kerner2001B,Kerner2002A,Kerner2002B,Kerner2002C,Kerner2002D}
		is out of scope of this mini-review and it can be found in the book~\cite{KernerBook}.}.

\subsection{Incommensurability of three-phase   theory and classical traffic-flow theories} 

Due to the criticism of classical traffic-flow theories made in Sec.~\ref{Crit_Cla_S}, a question arises:
\begin{itemize}
\item  May some of the classical traffic-flow theories be relatively easily adjusted to take into account the empirical evidence of the induced transition from free flow to synchronized flow?
  \end{itemize}

The explanation of traffic breakdown at a highway bottleneck by an F$\rightarrow$S transition in a metastable free flow at the bottleneck is the basic assumption of three-phase   theory~\cite{KernerBook,KernerBook2,Kerner1998E,Kerner1998C,Kerner1999B,Kerner1999D,Kerner2000D,Kerner2000A,Kerner2000B},
\cite{Kerner2001A,Kerner2001B,Kerner2002A,Kerner2002B,Kerner2002C,Kerner2002D,Kerner2003A,Kerner2004A}.
None of the classical traffic-flow theories   
incorporates metastable free flow with respect to an F$\rightarrow$S transition  at the bottleneck.
For this reason, 
the classical traffic-flow models cannot explain empirical
 induced F$\rightarrow$S  transition in   free flow at the bottleneck. However, the empirical induced 
F$\rightarrow$S  transition is the empirical evidence of the nucleation nature
of traffic breakdown (F$\rightarrow$S transition).
Therefore, the
three-phase   theory is incommensurable with all classical traffic flow theories~\cite{KKS2014A}. 
\begin{itemize}
\item  The existence in the three-phase   theory of the minimum highway capacity $C_{\rm min}$ 
 at which traffic breakdown (F$\rightarrow$S   transition) can still be induced at a highway bottleneck
 has no sense for the classical traffic and transportation theories.
  \end{itemize}
The term $\lq\lq$incommensurable"  has been introduced by Kuhn in his classical book~\cite{Kuhn_Str_Int}   to explain a paradigm shift in a scientific field.
This explains the title of Sec.~\ref{Int}.

\subsection{The objective of this mini-review \label{Rev_Ob_S}}

After publication of the mini-review~\cite{MiniReview} the author is often confronted with the following questions of many  researches: 
  \begin{description} 
\item [(i)]
There is the infinite number of capacities within some capacity range
in both 
  the classical understanding of stochastic highway capacity of free flow at
	highway bottlenecks~\cite{Brilon310,Brilon210,Brilon,BrilonISTTT2009,ach_Elefteriadou2014A,ach_ElefteriadouBook2014}
  and in the three-phase   
	theory~\cite{KernerBook,KernerBook2,Kerner1998E,Kerner1998C,Kerner1999B,Kerner1999D,Kerner2000D,Kerner2000A,Kerner2000B},
\cite{Kerner2001A,Kerner2001B,Kerner2002A,Kerner2002B,Kerner2002C,Kerner2002D,Kerner2003A,Kerner2004A}. How does the evidence of the nucleation nature of traffic breakdown resolve 
a highly controversial discussion in the field of the physics of vehicular traffic
associated with the understanding of stochastic highway capacity?
\item [(ii)] 
How to find the effect of automatic driving vehicles on stochastic highway capacity?
\item [(iii)]  What features should exhibit vehicle systems for automatic driving and other ITS  to  enhance     stochastic highway capacity, in particular, to decrease
the probability of traffic breakdown in  traffic and transportation networks?
\end{description} 
  Clearly, for a reliable analysis of the effect of automatic driving vehicles on traffic breakdown in vehicular traffic,
 traffic and transportation theories used for this analysis must firstly explain the nucleation 
nature of real traffic breakdown.

This explains the motivation for a new mini-review as follows.
In comparison with the mini-review~\cite{MiniReview},
the main new objectives of this article 
are as follows: 
\begin{itemize} 
\item  
We  
study the consequence of the failure of classical traffic-flow theories 
in the explanation of empirical traffic breakdown
  for an analysis of   the   effect of automatic driving vehicles  on traffic flow.
  We show that there is
a deep connection between the understanding of empirical stochastic highway capacity
and  a reliable analysis of the effect of automatic driving vehicles on traffic flow.
We explain why the classical   theories failed in the  understanding of  
stochastic highway capacity and why it is not possible to perform a reliable study of the effect of automatic driving vehicles and other  ITS on traffic flow with the use of the classical
traffic-flow theories.
\end{itemize} 
To reach these goals, in comparison with~\cite{MiniReview} the following 
new subjects will be considered in the mini-review:     
  \begin{description} 
\item [1.] We consider basic characteristics  
of 
traffic breakdown at a bottleneck 
 (Secs.~\ref{Theor_Pr_Sp_S}--\ref{Theor_P_Sp_S}).
\item [2.] 
We discuss why the effect of the
	cooperative driving through the use of V2V-communication 
	can increase the threshold flow rate for spontaneous traffic breakdown and the maximum capacity
	at a bottleneck (Sec.~\ref{Coop_Nature_S}).
\item [3.] We show how the discovering of the empirical nuclei for traffic breakdown at highway 
bottlenecks made in~\cite{Waves}
resolves  the controversial discussion about the nature of 
empirical stochastic highway capacity (Sec.~\ref{versus_Nature_S}). 
\item [4.]
We explain that driver behaviors assumed   in the three-phase theory to explain the empirical nucleation nature of traffic breakdown
	leads to the conclusion that human drivers
	do not exhibit  string instability in free flow, which is an important characteristic of the classical model
	of automatic driving vehicles as well as classical traffic flow models (Sec.~\ref{GM_S}).
\item [5.] 
With the use of simulations in the framework of the three-phase   theory,
we show that 
depending on parameters of automatic driving vehicles, 
  traffic flow that consists of a mixture of
human driving	and automatic driving vehicles ($\lq\lq$mixture traffic flow" for short)
	can  either decrease  or increase    the probability of traffic breakdown  at road 
	bottlenecks  (respectively, Secs.~\ref{Aut_Nature_S} and~\ref{P2_Aut_Human_S}). 
	\item [6.]	 
	We discuss briefly how dynamic motion rules of  future automatic driving vehicles can learn from
	the behavior of human driving vehicles in real traffic (Sec.~\ref{Aut_Hum_S}).
\end{description}

\section{Basic characteristics of 
   traffic breakdown in three-phase   theory   \label{Theor_Sp_S}}

In the three-phase theory, we distinguish the following
  basic characteristics of 
  traffic breakdown at a bottleneck~\cite{KernerBook,KernerBook2,Kerner1998E,Kerner1998C,Kerner1999B,Kerner1999D,Kerner2000D,Kerner2000A,Kerner2000B},
\cite{Kerner2001A,Kerner2001B,Kerner2002A,Kerner2002B,Kerner2002C,Kerner2002D,Kerner2003A,Kerner2004A}:
 \begin{itemize} 
\item The minimum highway capacity $C_{\rm min}$. 
\item The threshold flow rate for spontaneous traffic breakdown $q^{\rm (B)}_{\rm th}$. 
\item The maximum highway capacity $C_{\rm max}$.
\item A random time delay of traffic breakdown $T^{\rm (B)}$.
\item The probability of spontaneous  traffic breakdown $P^{\rm (B)}$.
\end{itemize}

 \subsection{Theoretical probability of spontaneous  traffic breakdown   \label{Theor_Pr_Sp_S}}
 
\begin{figure}
\begin{center}
\includegraphics*[width=13 cm]{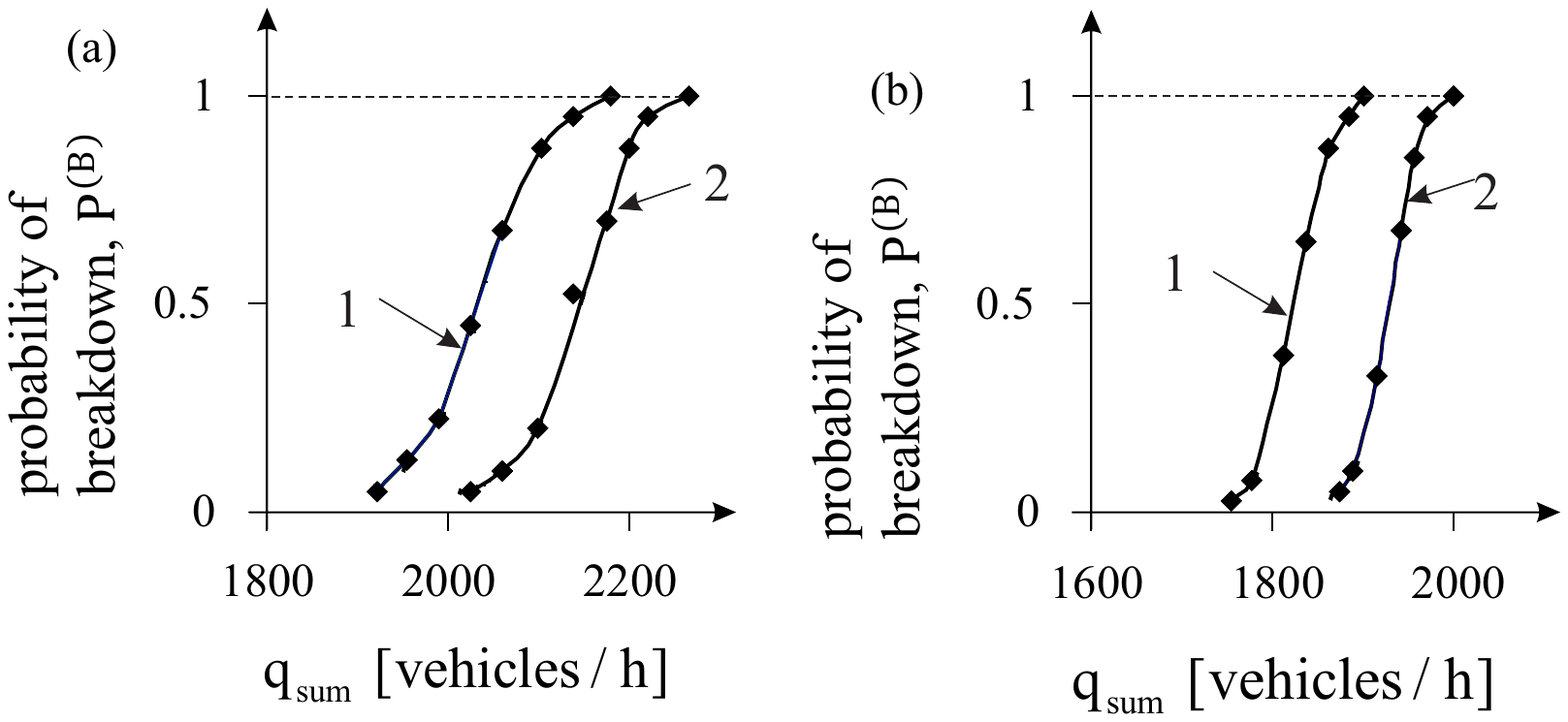}
\caption{Simulated flow-rate dependencies of probabilities of traffic breakdown at  an on-ramp bottleneck on single-lane road~\cite{KKW}: (a) On-ramp inflow rate $q_{\rm on}=$ 60 vehicles/h.
(b) On-ramp inflow rate $q_{\rm on}=$ 200 vehicles/h. Curves 1 and 2 in (a, b) are related, respectively, to two different values of
time interval for observing traffic flow $T_{\rm ob}=$ 30 (curves 1) and 15 min (curves 2).
$q_{\rm sum}=q_{\rm in}+q_{\rm on}$, where $q_{\rm in}$
is the flow rate in free flow on the main road upstream of the bottleneck.
 \label{Probability_KKW} } 
\end{center}
\end{figure}

  A theoretical  
     probability  of spontaneous traffic breakdown in the framework of the three-phase  
		theory was firstly found in 2002 (Fig.~\ref{Probability_KKW})~\cite{KKW}. 
		This flow-rate function of the breakdown probability
      is well  fitted by a function~\cite{KKW}:
\begin{equation}
P^{\rm (B)}(q_{\rm sum})=\frac{1}{1+ {\rm exp}[\alpha(q_{\rm P}-q_{\rm sum})]},
\label{Prob_For}
\end{equation}
where\footnote{Obviously, formula (\ref{Prob_For}) can be rewritten as follows (this equivalent form
for formula (\ref{Prob_For}) has been used in~\cite{KKW}; see caption to Fig.~18 of~\cite{KKW}):
       \begin{eqnarray}
P^{\rm (B)}(q_{\rm sum})=\left(1+ {\rm tanh}[\chi (q_{\rm sum}-q_{\rm P})] \right)/2,   \nonumber
 \end{eqnarray}
  where $\chi=\alpha/2$.} 
$\alpha$ and $q_{\rm P}$ are parameters\footnote{In particular, for an on-ramp bottleneck     
in  (\ref{Prob_For}) the flow rate $q_{\rm sum}=q_{\rm on}+q_{\rm in}$ is the flow rate downstream of the bottleneck,  $q_{\rm in}$
is the flow rate in free flow on the main road upstream of the bottleneck, and
  $q_{\rm on}$ is the on-ramp inflow rate
	that determines the bottleneck   strength; correspondingly,
 parameters $q_{\rm P}$ and $\alpha$ in (\ref{Prob_For})  
depend on $q_{\rm on}$.  In formula (\ref{Prob_For})  for an off-ramp bottleneck,  
 $q_{\rm sum}$     is the flow rate upstream of the off-ramp bottleneck, i.e., $q_{\rm sum}=q_{\rm in}$; the percentage of 
 vehicles leaving the main road to off-ramp at the off-ramp bottleneck $\eta=q_{\rm off}/q_{\rm in}$   determines the bottleneck   strength,
 $q_{\rm off}$ is the flow rate of vehicles leaving the main road to off-ramp at the off-ramp bottleneck;
 correspondingly,
 parameters $q_{\rm P}$ and $\alpha$ in (\ref{Prob_For})  
depend on $\eta$~\cite{KernerBook}.}. Formula (\ref{Prob_For}) is the result of the metastability of free flow with respect to an F$\rightarrow$S transition at the bottleneck incorporated in the KKW CA model~\cite{KKW}.
The theoretical  growing  flow-rate function for the breakdown probability
  (\ref{Prob_For})~\cite{KKW} explains
  empirical growing  flow-rate dependencies of the breakdown probability 
discovered firstly by Persaud {\it et al.}~\cite{Persaud1998B} and later found
in   other studies of real field traffic 
data~\cite{Lorenz2000A10,Brilon310,Brilon210,Brilon,BrilonISTTT2009,Koller2015A,Waves}.

  \subsection{Threshold flow rate for spontaneous traffic breakdown \label{Basic_expl_S}}
  
  For a qualitative analysis of conditions (\ref{C1}),
	(\ref{meta_F}), and the flow-rate function 
	of the breakdown probability $P^{\rm (B)}(q_{\rm sum})$ 
	(\ref{Prob_For})~\cite{KernerBook,KernerBook2,Kerner1998E,Kerner1998C,Kerner1999B,Kerner1999D,Kerner2000D,Kerner2000A,Kerner2000B},
\cite{Kerner2001A,Kerner2001B,Kerner2002A,Kerner2002B,Kerner2002C,Kerner2002D,Kerner2003A,Kerner2004A},  we recall firstly that 
	a nucleus  for traffic breakdown (F$\rightarrow$S transition) at a bottleneck is
a time-limited local disturbance in free traffic flow that occurrence leads to the breakdown.
Clearly that in free flow  there can be many time-limited local disturbances with different amplitudes, i.e., many different nuclei 
that lead to traffic breakdown at the bottleneck.   A   local disturbance with a minimum amplitude that   leads to the breakdown can be called a critical local 
disturbance. Respectively, the critical local disturbance determines a critical nucleus required for traffic breakdown at the bottleneck\footnote{It should be noted that in this qualitative consideration
we neglect the fact that in different $\lq\lq$realizations" of a study of traffic breakdown
in free flow at  given flow rates at the bottleneck there can be  different
 amplitudes of the critical nucleus that causes the breakdown. This means that in the reality 
for each given flow rates at the bottleneck
the amplitude of the critical disturbance (critical nucleus)   is a stochastic value. Thus,
the amplitude of the critical nucleus is determined with some probability only. \label{cr_nuc_foot}}.
	
We	can   assume that the larger is the flow rate in free flow, the smaller is the critical nucleus required to initiate spontaneous traffic breakdown
  in metastable free flow  at a bottleneck. Obviously, the probability of the occurrence of a small speed disturbance in free flow is considerably larger than the probability of the occurrence of a large disturbance.  This means that probability of the spontaneous occurrence of a nucleus for traffic breakdown $P^{\rm (B)}_{\rm nucleus}(q)$  is an increasing function of the flow rate. In accordance with (\ref{C1}), this explains the increasing flow rate function of the breakdown probability  $P^{\rm (B)}(q_{\rm sum})$  (\ref{Prob_For}).

\begin{figure}
\begin{center}
\includegraphics*[width=11 cm]{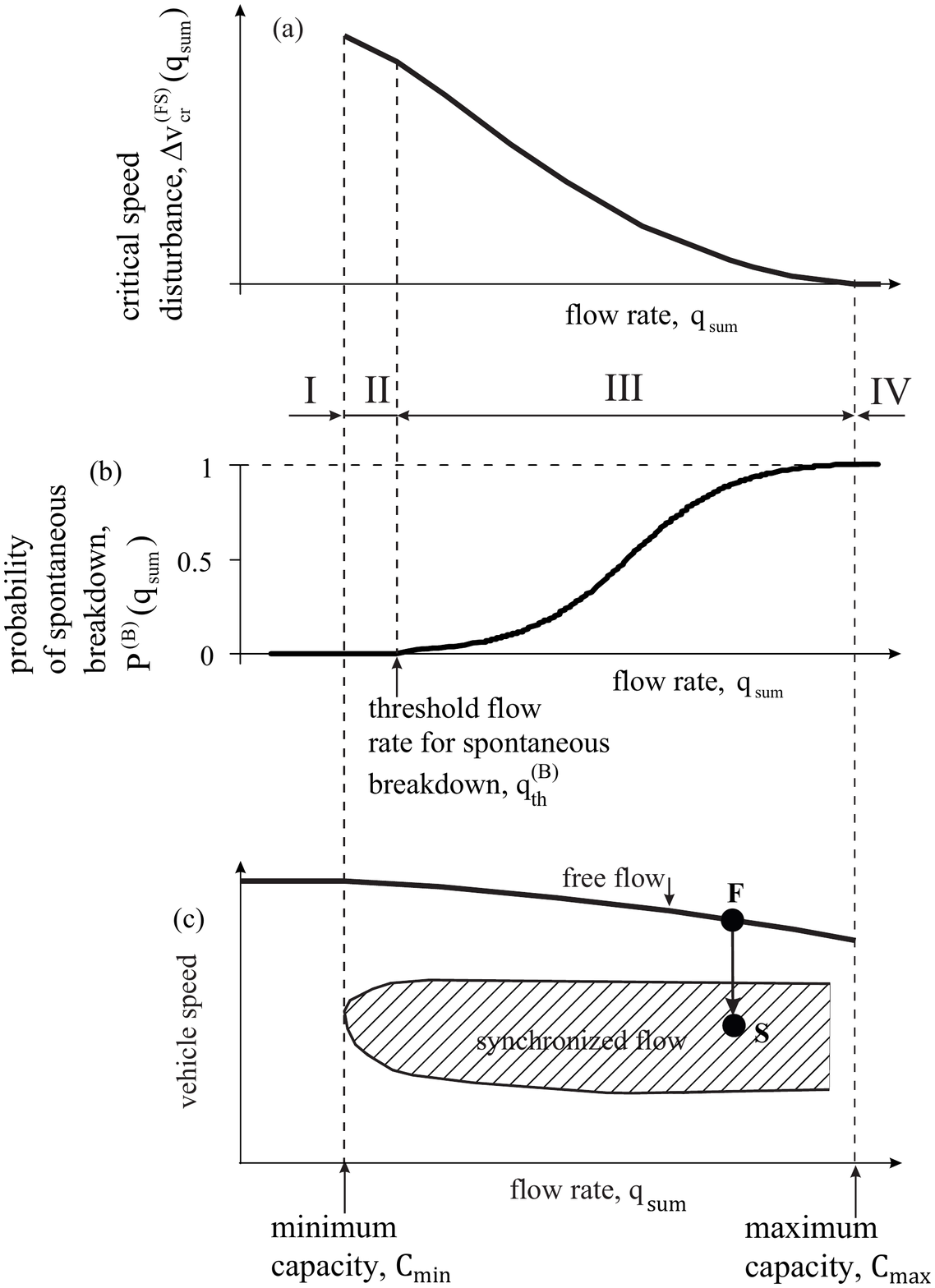}
\caption{Qualitative explanations of condition
 (\ref{C1})~~\cite{KernerBook,KernerBook2,Kerner1998E,Kerner1998C,Kerner1999B,Kerner1999D,Kerner2000D,Kerner2000A,Kerner2000B},
\cite{Kerner2001A,Kerner2001B,Kerner2002A,Kerner2002B,Kerner2002C,Kerner2002D,Kerner2003A,Kerner2004A}: 
(a) Qualitative flow rate dependence of function  $\Delta v^{\rm (FS)}_{\rm cr}(q_{\rm sum})$. (b) Breakdown probability  $P^{\rm (B)}(q_{\rm sum})$   related to formula (\ref{Prob_For}). 
(c) Z-characteristic for traffic breakdown; arrow in (c) denotes an F$\rightarrow$S transition
from a free flow state F to a synchronized flow state S.
 \label{Emp_Breakdown_nuc} } 
\end{center}
\end{figure}

As an example of this  qualitative discussion of condition (\ref{C1}), 
we assume that a nucleus for traffic breakdown at a highway bottleneck
occurs in free flow that is associated with a time-limited critical local decrease in the speed in an initial free flow at the bottleneck denoted by 
$\Delta v^{\rm (FS)}_{\rm cr}$ (Fig.~\ref{Emp_Breakdown_nuc}(a)). The larger the flow rate in free flow, the smaller should be the value $\Delta v^{\rm (FS)}_{\rm cr}$
 that initiates the breakdown in free flow. The related decreasing function  $\Delta v^{\rm (FS)}_{\rm cr}(q_{\rm sum})$, which is qualitatively shown in Fig.~\ref{Emp_Breakdown_nuc}(a), has indeed been found in simulations with Kerner-Klenov stochastic microscopic three-phase traffic flow 
model~\cite{KKl}\footnote{In accordance with explanations given in footnote~\ref{cr_nuc_foot}, at a given flow rate $q_{\rm sum}$
there can be  different critical amplitudes of a time-limited   local decrease in the speed in an initial free flow at the bottleneck that causes the breakdown: The function
$\Delta v^{\rm (FS)}_{\rm cr}(q_{\rm sum})$ shown in Fig.~\ref{Emp_Breakdown_nuc}(a) is related to {\it a given probability} of
   the amplitude of the critical nucleus $\Delta v^{\rm (FS)}_{\rm cr}$.}.

  At very small flow rates   
  \begin{equation}
  q_{\rm sum}<C_{\rm min},
\label{I}
\end{equation}
no traffic breakdown can occur (flow rate range I in Fig.~\ref{Emp_Breakdown_nuc}). Therefore, there can be no a time-limited speed disturbance in free flow at the bottleneck that can be a nucleus for the breakdown.
In flow rate range II,
satisfying condition
  \begin{equation}
C_{\rm min}\leq q_{\rm sum}<q^{\rm (B)}_{\rm th},
\label{II}
\end{equation}
there can be   a time-limited speed disturbance in free flow at the bottleneck that can be a nucleus for the breakdown; in (\ref{II}),
  $q^{\rm (B)}_{\rm th}$  is   a threshold flow rate for 
  spontaneous traffic 
  breakdown (Fig.~\ref{Emp_Breakdown_nuc} (a, b)). 
  We assume that under condition   (\ref{II})
 a very large value  $\Delta v^{\rm (FS)}_{\rm cr}$
 (large nucleus) is required for the breakdown, so   that the probability of {\it spontaneous occurrence} of such very large speed disturbance in free flow during a given time interval
 $T_{\rm ob}$
  is zero, i.e., $P^{\rm (B)}_{\rm nucleus}=0$. In accordance with (\ref{C1}), the probability of spontaneous traffic breakdown  
  \begin{equation}
P^{\rm (B)}=0.
\label{P_II}
\end{equation}
  This means that in this case 
  only induced traffic breakdown is possible. In flow rate range III
  satisfying condition
  \begin{equation}
q^{\rm (B)}_{\rm th}\leq q_{\rm sum} <C_{\rm max},
\label{III}
\end{equation}
due to the increase in the flow rate $q_{\rm sum}$ the value  $\Delta v^{\rm (FS)}_{\rm cr}(q_{\rm sum})$ required for the breakdown should decrease sharply. 
 For this reason, the probability of the spontaneous occurrence of such a speed disturbance 
during the time interval
 $T_{\rm ob}$ can satisfy conditions $0<P^{\rm (B)}_{\rm nucleus}(q_{\rm sum})<1$ and,
 therefore, in accordance with (\ref{C1}), the probability of spontaneous breakdown  
  \begin{equation}
0<P^{\rm (B)}(q_{\rm sum})<1.
\label{P_III}
\end{equation}
This consideration explains the sense of the threshold  flow rate $q^{\rm (B)}_{\rm th}$:
At $q_{\rm sum}=q^{\rm (B)}_{\rm th}$ the breakdown probability is very small but it is still larger than 
zero.
  This definition of the threshold flow rate $q^{\rm (B)}_{\rm th}$ for spontaneous traffic breakdown
 explains why in (\ref{II}) we assume that at any flow rate $q_{\rm sum}<q^{\rm (B)}_{\rm th}$
the probability of spontaneous  traffic breakdown during the time interval
 $T_{\rm ob}$ is equal to zero (\ref{P_II}).   
  In flow rate range IV satisfying condition
  \begin{equation}
  q_{\rm sum} \geq C_{\rm max},
\label{IV}
\end{equation}
the value  $\Delta v^{\rm (FS)}_{\rm cr}(q_{\rm sum})$ required for the breakdown is as small as zero; therefore, during the time interval
 $T_{\rm ob}$
the probability of the nucleus occurrence  $P^{\rm (B)}_{\rm nucleus}=1$,
 and,
 therefore, in accordance with (\ref{C1}), the probability of spontaneous breakdown  
  \begin{equation}
 P^{\rm (B)}(q_{\rm sum})=1.
\label{P_IV}
\end{equation}

   \begin{figure}
\begin{center}
\includegraphics*[width=10 cm]{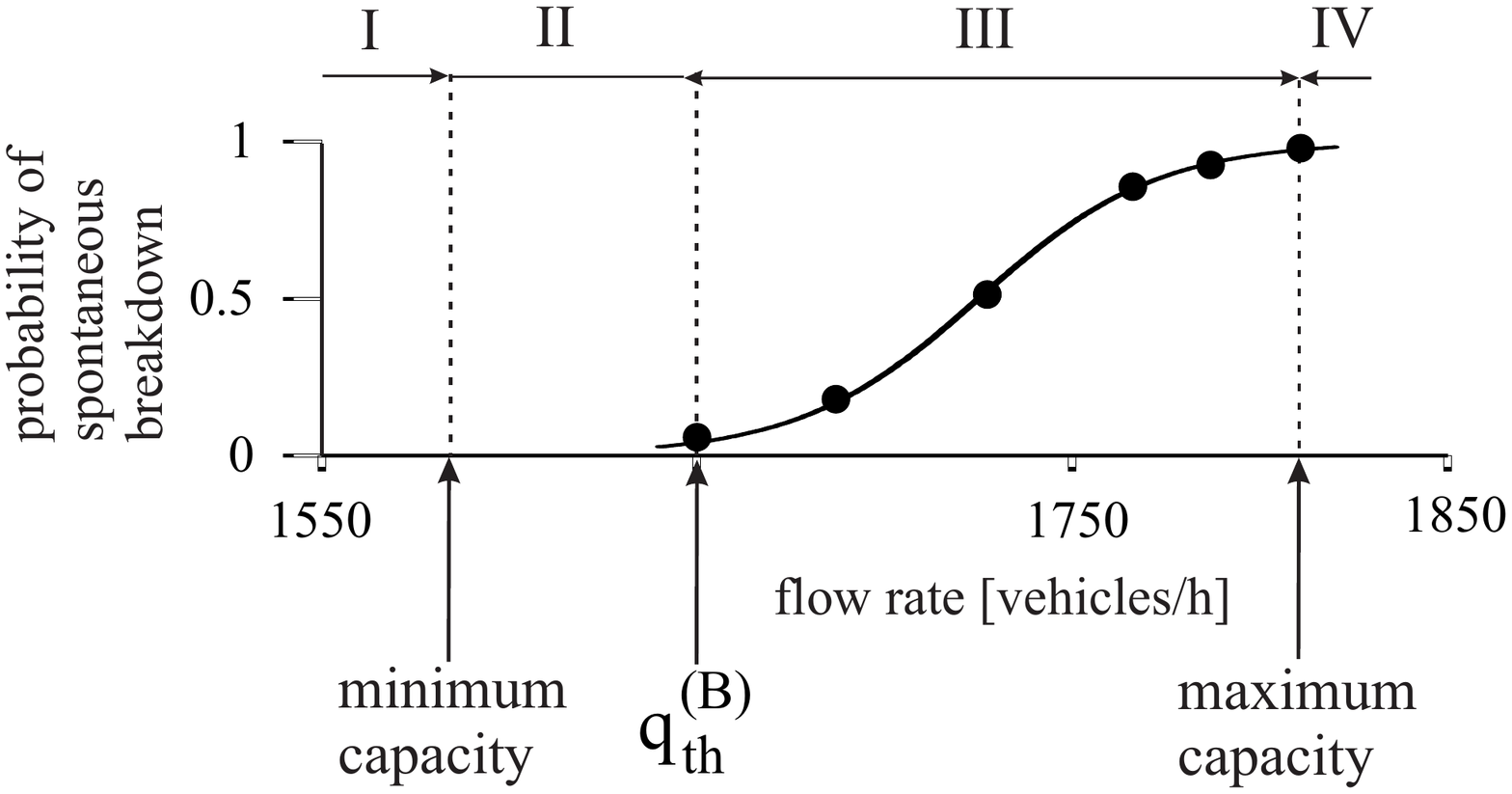}
\end{center}
\caption[]{Simulations of flow rate ranges I--IV (\ref{I})--(\ref{IV}) for spontaneous and induced traffic breakdowns at on-ramp bottleneck:  Typical
dependence of the probability of spontaneous traffic breakdown $P^{\rm (B)}(q_{\rm sum})$ on the flow rate   $q_{\rm sum}$  calculated with the use
of the KKSW CA model for different
flow rates $q_{\rm in}$ in free flow upstream of the bottleneck at given   $q_{\rm on}=$ 400 vehicles/h.
Function $P^{\rm (B)}(q_{\rm sum})$ is related to formula (\ref{Prob_For}). 
  Calculated minimum capacity
$C_{\rm min}=$ 1585 vehicles/h. Calculated threshold flow rate $q^{\rm (B)}_{\rm th}\approx 1650$ vehicles/h.
Calculated maximum capacity $C_{\rm max}=$ 1810 vehicles/h. $T_{\rm ob}=$ 30 min.
Taken from~\cite{KKS2014A}.
 }
\label{Induced_KKSW_1}
\end{figure}

        \begin{figure}
\begin{center}
\includegraphics*[width=13 cm]{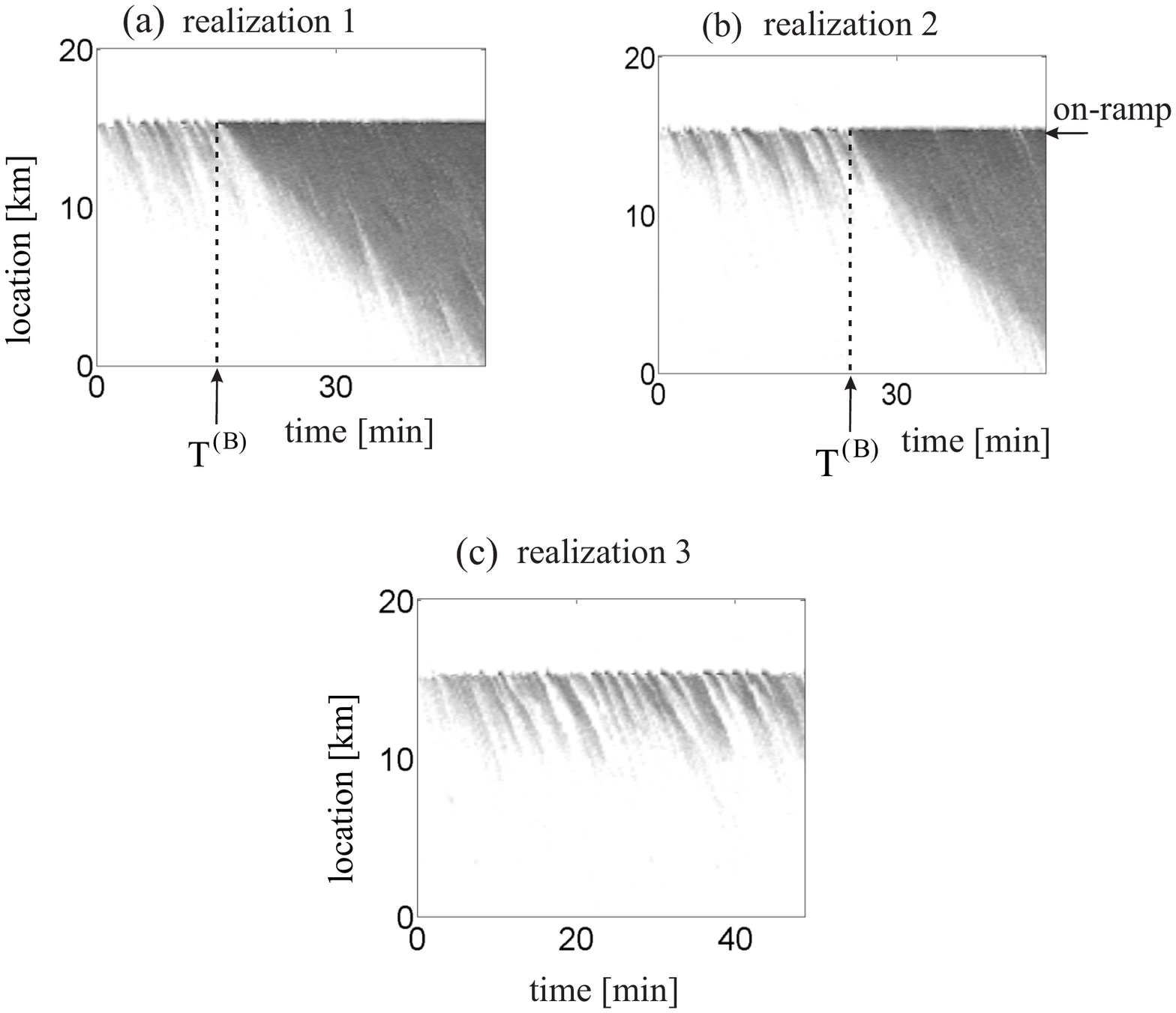}
\end{center}
\caption[]{Spontaneous traffic breakdown in   KKSW CA model under conditions (\ref{III}):    Speed   data
presented  by regions with variable shades of gray   (in white regions the speed
is  equal to or higher than 120  km/h, in black regions the speed is zero). 
Three different simulation realizations 1--3 in (a--c) are related to $P^{\rm (B)}=$ 0.775 calculated at $q_{\rm in}=$ 1364 vehicles/h and  $q_{\rm on}=$ 400 vehicles/h;
different realizations are related to different initial
values (at $t=0$) of random model variables.
 Location of the on-ramp bottleneck is $x_{\rm on}=$ 15 km.
   Taken from~\cite{KKS2014A}.
  }
\label{Probability_KKSW}
\end{figure}

This qualitative discussion of the basic assumption of the three-phase  theory
  (\ref{C1})~\cite{KernerBook,KernerBook2,Kerner1998E,Kerner1998C,Kerner1999B,Kerner1999D,Kerner2000D,Kerner2000A,Kerner2000B},
\cite{Kerner2001A,Kerner2001B,Kerner2002A,Kerner2002B,Kerner2002C,Kerner2002D,Kerner2003A,Kerner2004A}
  is confirmed by numerical simulations made with the use of 
 the KKSW CA model  presented in Figs.~\ref{Induced_KKSW_1} and~\ref{Probability_KKSW}~\cite{KKS2014A}.
 Simulations show that under condition (\ref{I}) no traffic breakdown can be either induced or occur spontaneously.
Under condition (\ref{II}), traffic breakdown  can be induced at a bottleneck {\it only}
(Fig.~\ref{Induced_KKSW_1}). 
Under condition (\ref{III}),
the breakdown can either be induced  or it can occur spontaneously (Figs.~\ref{Probability_KKSW} (a, b)). 
  There is a time delay for spontaneous traffic breakdown  $T^{\rm (B)}$ that is a random value
 for different simulation realizations (compare values of $T^{\rm (B)}$ for spontaneous traffic breakdown occurring in two different simulation realizations 1 and 2
 that are related to the same set of the flow rates   $q_{\rm on}$ and  
 $q_{\rm in}$  in Fig.~\ref{Probability_KKSW} (a, b)).  
 Because under condition (\ref{III}) we get $0<P^{\rm (B)}(q_{\rm sum})<1$,
 in some of the simulation realizations no spontaneous breakdown occurs during a chosen observation time for traffic variables $T_{\rm ob}$, as shown for realization 3   in Fig.~\ref{Probability_KKSW} (c).
 In this case,
 traffic breakdown can nevertheless be induced during the observation time $T_{\rm ob}$.   
Under condition (\ref{IV}), traffic breakdown occurs
 spontaneously in each of the simulation realizations, i.e., the breakdown probability $P^{\rm (B)}=1$.

\subsection{Accuracy of determination of
characteristics of probability of traffic breakdown   
			\label{Real_Prob_S}}

As in a study of the flow-rate dependence of the
  empirical breakdown probability  $P^{\rm (B)}$~\cite{Persaud1998B,Koller2015A},
   in numerical calculations of the breakdown probability  $P^{\rm (B)}(q_{\rm sum})$~\cite{KKW} only a finite number $N$ of simulation realizations (runs) can be made
   for the calculation of the value $P^{\rm (B)}$ for each given flow rate $q_{\rm sum}$. In accordance 
   with the definition of the threshold flow rate $q^{\rm (B)}_{\rm th}$, the smallest
   value of the breakdown probability $P^{\rm (B)}$ that is still larger than zero reaches at the flow rate $q_{\rm sum}=q^{\rm (B)}_{\rm th}$. Thus,  the smallest value of the breakdown probability satisfying condition (\ref{P_III})
    is given by formula
   \begin{eqnarray}
    P^{\rm (B)}\mid_{ q_{\rm sum}=q^{\rm (B)}_{\rm th}} = \frac{1}{N}.  
    \end{eqnarray}
  In other words, the larger the number $N$ of simulation realizations (runs), the more exactly the threshold flow rate $q_{\rm sum}=q^{\rm (B)}_{\rm th}$
  can be calculated. Correspondingly, an approximate value of  $C_{\rm max}$ is found from   condition  
   \begin{eqnarray}
    P^{\rm (B)}\mid_{ q_{\rm sum}=C_{\rm max}} = \frac{N-1}{N}.  
    \end{eqnarray}

  \subsection{Dependence of    characteristics of breakdown probability on heterogeneity of traffic flow  
			\label{Theor_P_Sp_S}}
  
\begin{figure}
\begin{center}
\includegraphics*[width=12 cm]{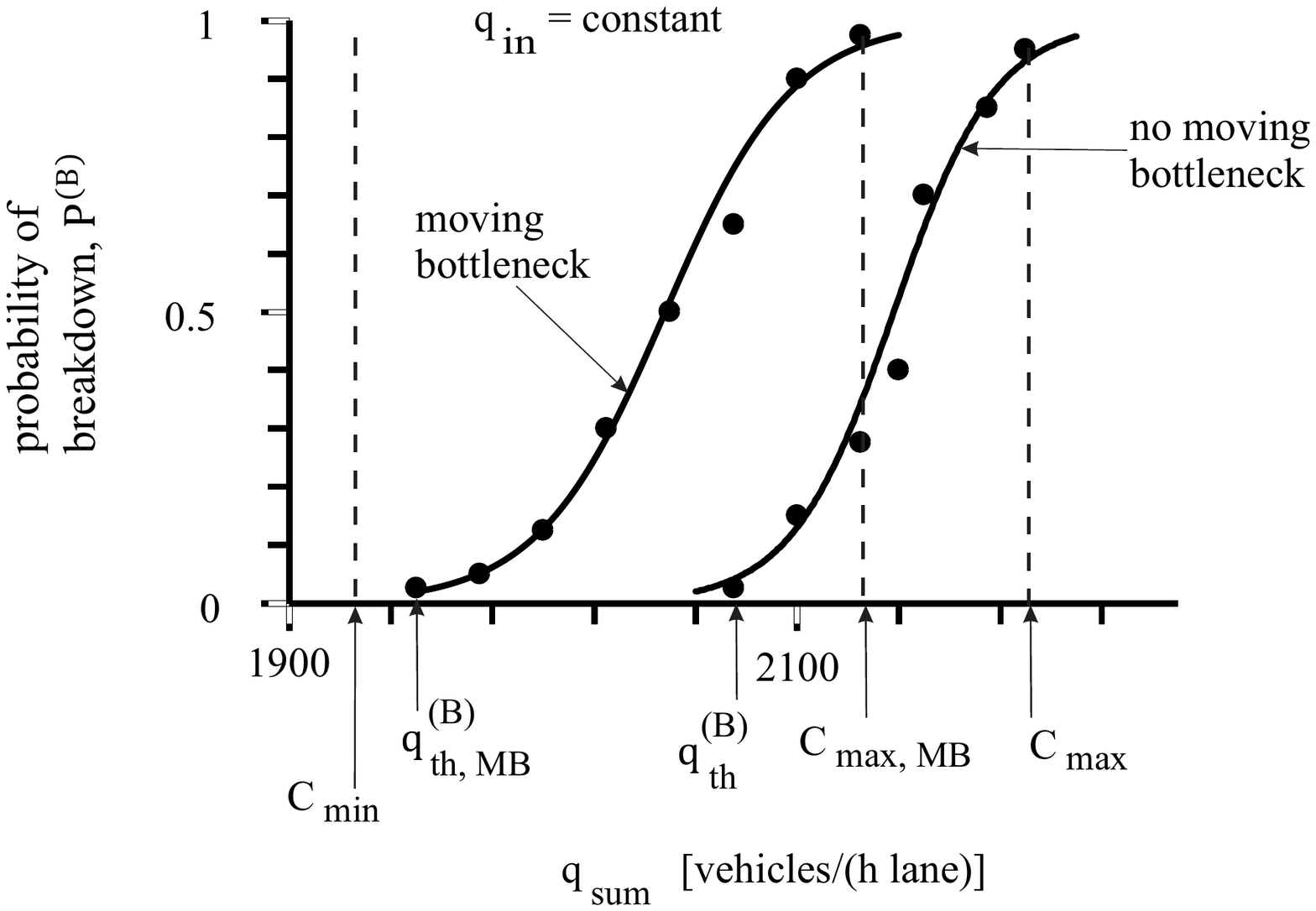}
\caption{Simulations of probabilities of traffic breakdown in traffic flows with the moving bottleneck and without moving bottlenecks   with Kerner-Klenov
stochastic three-phase traffic flow model on a two-lane road with on-ramp bottleneck:
 Probabilities of traffic breakdown $P^{\rm (B)}(q_{\rm sum})$ in traffic flow of identical drivers and vehicles  with moving bottleneck (left curve
labeled by $\lq\lq$moving bottleneck") and without moving bottlenecks (right curve
labeled by $\lq\lq$no moving bottleneck") as  functions of the flow rate downstream
of the bottleneck $q_{\rm sum}=q_{\rm in}+0.5 q_{\rm on}$  that is varied through the change in the
on-ramp inflow rate $q_{\rm on}$ at constant 
$q_{\rm in}=$ 1800 vehicles/h.
Functions $P^{\rm (B)}(q_{\rm sum})$ are described by formula (\ref{Prob_For}). $C_{\rm min}\approx$ 1925 vehicles/(h lane).
Taken from~\cite{Waves}.
 \label{Model_id} } 
\end{center}
\end{figure}

	  As shown in real field traffic data~\cite{Waves},   empirical spontaneous
	traffic breakdown  is caused by the propagation of
  a wave in free flow through 
	a permanent speed disturbance localized at a bottleneck. The wave  is associated with
	slow moving vehicles in heterogeneous traffic flow. The slow moving vehicles
  can be considered $\lq\lq$moving bottleneck". Therefore,
   to 
  explain the basic importance of  the function $P^{\rm (B)}(q_{\rm sum})$   for transportation science, we consider    simulations of   traffic breakdowns  in a heterogeneous
   traffic flow   with a moving bottleneck  and
   in traffic flow   without moving bottlenecks   (Fig.~\ref{Model_id}).

   We have found that
  at any chosen set of the  flow rates $(q_{\rm in}, \ q_{\rm on})$,
    the moving bottleneck results in the increase
   in the probability for traffic breakdown in comparison with the breakdown probability in traffic flow without moving bottlenecks.
   In other words, the function $P^{\rm (B)}(q_{\rm sum})$ for traffic flow with the moving bottleneck is shifted
  to the left in the flow rate axis in comparison with the function $P^{\rm (B)}(q_{\rm sum})$    
for traffic flow without moving bottlenecks (Fig.~\ref{Model_id}).
  
	We have also found that the moving bottleneck results in the decrease in both the maximum capacity $C_{\rm max}$  and the threshold flow rate for spontaneous traffic breakdown $q^{\rm (B)}_{\rm th}$. To distinguish   the cases of traffic flows with the moving bottleneck and without
 moving bottlenecks, 
 we denote the maximum capacity  $C_{\rm max}$ and the threshold flow rate   $q^{\rm (B)}_{\rm th}$
 for traffic flow with the moving bottleneck  by $C_{\rm max, \ MB}$ and $q^{\rm (B)}_{\rm th, \ MB}$, respectively (Fig.~\ref{Model_id}).
   
   In contrast with values $C_{\rm max}$ and $q^{\rm (B)}_{\rm th}$, the minimum capacity $C_{\rm min}$ does not depend on whether there is a moving bottleneck in traffic flow or not.
   This is because at the given flow rate $q_{\rm in}$
    the minimum capacity $C_{\rm min}$ shown in Fig.~\ref{Model_id}  determines
		the smallest flow rate $q_{\rm on}$
   at which traffic breakdown can still be {\it induced} at the bottleneck.

	\subsection{Effect of cooperative vehicles on breakdown probability \label{Coop_Nature_S}}
	
	Wireless vehicle communication, which is the basic technology for ad-hoc vehicle networks, is one of the most important scientific fields of  ITS. This is because there can be many   applications of ad-hoc vehicle networks for so-called $\lq\lq$cooperative driving" in vehicular traffic, including systems for danger warning, adaptive assistance systems, traffic information, improving of traffic flow characteristics, 
	etc. (see, e.g.,~\cite{Sklar1997A,ChenSch2007A,Choffnes2005A,Hartenstein2010A,Sepulcre2011A,Schmidt-Eisenlohr2007A,Torrent-Moreno2004A,Torrent-Moreno2009A,Torrent-Moreno2006A}). 
	As emphasized in Sec.~\ref{Crit_Cla_S},
	the classical traffic flow models used in all known traffic simulation tools cannot explain features of traffic breakdown (F$\rightarrow$S transition) as observed in real measured traffic data. Therefore, these traffic simulation tools cannot be used for the reliable analysis of the effect of an ad-hoc network on traffic flow. For this reason, we review briefly only results of 
	the application of traffic flow models in the framework of the three-phase theory
	for the analysis of the effect of cooperative vehicles on traffic flow.
	
	 Davis was one of the first who  applied hypotheses of the three-phase theory
	for simulations of the cooperative merging at an on-ramp bottleneck to study the prevention of
	the formation of synchronized flow at the bottleneck~\cite{Davis2006b}. In~\cite{Brakemeier2007A}, 
	for a study of cooperative driving we  have
	developed
  $\lq\lq$a united network model" that incorporates both a model of ad-hoc network and  
	the Kerner-Klenov three-phase microscopic stochastic traffic flow 
	model.

	  \begin{figure}
\begin{center}
\includegraphics*[width=11 cm]{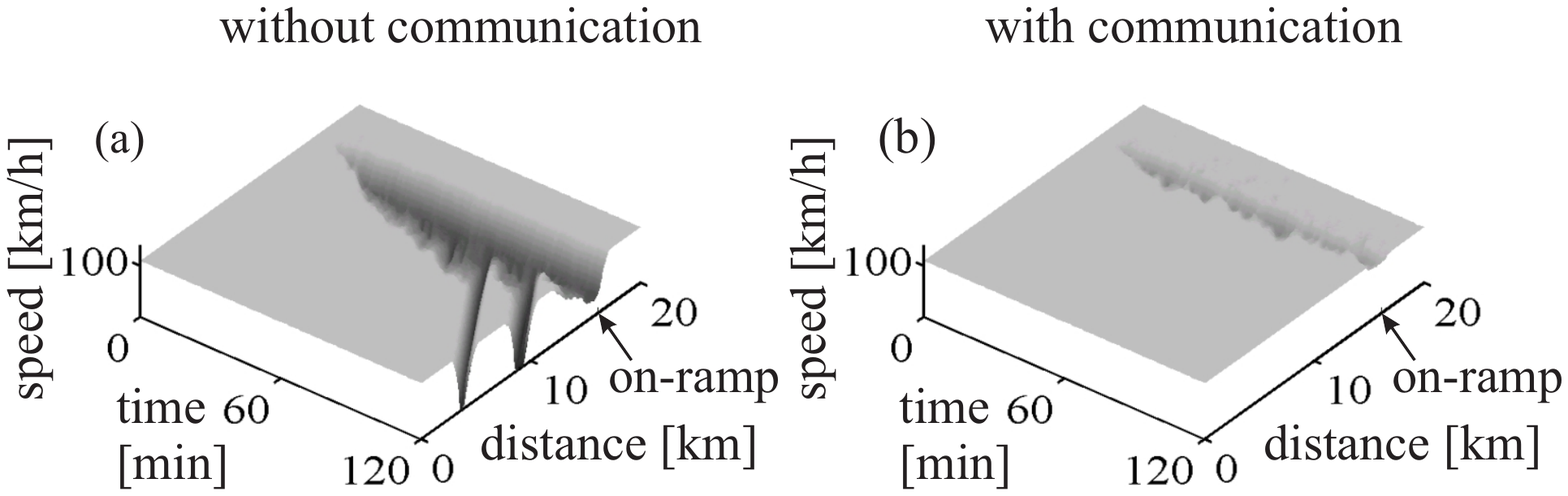}
\end{center}
\caption[]{Simulations of prevention of traffic breakdown at on-ramp bottleneck through
vehicle communication:  Speed in time and space without communication (a)
 and with vehicle communication (b). 
Taken from~\cite{Brakemeier2009A}.
 \label{ad-hoc5} }
\end{figure}

\begin{figure}
\begin{center}
\includegraphics*[width=11 cm]{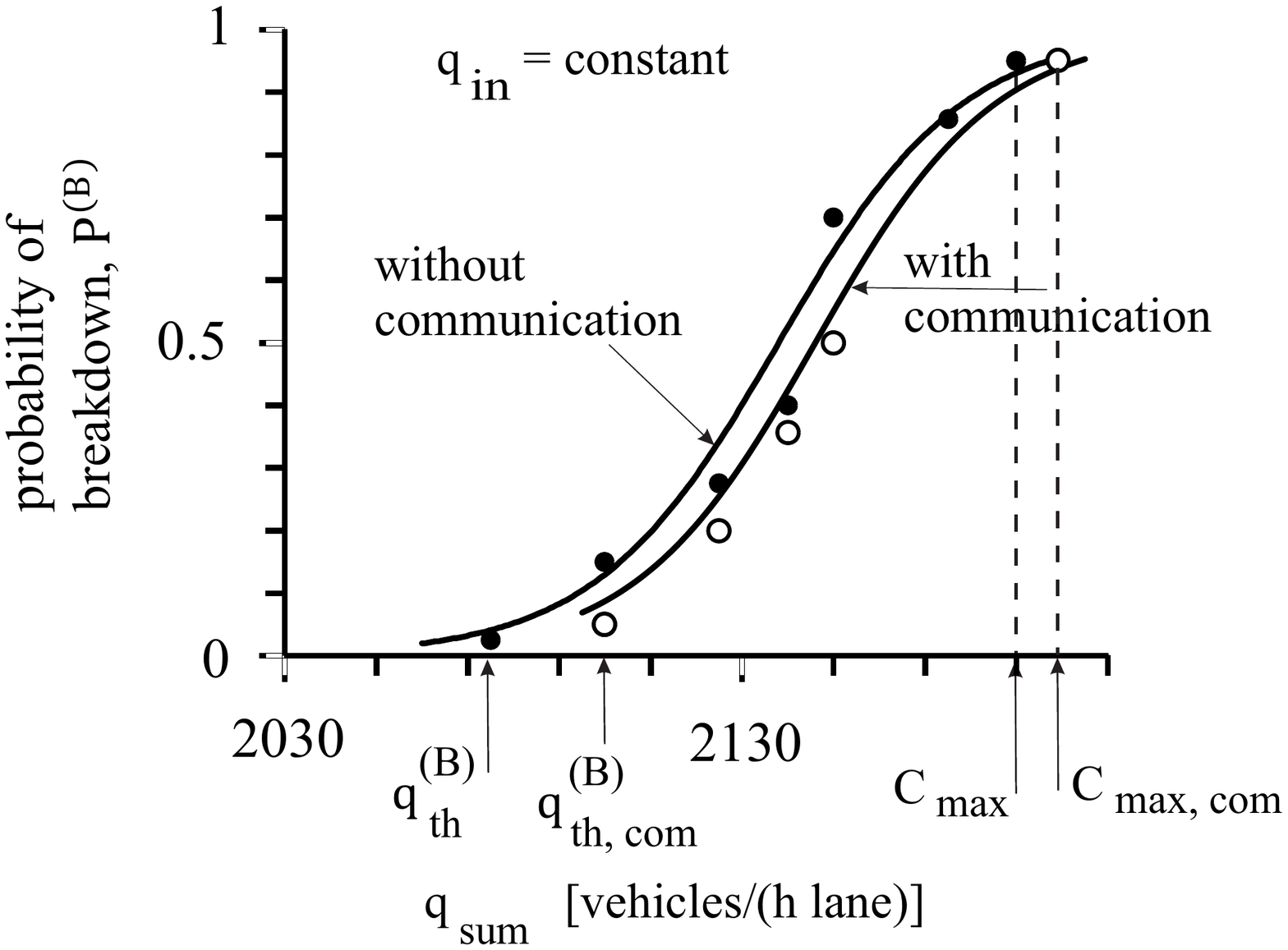}
\caption{Simulations of flow-rate functions of 
probabilities of traffic breakdown $P^{\rm (B)}(q_{\rm sum})$  made
   with Kerner-Klenov
stochastic three-phase   model on a two-lane road with on-ramp bottleneck: 
Traffic flows with V2V-communication  (left curve
denoted by $\lq\lq$with communication") and without V2V-communication (right curve
denoted by $\lq\lq$without communication").
The flow rate downstream
of the bottleneck $q_{\rm sum}=q_{\rm in}+0.5 q_{\rm on}$    is varied through the change in the
on-ramp inflow rate $q_{\rm on}$ at constant 
$q_{\rm in}=$ 1800 vehicles/h. $C_{\rm min}\approx$ 1925 vehicles/(h lane).
Function $P^{\rm (B)}(q_{\rm sum})$ denoted by $\lq\lq$without communication" is the same as that 
denoted by $\lq\lq$no moving bottleneck" in Fig.~\ref{Model_id}.
Functions $P^{\rm (B)}(q_{\rm sum})$ are described by formula (\ref{Prob_For}).
\label{Model_com} } 
\end{center}
\end{figure}

In simulations
	based on this model~\cite{Brakemeier2009A}   
(Fig.~\ref{ad-hoc5}), 
  we assume that   vehicles moving
 in the on-ramp lane send  a message for neighbor vehicles moving in the right road lane
when the vehicle intends to merge  from the on-ramp onto the main 
road.
We assume that the following vehicle in the right lane   
  increases a time headway for the vehicle merging to satisfy
	a safe gap between the merging vehicle and the following vehicle in the right lane
	of the main road. Simulations show 
	that in comparison with the case in which
  no V2V-communication is applied and traffic breakdown occurs (Fig.~\ref{ad-hoc5} (a))
  this change in driver behavior of communicating vehicles decreases   disturbances in free flow at the bottleneck. This results in the prevention of traffic breakdown
  (Fig.~\ref{ad-hoc5} (b)).

The effect of the cooperative merging of vehicles from the on-ramp onto the main road on the
breakdown probability is shown in Fig.~\ref{Model_com}.
We can see that the cooperative merging (that is the same as that in 
	Fig.~\ref{ad-hoc5} (b))
	leads to  slight increases in the threshold flow rate for spontaneous traffic breakdown
	(spontaneous F$\rightarrow$S transition) as well as   in the maximum capacity
	denoted for the case of the cooperative merging by
	$q^{\rm (B)}_{\rm th, \ com}$ and $C_{\rm max, \ com}$
	in comparison with the threshold flow rate $q^{\rm (B)}_{\rm th}$ and the maximum capacity $C_{\rm max}$
	related to traffic flow without V2V-communication.
	The minimum capacity
	$C_{\rm min}$ is not affected through the cooperative merging.
	The physics of these results is as follows. The cooperative merging decreases the mean amplitude of speed disturbances occurring
	through the vehicle merging from the on-ramp onto the main road.
	For this reason, the cooperative merging increases the threshold flow rate for the spontaneous
	traffic breakdown and
	the maximum capacity. The  minimum capacity
	$C_{\rm min}$ determines the threshold for the {\it induced} traffic breakdown.
	The cooperative merging does not affect  the possibility of the induced traffic breakdown,
	therefore, no change in $C_{\rm min}$ has been found.

 A study of the effect of moving bottlenecks
 and cooperative systems   on the breakdown probability presented 
above (Figs.~\ref{Model_id} and~\ref{Model_com})   shows
that the characteristics of the breakdown probability  $P^{\rm (B)}(q_{\rm sum})$ are
 the basis characteristics of  traffic flow. Therefore, we can make the following conclusions.
 \begin{itemize}
   \item
 A proof of whether    ITS improve traffic flow or not can be made through   an analysis of whether there is a shift of the flow-rate function of the breakdown probability  
  $P^{\rm (B)}(q_{\rm sum})$ to the larger flow rates $q_{\rm sum}$ or not.
 The larger this shift  is, the more the effect of ITS on the increase in stochastic highway capacity. 
 \end{itemize}
  We will use this criterion for an analysis of 
 the effect of automatic driving vehicle on traffic flow in Secs.~\ref{Aut_Nature_S} and~\ref{P2_Aut_Human_S}.
However, before we consider the nature of empirical stochastic highway capacity.

\section{Stochastic highway capacity:
Classical theory  versus 
  three-phase   theory \label{versus_Nature_S}} 

\subsection{Classical understanding of stochastic highway capacity  \label{Cl_Cap_S}}

 The   classical understanding of   highway capacity
  is defined through the occurrence of traffic breakdown at a bottleneck: The highway capacity is equal to the flow rate in an initially free flow at the bottleneck at which traffic breakdown is observed at 
  the 
	bottleneck~\cite{May,Hall1986A,Hall1987A,Hall1991A,Hall1992A10,Elefteriadou1995A,Persaud1998B,Lorenz2000A10},
	\cite{Brilon310,Brilon210,Brilon,BrilonISTTT2009,ach_Elefteriadou2014A,ach_ElefteriadouBook2014}.  
  
  As above-explained, 
  empirical traffic breakdown exhibits the probabilistic nature
   (Sec.~\ref{Ach_S})~\cite{Elefteriadou1995A,Persaud1998B,Lorenz2000A10},
	~\cite{Brilon310,Brilon210,Brilon,BrilonISTTT2009,ach_Elefteriadou2014A,ach_ElefteriadouBook2014}.
 Respectively, Brilon~\cite{Brilon310,Brilon210,Brilon,BrilonISTTT2009,ach_Elefteriadou2014A,ach_ElefteriadouBook2014}
  has introduced the following definition for  stochastic highway capacity that is in agreement with   
 the classical capacity definition:
 Brilon's stochastic highway capacity $C$  is equal to the flow rate  $q_{\rm sum}$ in an initially free flow at the bottleneck at which traffic breakdown is observed at the bottleneck. At any time instant, there is a particular value of stochastic capacity of free flow at the bottleneck. 
However, as long as free flow is observed at the bottleneck, this particular value of stochastic capacity cannot be measured. Therefore, stochastic capacity is defined through a capacity distribution function
$F^{\rm (B)}_{\rm C}(q_{\rm sum})$~\cite{Brilon310,Brilon210,Brilon,BrilonISTTT2009}:
\begin{equation}
F^{\rm (B)}_{\rm C}(q_{\rm sum})=p(C\leq q_{\rm sum}),
\label{A1}
\end{equation}  										 
where  $p(C\leq q_{\rm sum})$ is the probability that stochastic highway capacity  $C$ is equal to or smaller than the flow rate  $q_{\rm sum}$ in free flow at a highway bottleneck. 

Thus the basic theoretical assumption of the classical understanding of stochastic highway capacity is that traffic breakdown is observed at a time instant $t$ at which the flow rate $q_{\rm sum}$  reaches the capacity $C(t)$. This means that the flow rate function of the probability of traffic breakdown $P^{\rm (B)}(q_{\rm sum})$ should be determined by the capacity distribution function
 $F^{\rm (B)}_{\rm C}(q_{\rm sum})$~\cite{Brilon310,Brilon210,Brilon,BrilonISTTT2009}:  
\begin{equation}
P^{\rm (B)}(q_{\rm sum})=F^{\rm (B)}_{\rm C}(q_{\rm sum}).
\label{A2}
\end{equation}  									 

It must be noted that the breakdown probability function   found in empirical observations  is the {\it empirical evidence}. However, condition (\ref{A2}) is a {\it theoretical hypothesis} only. 
This is because in contrast with the breakdown probability function $P^{\rm (B)}(q_{\rm sum})$, the 
capacity distribution function $F^{\rm (B)}_{\rm C}(q_{\rm sum})$ cannot be
 measured~\cite{Brilon310,Brilon210,Brilon,BrilonISTTT2009,ach_Elefteriadou2014A,ach_ElefteriadouBook2014}.

  \begin{figure}
\begin{center}
\includegraphics*[width=12 cm]{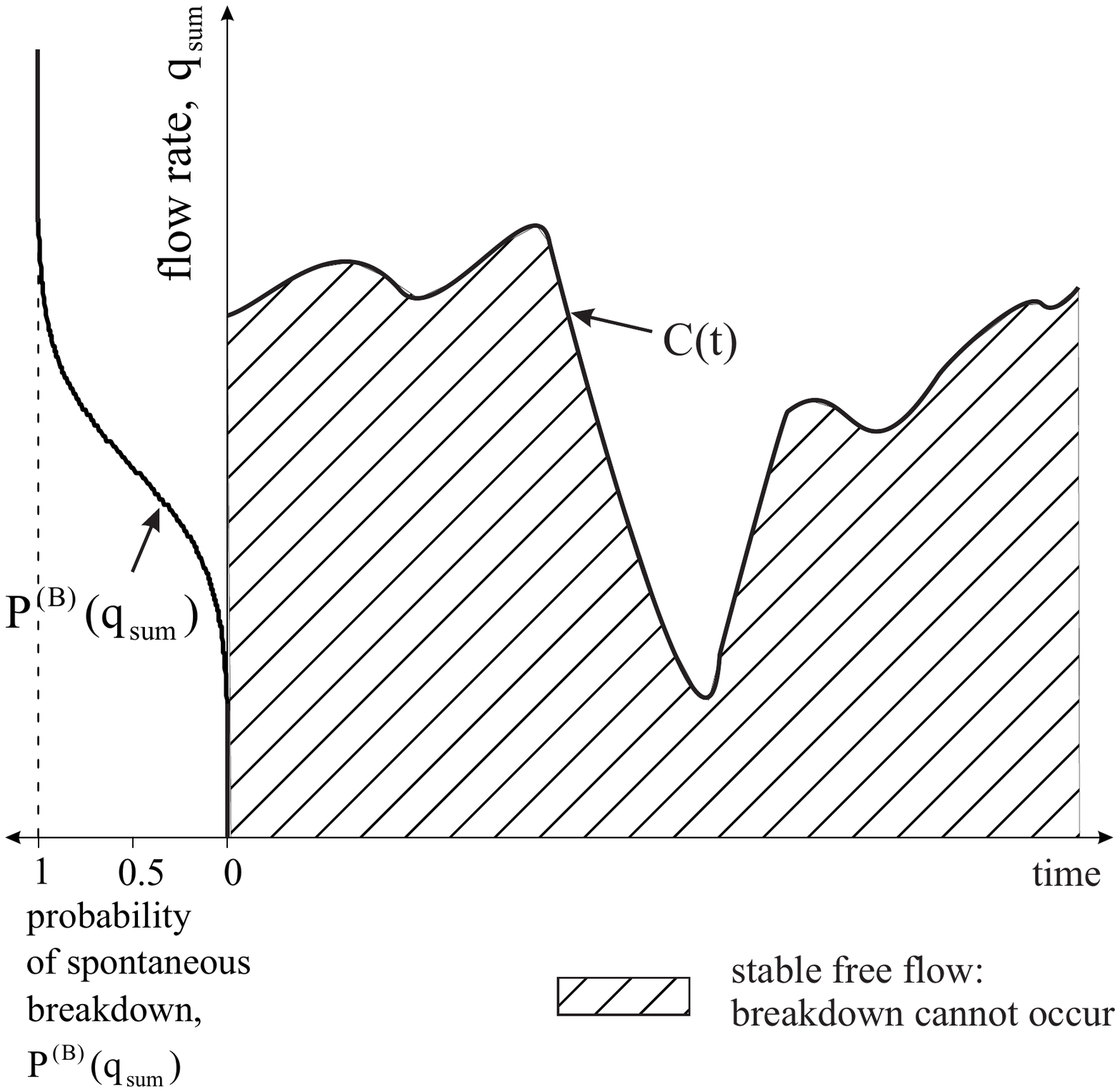}
\caption{Qualitative explanation of Brilon's  stochastic highway capacity (\ref{A2})
 of free flow at a highway bottleneck: Figure left is the flow rate function of the probability of the spontaneous breakdown  $P^{\rm (B)}(q_{\rm sum})$.
Figure right is a hypothetical fragment of stochastic highway capacity $C(t)$ over time.
 \label{Capacity_short_sw1} } 
\end{center}
\end{figure} 

This understanding of stochastic capacity of free flow at a bottleneck  is
qualitatively illustrated in Fig.~\ref{Capacity_short_sw1}, right, where
  we show a qualitative hypothetical fragment of the time-dependence of stochastic capacity $C(t)$. Left in Fig.~\ref{Capacity_short_sw1}, a qualitative flow rate dependence of the probability of spontaneous traffic breakdown  $P^{\rm (B)}(q_{\rm sum})$ is shown.    
  The stochastic capacity $C(t)$ can stochastically change over time (Fig.~\ref{Capacity_short_sw1}). 
It is often assumed that a stochastic behavior of highway capacity is associated 
with a stochastic change in traffic parameters over time. Examples of the traffic parameters, which can indeed be stochastic time-functions in real traffic, are weather, mean driver's characteristics (e.g., mean driver reaction time), share of long vehicles, etc.

Below we will show that the classical understanding of stochastic highway capacity (\ref{A2}), which is generally accepted in transportation research community,
contradicts basically the nucleation nature of traffic breakdown in real traffic (Sec.~\ref{Nuc_Nature_S}).

\subsection{Characteristics of stochastic highway capacities in three-phase   theory \label{Inf_Nature_S}}

  \begin{figure}
\begin{center}
\includegraphics*[width=14 cm]{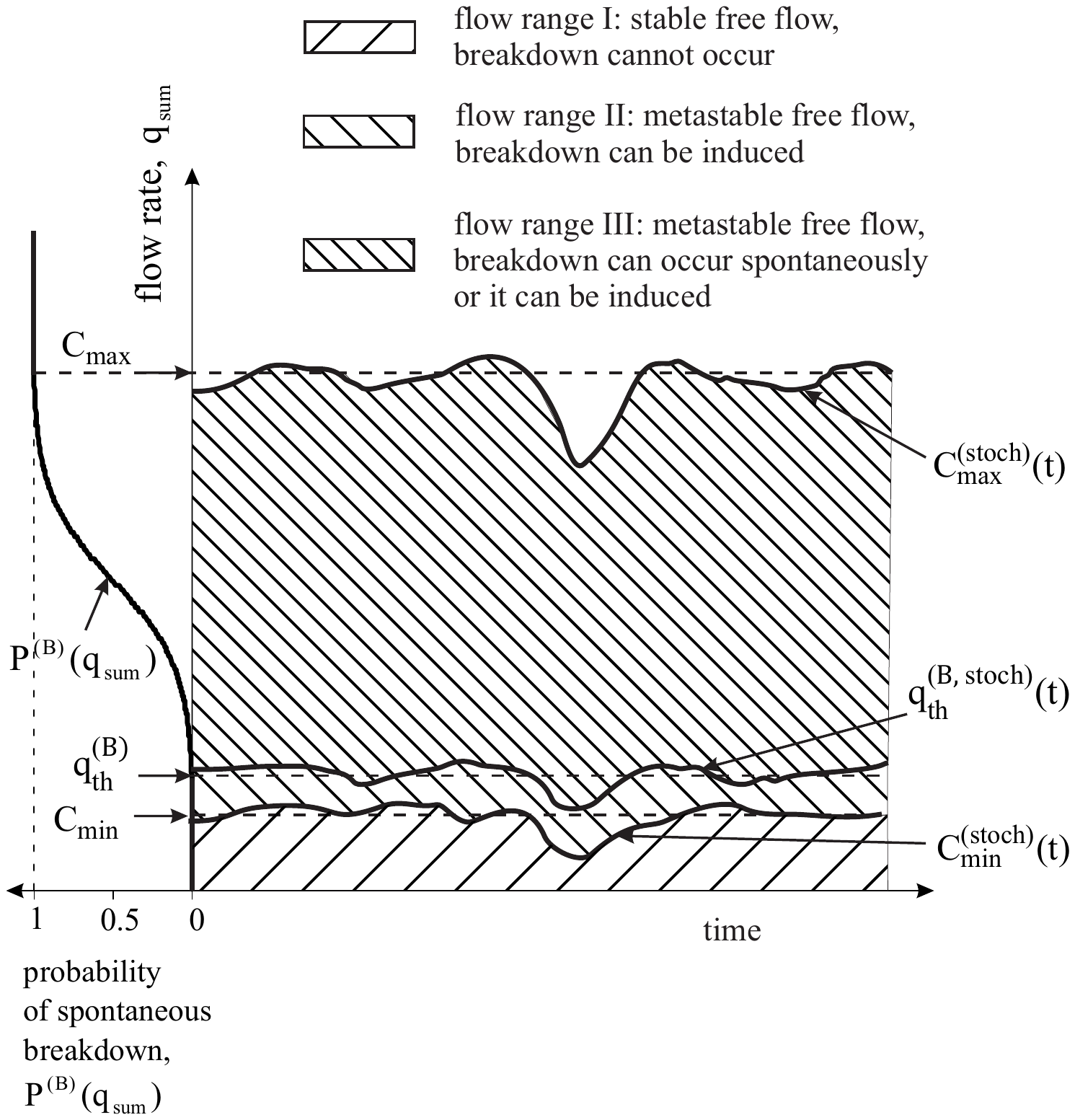}
\caption{Qualitative explanation of the infinite number of capacities of free flow at a highway bottleneck in three-phase traffic theory. Probability function for traffic breakdown $P^{\rm (B)}(q_{\rm sum})$  (figure left) is the same as that shown in Figs.~\ref{Capacity_short_sw1} (left) and~\ref{Emp_Breakdown_nuc}(b). Flow rate regions I, II, and III mentioned in labeling are, respectively, the same as those shown in Fig.~\ref{Emp_Breakdown_nuc} (b) and explained in Sec.~\ref{Basic_expl_S}.  
 \label{CapacityFS_short_sw1} } 
\end{center}
\end{figure}

It must be noted that the maximum capacity  $C_{\rm max}$, the minimum capacity  $C_{\rm min}$, and the value  $q^{\rm (B)}_{\rm th}$ depend on traffic parameters, like weather, mean driver's characteristics (e.g., mean driver reaction time), share of long vehicles, etc. In real traffic flow, these traffic parameters change over time. For this reason, the values  $C_{\rm max}$,  $C_{\rm min}$, and $q^{\rm (B)}_{\rm th}$  change also over time.
Moreover, in real traffic flow, the traffic parameters are stochastic time functions. Therefore, in real traffic flow we should consider some stochastic maximum capacity  $C^{\rm (stoch)}_{\rm max}(t)$,    stochastic minimum capacity  $C^{\rm (stoch)}_{\rm min}(t)$, and a stochastic threshold flow rate $q^{\rm (B, \ stoch)}_{\rm th}(t)$  whose time dependence is determined by stochastic characteristics of traffic parameters. Qualitative hypothetical fragment of these time-functions within a time interval is shown in Fig.~\ref{CapacityFS_short_sw1} (right).

Stochastic functions $C^{\rm (stoch)}_{\rm max}(t)$,  $C^{\rm (stoch)}_{\rm min}(t)$, and $q^{\rm (B, \ stoch)}_{\rm th}(t)$  shown in Fig.~\ref{CapacityFS_short_sw1} 
are qualitative hypothetical functions that cannot be measured in empirical observations. Only their mean values (respectively,  $C_{\rm max}$,  $C_{\rm min}$, and $q^{\rm (B)}_{\rm th}$) can be found in empirical studies of measured traffic data. In particular, the mean values  $C_{\rm max}$ and  $q^{\rm (B)}_{\rm th}$ can be found from an empirical study of the flow rate function of the breakdown probability  $P^{\rm (B)}(q_{\rm sum})$ (Fig.~\ref{Emp_Breakdown_nuc} (b)).

It must be noted that in empirical observations the mean value of the minimum capacity $C_{\rm min}$ can be found from a study of a finite number of different days at which induced
traffic breakdowns have been observed at a given bottleneck. The value $C_{\rm min}$
is related to these empirical days of observations only. In other words, it can occur that at another day, which is not within the days 
used for the calculation of  $C_{\rm min}$, traffic breakdown at this bottleneck can be induced at a smaller flow rate $q_{\rm sum}$ than  the minimum capacity and    
the minimum flow rate at which traffic breakdown was induced at this bottleneck in all 
earlier observations. A similar comment is related to the physical meaning of the mean value of  $q^{\rm (B)}_{\rm th}$ and $C_{\rm max}$. 
To explain this, we should note that with a finite number of measurements it is not possible to find some $\lq\lq$exact value" of the minimum flow rate at which traffic breakdown can occur. 
In other words, strictly speaking, mean values  $C_{\rm max}$,  $C_{\rm min}$, and $q^{\rm (B)}_{\rm th}$  are valid only for the days of the observing of traffic breakdown that have been used for the calculations of these mean values. 

From Fig.~\ref{CapacityFS_short_sw1} we can see that in the three-phase  theory traffic breakdown cannot occur spontaneously at $\lq\lq$any flow rate". Indeed, at any time instant at which the flow rate
$q_{\rm sum}$ in free flow 
is smaller than the minimum capacity
  $C^{\rm (stoch)}_{\rm min}(t)$, no traffic breakdown can occur at the bottleneck. When the flow rate $q_{\rm sum}$ satisfies conditions (\ref{II}), 
specifically, 
\begin{equation} 
C^{\rm (stoch)}_{\rm min}(t)\leq q_{\rm sum}(t) < q^{\rm (B, \ stoch)}_{\rm th}(t),
\label{P_meta}
\end{equation}
traffic breakdown can be induced only. 
Only under conditions
\begin{equation} 
q^{\rm (B, \ stoch)}_{\rm th}(t)\leq q_{\rm sum}(t)< C^{\rm (stoch)}_{\rm max}(t)
\label{P_meta2}
\end{equation} 
 traffic breakdown can occur spontaneously with some probability  $0<P^{\rm (B)}<1$ during a given observation time.

Thus, we can see in Fig.~\ref{CapacityFS_short_sw1} that in accordance with the highway capacity definition made in three-phase theory, under conditions
\begin{equation} 
C^{\rm (stoch)}_{\rm min}(t)\leq q_{\rm sum}(t)< C^{\rm (stoch)}_{\rm max}(t)
\label{P_meta3}
\end{equation}
  at any time instant there is the infinite number of highway capacities
  at which traffic breakdown can occur with some probability or can be induced at the bottleneck. 

\subsection{Infinite number of stochastic highway capacities 
in  the classical theory and  the three-phase theory  \label{Cl_Nature_S}} 
 
 \begin{figure}
\begin{center}
\includegraphics*[width=13 cm]{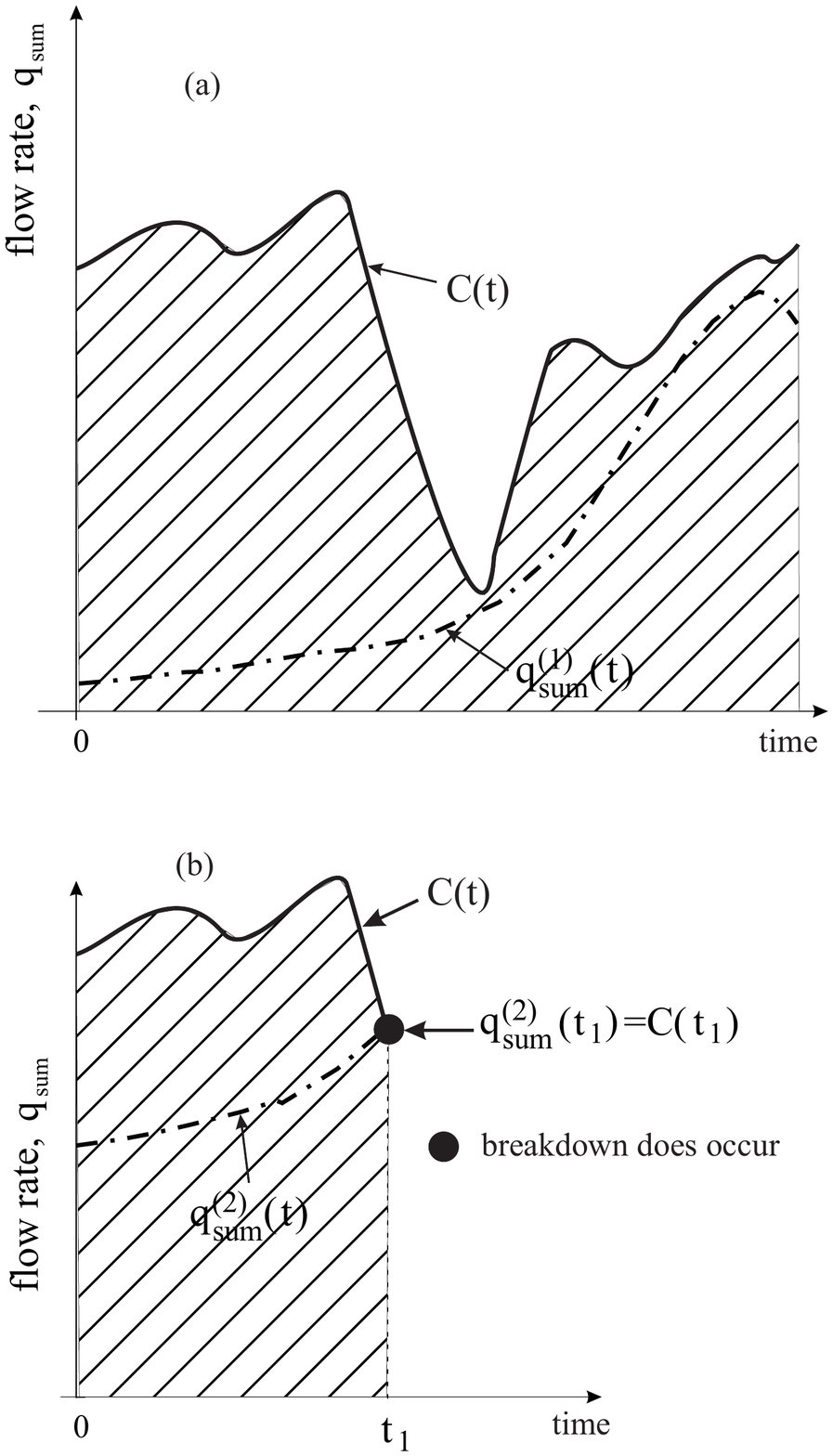}
\caption{Qualitative explanation of traffic breakdown with the use of Brilon's  stochastic highway capacity of free flow at a highway bottleneck. The fragment of the hypothetical time-function of  Brilon's  stochastic highway capacity $C(t)$  is taken from Fig.~\ref{Capacity_short_sw1} (right).
 \label{Capacity_short_sw} } 
\end{center}
\end{figure}

In accordance with the classical definition of stochastic capacity (\ref{A1}), (\ref{A2}), no traffic breakdown can occur, when the time dependence of the flow rate is given by a hypothetical time dependence 
$q^{(1)}_{\rm sum}(t)$ (Fig.~\ref{Capacity_short_sw}(a)). This is because  the following condition 
\begin{equation}
q^{(1)}_{\rm sum}(t)<C(t).
\label{A2_1}
\end{equation}
is satisfied at all time instants  shown in
Fig.~\ref{Capacity_short_sw}(a). 

In contrast, for another hypothetical time dependence $q^{(2)}_{\rm sum}(t)$
(Fig.~\ref{Capacity_short_sw}(b)) traffic breakdown should occur at time instant  $t_{1}$ at which  
\begin{equation}
q^{(2)}_{\rm sum}(t_{1})=C(t_{1}).
\label{A2_2}
\end{equation}
This is because under condition (\ref{A2_2}) the flow rate $q^{(2)}_{\rm sum}(t)$
is equal to the capacity value (Fig.~\ref{Capacity_short_sw}(b)).

In other words, the classical understanding of a particular value of stochastic capacity can be explained as follows: At a {\it given time instant}
 no traffic breakdown can occur at a highway bottleneck if the flow rate $q_{\rm sum}$ in free flow at the bottleneck at the time instant is smaller than the value of the capacity $C$ at {\it this time instant}. 
 The basic importance of the words $\lq\lq$at a given time instant" in the capacity definition is as follows: 
 Because Brilon's stochastic capacity $C(t)$  changes stochastically 
over time (Fig.~\ref{Capacity_short_sw1}),   at a {\it given time 
 instant} traffic breakdown can occur at the flow rate $q_{\rm sum}$ that is smaller than the value of the stochastic capacity was at {\it another time instant}.

In the classical understanding of stochastic capacity, free flow is stable under condition (\ref{A2_1}). 
This means that no traffic breakdown can occur or be induced at the bottleneck at long as the flow rate in free flow at the bottleneck
 is smaller than the stochastic capacity. This contradicts to the empirical fact that traffic breakdown can be induced 
 at the bottleneck due to the upstream propagation of a localized congested pattern
 (Fig.~\ref{OnRampInd220301}(b)). 

This is because stochastic highway capacity {\it cannot}  depend on whether there is a congested pattern, which has occurred outside of the bottleneck and independent of the bottleneck existence, or not. Indeed, the empirical evidence of induced traffic breakdown is the empirical proof that at a given flow rate at a bottleneck there can be one of two different traffic states at the bottleneck: (i) A   state F 
(free flow) and (ii) a   state S (synchronized flow). 
Due to the upstream propagation of a localized congested pattern, a transition from the state F to the state S, i.e., traffic breakdown is induced.
The induced traffic breakdown is impossible to occur under the classical understanding of stochastic highway capacity. This is because in this classical understanding of stochastic highway capacity,
 free flow is stable under condition
 (\ref{A2_1}), i.e., no traffic breakdown can occur (Fig.~\ref{Capacity_short_sw}(a)).

In contrast with the   classical understanding of stochastic highway capacity, the evidence of the empirical induced breakdown means that free flow is in a metastable state with respect to the breakdown. The metastability of free flow at the bottleneck should exist for all flow rates at which traffic breakdown can be induced at the bottleneck
 as observed in real traffic (Fig.~\ref{OnRampInd220301}(b)). This empirical evidence of the metastability of free flow at the bottleneck contradicts fundamentally the concept of Brilon's stochastic capacity, in which free flow is stable under condition (\ref{A2_1}).
This explains why the generally accepted classical understanding of stochastic highway capacity  has failed.

   \begin{figure}
\begin{center}
\includegraphics*[width=13 cm]{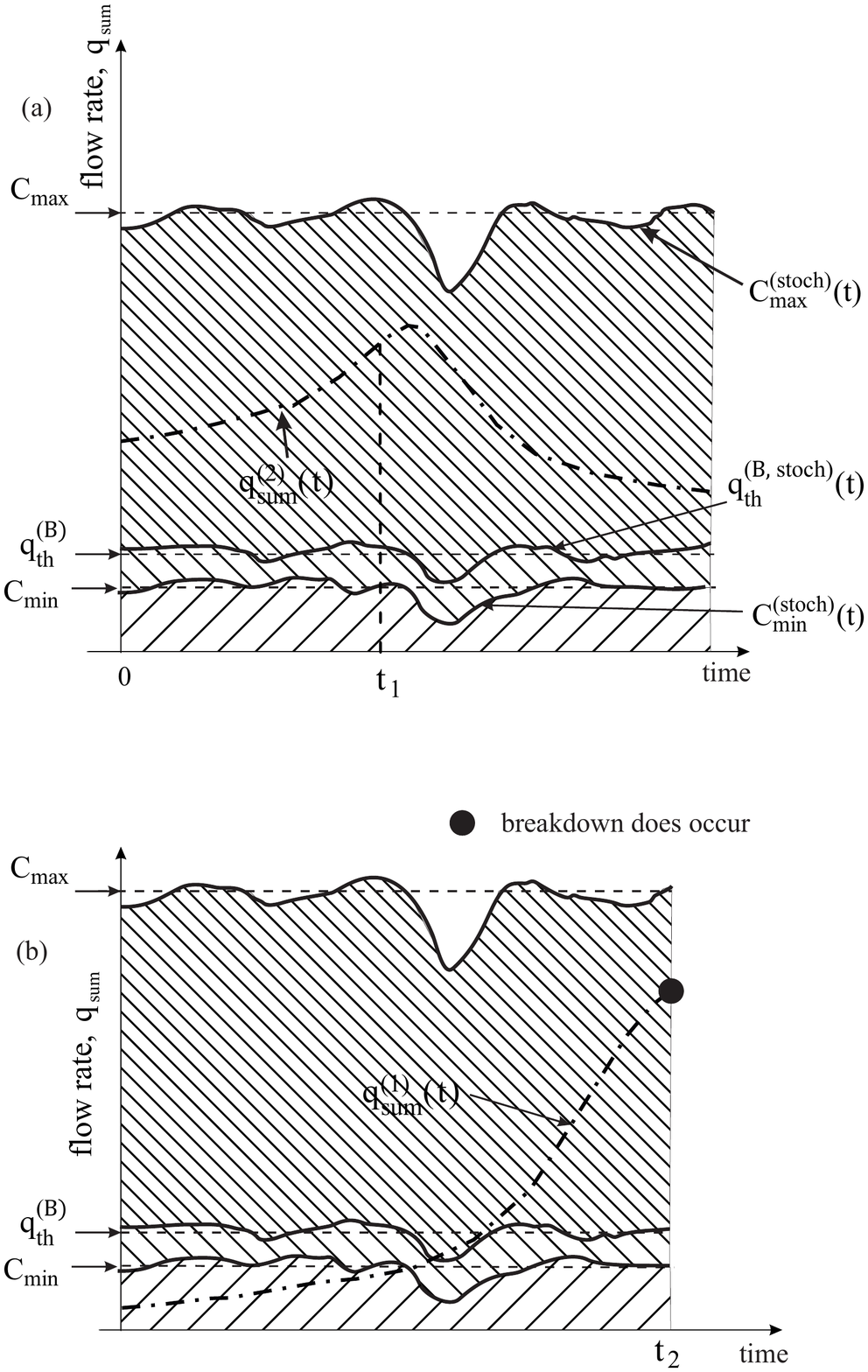}
\caption{Qualitative explanation of traffic breakdown with the use of the infinite number of capacities of free flow at a highway bottleneck of three-phase  theory. 
Hypothetical time-functions $C^{\rm (stoch)}_{\rm max}(t)$,  $C^{\rm (stoch)}_{\rm min}(t)$, and $q^{\rm (B, \ stoch)}_{\rm th}(t)$  are taken from Fig.~\ref{CapacityFS_short_sw1}. Hypothetical time functions of the flow rates    $q^{(2)}_{\rm sum}(t)$ in (a) and  $q^{(1)}_{\rm sum}(t)$ in (b) as well as time instant  $t_{1}$ in (a) are, respectively, the same as those in Fig.~\ref{Capacity_short_sw}. 
 \label{CapacityFS_short_sw} } 
\end{center}
\end{figure}

The classical understanding of   stochastic highway capacity   is based on the assumption that the 
empirical probability of traffic breakdown is determined by the capacity distribution function, i.e.,   condition (\ref{A2}) is valid.
In contrast, the assumption of the three-phase   theory about
 the metastability of traffic breakdown with respect to traffic breakdown (\ref{C1}) 
 is based on the empirical evidence that traffic breakdown can be induced at a bottleneck. 
 In both  the classical theory and three-phase   theory there is the infinite number of stochastic capacities.
However, in the classical understanding of  stochastic highway capacity at a {\it given time instant} 
there is {\it only one}
value of capacity (Fig.~\ref{Capacity_short_sw1}) that we do not know because the capacity is a stochastic value.

Contrarily, in the three-phase  theory at {\it any given time instant} there is the infinite number of stochastic capacities
within some capacity range between  minimum $C^{\rm (stoch)}_{\rm min}$ and  maximum capacities $C^{\rm (stoch)}_{\rm max}$.
  We cannot measure values of $C^{\rm (stoch)}_{\rm max}(t)$
and  $C^{\rm (stoch)}_{\rm min}(t)$ because   $C^{\rm (stoch)}_{\rm max}(t)$
and  $C^{\rm (stoch)}_{\rm min}(t)$ are stochastic values. However,
due to the empirical evidence of the possibility of induced traffic breakdown,
we know   that $C^{\rm (stoch)}_{\rm max}(t)>
 C^{\rm (stoch)}_{\rm min}(t)$ (Fig.~\ref{CapacityFS_short_sw1}). This emphasizes
  a crucial difference between
	the sense of the term {\it infinite number of stochastic capacities}
in the classical theory and the three-phase theory.
 
Thus the observation of empirical induced breakdowns proves that condition (\ref{A2}) of Brilon's stochastic capacity  cannot be valid for real traffic. However, the following question arises:
\begin{itemize}
\item	What are the consequences of this controversial understanding of the nature of traffic breakdown? 
\end{itemize}

With the use of Fig.~\ref{CapacityFS_short_sw}, we can qualitatively illustrate the basic difference between the classical understanding of   stochastic highway capacity 
 and the understanding of the infinite number of stochastic highway capacities made in the three-phase  theory.
In the classical understanding of stochastic capacity (\ref{A2}), for the hypothetical time dependence of the flow rate  $q^{(2)}_{\rm sum}(t)$ 
 shown in Fig.~\ref{Capacity_short_sw}(b), traffic breakdown has occurred at time instant $t_{\rm 1}$  at which  condition (\ref{A2_2}) is satisfied, i.e., when the flow rate is equal to the capacity value. 
 In contrast, in the three-phase  theory for the same time dependence of the flow rate   $q^{(2)}_{\rm sum}(t)$, for which conditions 
 (\ref{III})  are satisfied, no breakdown should be necessarily occur both at time instant  $t_{\rm 1}$ and for a later time interval (Fig.~\ref{CapacityFS_short_sw}(a)). 

In the classical understanding of stochastic capacity  (\ref{A2}), for the hypothetical time dependence of 
the flow rate  $q^{(1)}_{\rm sum}(t)$  shown in Fig.~\ref{Capacity_short_sw}(a), traffic breakdown could not occur because for all time instants  condition (\ref{A2_1}) is satisfied. 
In contrast, in the three-phase  theory for the same time dependence of the flow rate  $q^{(1)}_{\rm sum}(t)$
 traffic breakdown can occur spontaneously with some probability as this is shown for time instant   $t_{\rm 2}$ in Fig.~\ref{CapacityFS_short_sw} (b).

Because the classical understanding of stochastic highway capacity (\ref{A1}), (\ref{A2}) contradicts the empirical nucleation nature of real traffic breakdown, 
the understanding of stochastic highway capacity made in~\cite{Brilon310,Brilon210,Brilon,BrilonISTTT2009,ach_Elefteriadou2014A,ach_ElefteriadouBook2014}   cannot be used for reliable highway design and highway operations.

   \section{Enhancement of vehicular traffic through automatic driving vehicles \label{Aut_Cl_S}}

 For  a probabilistic analysis of the effect of automatic driving vehicles on traffic flow, we consider a simple case of vehicular traffic on a single-lane road with an on-ramp bottleneck.
 On the single-lane road, no  vehicles can pass.
  For this reason, automatic driving can be achieved through the use of an adaptive cruise control (ACC) in a vehicle:
   An ACC-vehicle follows a preceding vehicle automatically based on some ACC dynamics rules of motion. Depending on the   dynamic behavior of the preceding vehicle, these ACC-rules determine
   either automatic acceleration or automatic deceleration of the ACC-vehicle or else the maintaining of time-independent speed.  The preceding vehicle
    can be either a human driving vehicle or an automatic driving vehicle through an ACC-system in the vehicle.

  \subsection{Classical model of ACC \label{ACC_Cl_S}}
 
 There can be many different ACC dynamics rules of motion behind the preceding vehicle
(e.g.,~\cite{TreiberD,Treiber,KestingPhil2010,Kerner2003E,Kerner2003G,Davis2004B9,Davis2006,Davis2008A9,Davis2014C,Levine1966A,Becker1994A,Becker1994B,Cheng1997A},

\cite{Demir1998A,Demir2002A,Donikian1998A,Hogema1996A,McDonald1997A,Sala1996A,VanArem1996A,VanArem1997A,Liang1999A,Liang2000A},

\cite{Treiber2001AA,Mueller1992A,Noecker1994A,Hiller2003A,ISO_15622,Ourulingesh2004A,Ioannou1993A,Shladover1995A,Rajamani2012A,Swaroop1996A,Swaroop2001A,Kerner20059,Arem2006A,Kukuchi2003A,Bose2003A,Sathiyan2013A}).
 We limit the consideration by  a classical model of ACC-vehicle. 
In the classical ACC model, acceleration (deceleration) $a^{\rm (ACC)}(t)$ of
the ACC vehicle is  determined  by  current values of the space gap to a preceding vehicle $g(t)$ 
and the relative speed $\Delta v(t)=v_{\ell}(t)-v(t)$ measured by the ACC vehicle as well as by a   desired  space gap $g^{\rm (ACC)}=v(t) \tau^{\rm (ACC)}_{\rm d}$,
where $v(t)$ is the speed of the ACC-vehicle, $v_{\ell}(t)$ is the speed of the preceding vehicle,
 and $\tau^{\rm (ACC)}_{\rm d}$ is a desired net time gap (desired time headway) of the ACC-vehicle to the 
 preceding 
vehicle(e.g.,~\cite{TreiberD,Treiber,KestingPhil2010,Shrivastava2002A,Zhou2005A,Kesting2008A,Ngoduy2012A,Ngoduy2013A,Shladover2012A,Shladover2002A,Suzuki2003A},

\cite{Lin2009A,Martinez2007A,Papageorgiou2015A,Papageorgiou2015B,Papageorgiou2015C,Kerner2003E,Kerner2003G,Davis2004B9,Davis2006,Davis2008A9,Davis2014C},

\cite{Levine1966A,Becker1994A,Becker1994B,Cheng1997A,Demir1998A,Demir2002A,Donikian1998A,Hogema1996A,McDonald1997A,Sala1996A,VanArem1996A,VanArem1997A},

\cite{Liang1999A,Liang2000A,Treiber2001AA,Shladover1995A,Rajamani2012A,Swaroop1996A,Swaroop2001A,Kerner20059,Arem2006A,Kukuchi2003A,Bose2003A,Sathiyan2013A}):  
\begin{equation}
a^{\rm (ACC)}(t) = K_{1}(g(t)-v(t)\tau^{\rm (ACC)}_{\rm d})+K_{2} (v_{\ell}(t)-v(t)),
 \label{ACC_General}
 \end{equation} 
where  $K_{1}$
and $K_{2}$ are   coefficients of   ACC adaptation. 

\begin{figure}
\begin{center}
\includegraphics*[width=12 cm]{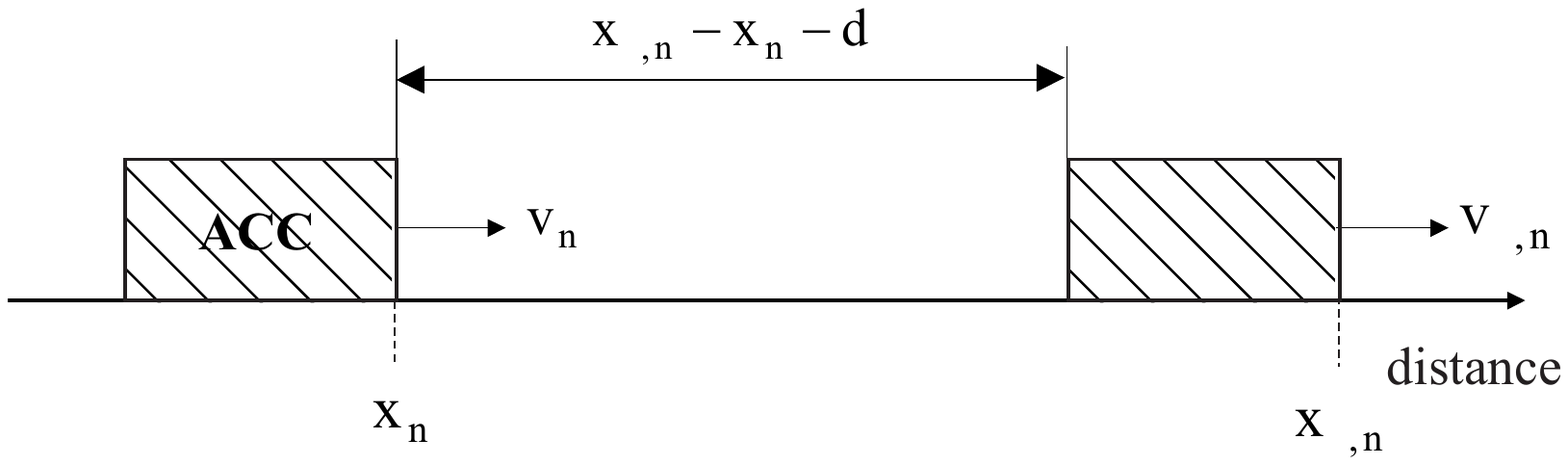}
\end{center}
\caption[]{Model of the ACC vehicle
\label{ACC_Model} }
\end{figure}

All simulations of human driving vehicles
 presented below are made with Kerner-Klenov microscopic stochastic three-phase traffic flow model~\cite{KKl}.
 Because in the Kerner-Klenov   model  
discrete time step is used (Appendix~\ref{App2}), we use in the classical ACC-model the discrete time $t=n\tau$, where $n=0,1,2,..$; $\tau=$1 s is   time step.
Therefore, the space gap to a preceding vehicle is equal to $g_{n}=x_{\ell, n}- x_{n}-d$
and the relative speed is given by $\Delta v_{n}= v_{\ell, n}-v_{n}$ (Fig.~\ref{ACC_Model}),
  where $x_{n}$ and $v_{n}$ are coordinate and speed of the ACC-vehicle, $x_{\ell, n}$ and $v_{\ell, n}$ are coordinate
and speed of the preceding vehicle, $d$ is   the vehicle length that is assumed the same for all automatic driving and human driving vehicles.
Respectively, the current net time gap (time headway) between ACC-vehicle and the preceding
vehicle calculated by  ACC-vehicle is equal to
$\tau^{\rm (net)}_{n}=  g_{n}/v_{n}$.
Correspondingly, the classical model of the dynamics of ACC-vehicle (\ref{ACC_General}) can be rewritten as follows: 
 \begin{equation}
 a^{\rm (ACC)}_{n} = K_{1}(g_{n}-v_{n}\tau^{\rm (ACC)}_{\rm d})+K_{2} (v_{\ell,n}-v_{n}).
\label{ACC_dynamics_Eq}
\end{equation}
  Coefficients of   ACC adaptation  $K_{1}$
and $K_{2}$
  describe the dynamic adaptation
of the ACC vehicle when either the space gap is different from the desired one $v_{n}\tau^{\rm (ACC)}_{\rm d}$:
\begin{equation}
g_{n}\neq v_{n}\tau^{\rm (ACC)}_{\rm d}
\label{Gap_ACC_steady_For}
\end{equation}
or   the vehicle speed  is different from the speed of the preceding vehicle:
\begin{equation}
v_{n}\neq v_{\ell,n}.
\end{equation}
If in contrast
\begin{equation}
v_{n}=v_{\ell,n}
\label{Speed_ACC_steady_For}
\end{equation}
and the condition 
\begin{equation}
g_{n}= v_{n}\tau^{\rm (ACC)}_{\rm d}
\label{Gap_ACC_steady_For_e}
\end{equation}
 is satisfied,
from (\ref{ACC_dynamics_Eq})
  we obtain that 
\begin{equation}
 a^{\rm (ACC)}_{n} = 0,
\end{equation}
i.e., the ACC vehicle moves with a time-independent speed.

The physics of the dynamic equation for the ACC vehicle (\ref{ACC_dynamics_Eq}) is as follows.
It can be seen that the current time headway $\tau^{\rm (net)}_{n}=g_{n}/v_{n}$ in
(\ref{ACC_dynamics_Eq})
is compared with
the desired time headway $\tau^{\rm (ACC)}_{\rm d}$.
If $\tau^{\rm (net)}_{n}>\tau^{\rm (ACC)}_{\rm d}$, then the ACC vehicle automatically accelerates to reduce
the time headway to the desired value $\tau^{\rm (ACC)}_{\rm d}$.
If $\tau^{\rm (net)}_{n}<\tau^{\rm (ACC)}_{\rm d}$, then the ACC vehicle decelerates automatically
to increase the time headway.
Moreover, the acceleration and   deceleration of the ACC vehicle
depend on the current difference between the speed of the ACC vehicle and the
preceding vehicle.
If the preceding vehicle has  a higher speed than   the ACC vehicle,
i.e., when $v_{\ell, n}>v_{n}$,  
the ACC vehicle accelerates. Otherwise, if $v_{\ell, n}<v_{n}$ the ACC vehicle decelerates.

In   simulations of traffic flow discussed  below,   there are
vehicles that have no ACC system (human driving vehicles) and  ACC-vehicles (automatic driving vehicles);
we call this traffic flow as $\lq\lq$mixture traffic flow". In mixture traffic flow,
the ACC vehicles are randomly distributed on the road between human driving vehicles.
The percentage of automatic driving vehicles denoted by   $\gamma$ is the same value
in traffic flow  upstream of the bottleneck and 
in the on-ramp inflow onto the main road.

The ACC vehicles move in accordance with Eq.  (\ref{ACC_dynamics_Eq}) where, in
addition, 
 the following  formulas are used:
\begin{equation}
\label{next1_ACC}
 v^{\rm (ACC)}_{{\rm c}, n} = v_{n}+\tau \max(-b_{\rm ACC},
\min(\lfloor a^{\rm (ACC)}_{n} \rfloor, a_{\rm ACC})),  
\end{equation}
\begin{equation}
 v_{n+1} = \max(0, \min({v_{\rm free}, v^{\rm (ACC)}_{{\rm c},n}, v_{{\rm s},n} })),
\label{next2_ACC}
\end{equation}
where we have taken into account that we use a discrete-in-space version of the Kerner-Klenov
model (Appendix~\ref{App2}), $\lfloor z \rfloor$ denotes the 
integer part of $z$.
Through the use of formula (\ref{next1_ACC}),  acceleration and   deceleration
of the ACC vehicles are limited by 
some maximum   acceleration $a_{\rm ACC}$ and  maximum deceleration
$b_{\rm ACC}$, respectively.
 Owing to the formula  (\ref{next2_ACC}),
the speed of the ACC vehicle $v_{n+1}$
 at   time step $n+1$ is limited by the 
maximum speed in free flow $v_{\rm free}$  and by the safe speed
 $v_{{\rm s},n}$  to avoid
 collisions between vehicles\footnote{Simulations show that formulas
(\ref{next1_ACC}), (\ref{next2_ACC}) do not influence on the dynamics
of the ACC vehicles (\ref{ACC_General}) in {\it free flow} outside of the bottleneck.
However, due to vehicle merging at the on-ramp bottleneck the time headway by merging
can be considerably smaller than $\tau^{\rm (ACC)}_{\rm d}$. Therefore,
formulas
(\ref{next1_ACC}), (\ref{next2_ACC}) allows us to avoid collisions
of the ACC vehicle with the preceding vehicle in such dangerous situations. Moreover,
very small values of time headway can occur in congested conditions; formulas
(\ref{next1_ACC}), (\ref{next2_ACC}) prevent vehicle collisions in these cases also.
 While   working at the Daimler Company,
 the author was lucky to take part in the development of real ACC vehicles,
which are on the market; 
  to avoid
collisions in dangerous simulations,
 dynamics rules of all   real ACC vehicles include some safety
dynamic rules that can be 
 similar to (\ref{next1_ACC}), (\ref{next2_ACC}).}.
The maximum speed in free flow $v_{\rm free}$ and the safe speed
 $v_{{\rm s},n}$ are chosen, respectively, the same as those in the
 microscopic model of human driving vehicles (Appendix~\ref{App2}).
It should be noted that the model of ACC-vehicle merging from the on-ramp
onto the main road is similar to that for for human driving vehicles 
(see Sec.~\ref{On_ACC_Bott_Sec} of  Appendix~\ref{App2}).

An important characteristic of the ACC-vehicles is a stability of a platoon of the
ACC-vehicles called {\it string stability}. Liang and Peng~\cite{Liang1999A}
have found that for a string stability of the ACC vehicles
coefficients of   ACC adaptation  $K_{1}$ and $K_{2}$ in (\ref{ACC_General}) 
and the desired time headway of the ACC vehicles
$\tau^{\rm (ACC)}_{\rm d}$ should satisfy 
condition~\cite{Liang1999A}
\begin{equation}
K_{2}>\frac{2-K_{1}(\tau^{\rm (ACC)}_{\rm d})^{2}}{2\tau^{\rm (ACC)}_{\rm d}}.
\label{String_stability}
\end{equation}
Below to limit the analysis, we consider the effect of the ACC vehicles on traffic flow
{\it only} for  a relatively short desired time headway of the ACC vehicles
$\tau^{\rm (ACC)}_{\rm d}=$ 1.1 s. However,  we will use different sets of
coefficients  $K_{1}$ and $K_{2}$ of   ACC 
adaptation\footnote{It should be noted that condition
 (\ref{String_stability}) has been derived in~\cite{Liang1999A} for
continuum time used in Eq.~(\ref{ACC_General}). In contrast, as above-mentioned,
in all simulations below we use Eq.~(\ref{ACC_dynamics_Eq}) with  discrete time $t=n\tau$,  
 $n=0,1,2,..$. This could alter the rules for string stability of an ACC-vehicle platoon. For this reason,
we have made numerical simulations of string stability of ACC-vehicle platoons moving on a circular road
(not shown in this article). We have found that at least for
    the sets of
coefficients  $K_{1}$ and $K_{2}$ of   ACC adaptation in Eq. (\ref{ACC_dynamics_Eq}), which have been used in this article, an ACC-vehicle platoon is stable with respect to small  
disturbances, when condition (\ref{String_stability})
 of Liang and Peng~\cite{Liang1999A} is satisfied,
 whereas the ACC-vehicle platoon is unstable  with respect to small  
disturbances, when
condition (\ref{String_stability}) is not satisfied.
Therefore, when string stability
 of   ACC-vehicle platoons for different sets of
coefficients  $K_{1}$ and $K_{2}$  of   ACC adaptation
 in Eq. (\ref{ACC_dynamics_Eq}) is discussed below, we will refer
 to condition (\ref{String_stability}).}.

   \subsection{String instability versus S$\rightarrow$F instability of three-phase theory \label{GM_S}}
	
\begin{figure} 
\begin{center}
\includegraphics*[width=11 cm]{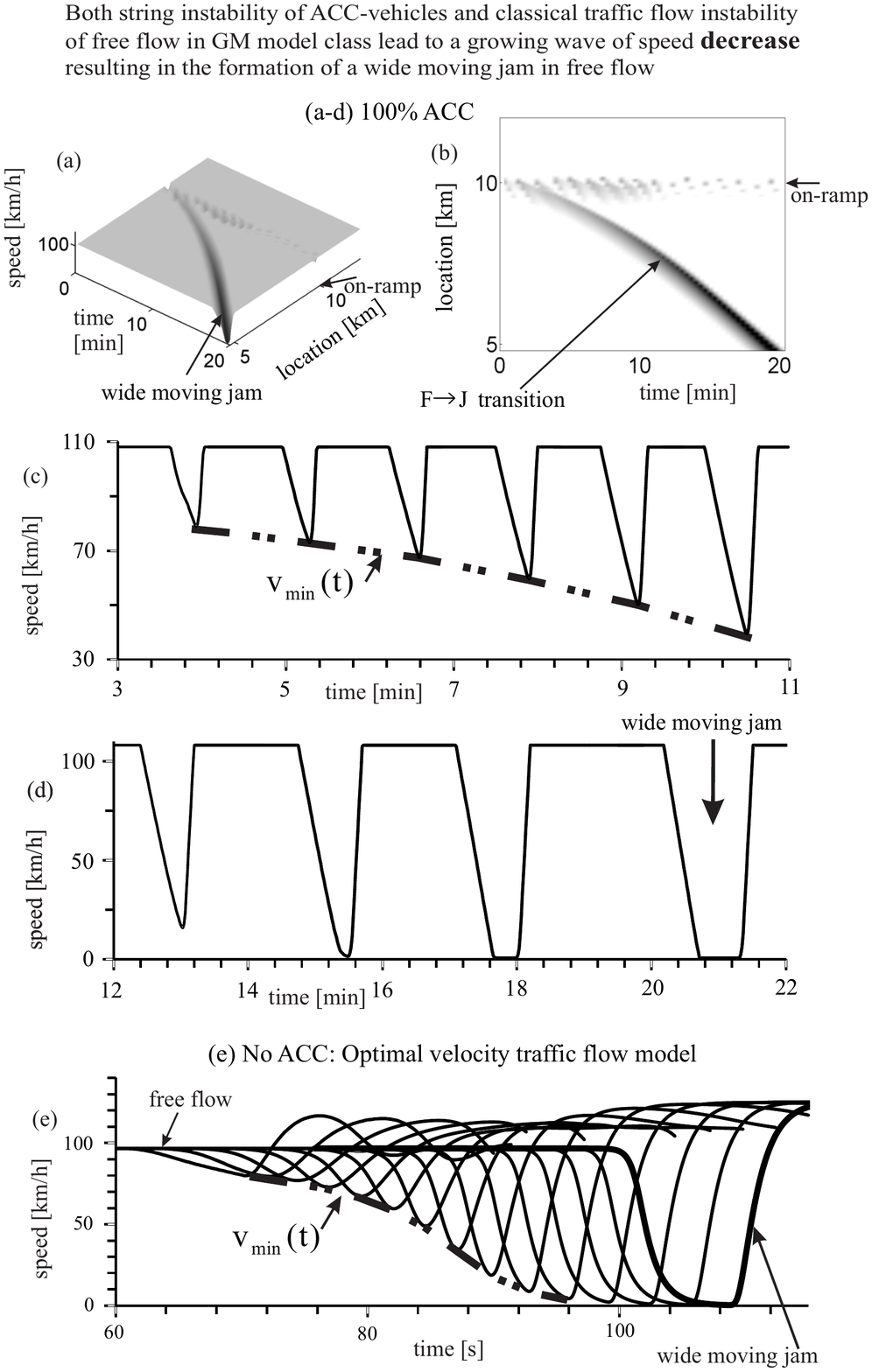}
 \end{center}
\caption{Comparison of  string instability in traffic flow of
$\gamma=100\%$ of  ACC-vehicles at on-ramp bottleneck  on single-lane road
simulated with classical ACC-model of Sec.~\ref{ACC_Cl_S} (a--d) with classical
traffic flow instability in traffic flow without ACC-vehicles
simulated with OV model~\cite{B1994A10} belonging to
 the GM model class (see references in  
	reviews~\cite{Brackstone,Gazis,Chowdhury,Helbing,Nagatani,Nagel,Maerivoet,Hesham10,TreiberD,Treiber,MiniReview})  (e):
(a, b) Vehicle speed in space and time (a) and the same speed data presented by regions with variable shades of gray (b) 
	(in white regions the speed is larger than 105 km/h, in black regions the speed is equal to zero). 
(c--e) Microscopic vehicle speed along  vehicle trajectories  (only a few of the
subsequent trajectories are shown)  as time-functions.
In (a--d), ACC-parameters are
$\tau^{\rm (ACC)}_{\rm d}=1.1$ s,
coefficients  
$(K_{1}, \ K_{2})=$ (0.5 $s^{-2}$, 0.2 $s^{-1}$) in (\ref{ACC_dynamics_Eq})
do not satisfy condition
  (\ref{String_stability}) for string stability; 
	values $a_{\rm ACC}=b_{\rm ACC}=3 \ {\rm m}/{\rm s}^2$; the
	flow rates   are
$q_{\rm on}=50$ vehicles/h, $q_{\rm in}=2609$ vehicles/h ($q_{\rm sum}=q_{\rm on}+q_{\rm in}=2659$ vehicles/h).
Figure (e) is taken from~\cite{Kerner2015B}.
}
\label{Instability_Com1}
\end{figure}

	To understand the effect of the ACC vehicles on traffic flow discussed below, firstly
	we should understand a crucial difference between the dynamic behaviors of the ACC-vehicles
	(\ref{ACC_dynamics_Eq}) and the dynamic behavior of manual driving vehicles in the three-phase theory.
	As above-mentioned, the platoon of the ACC-vehicles exhibits a string instability, when
	(\ref{String_stability})   is {\it not} satisfied. For the 
	string instability of traffic flow consisting of 100$\%$ ACC-vehicles, we find a known result
	   that the string instability is a growing wave of local {\it decrease} in speed of ACC-vehicles
  (Fig.~\ref{Instability_Com1} (a--c)). 
	
\begin{figure} 
 \begin{center}
\includegraphics[width=11 cm]{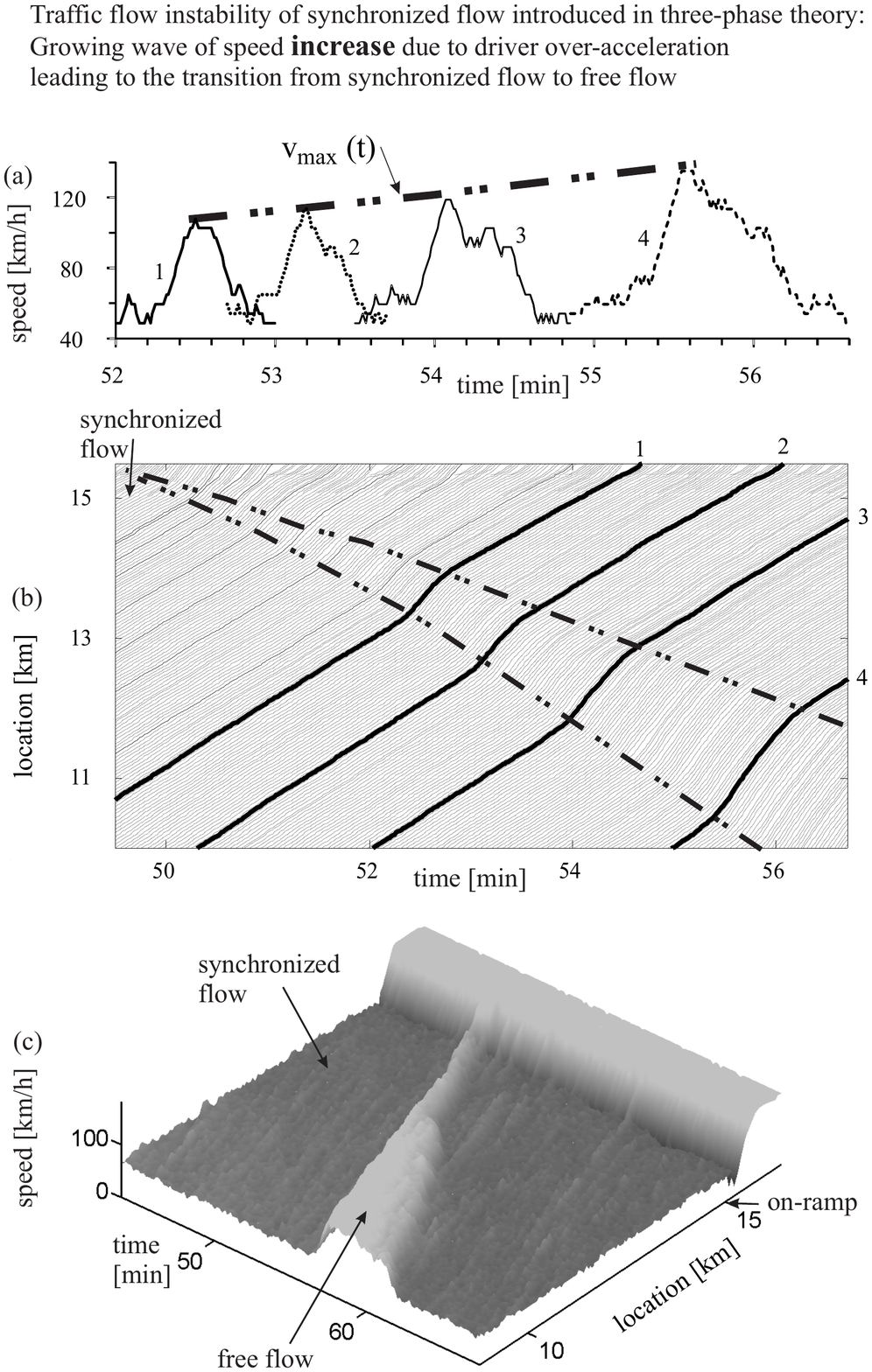}
 \end{center}
\caption{Simulations of S$\rightarrow$F instability
 introduced in three-phase theory that  governs an F$\rightarrow$S transition (traffic breakdown)
at on-ramp bottleneck~\cite{Kerner2015B}:
(a, b)
 Vehicle speed along trajectories as time-functions (a)  and
vehicle trajectories in space and time (b);
 numbers of   trajectories in (b) are related to trajectories labeled in (a) 
by the same numbers. (c) Speed 
in space and time.
 }
\label{S_F_F_J}
\end{figure}

	Qualitatively the same 
	growing wave of local   decrease  in speed occurs due to the classical instability 
	in a platoon of
	manual driving vehicles moving at free flow speed in accordance with 
	rules of a traffic flow model of the GM model class.
	As in traffic flow models of the GM model class, this    
	growing wave caused by the string instability of the ACC-vehicles 
	leads to the emergence of a wide moving jam in traffic flow of the ACC-vehicles, i.e.,
	to an F$\rightarrow$J transition
	(compare Fig.~\ref{Instability_Com1} (a--d) for string instability of the ACC-vehicles
	with well-known results shown in Fig.~\ref{Instability_Com1} (e) for the classical traffic flow instability of the GM model class).
Thus we can make the conclusions:
	\begin{itemize}
	\item
	As the classical
	traffic flow instability  of the GM model class (Fig.~\ref{Instability_Com1} (e)), string instability of the ACC-vehicles  (Fig.~\ref{Instability_Com1} (a--c))
	is a growing wave of speed reduction in free flow. 
	\item As the classical traffic flow instability (Fig.~\ref{Instability_Com1} (e)), 
	the string instability of the ACC-vehicles
	leads to the formation of wide moving jams 
	in free flow (F$\rightarrow$J transition) (Fig.~\ref{Instability_Com1} (a, b, d)).
	\end{itemize}

	In contrast with the ACC-vehicles and with the classical
	traffic flow instability  of the GM model class, as explained and proven in a recent
	paper~\cite{Kerner2015B}, in the three-phase theory there is {\it no} string instability
	   in a platoon of manual driving vehicles moving at free flow speed. Rather than
		string instability, traffic breakdown in traffic flow that consists only of
		manual driving vehicles is associated with the existence of
		an S$\rightarrow$F instability introduced in the three-phase 
		theory.
		As proven~\cite{Kerner2015B}, the S$\rightarrow$F instability
	governs the metastability of free flow with respect to an F$\rightarrow$S transition
	(traffic breakdown) at a bottleneck as observed in all real field traffic data.
	As explained in details in~\cite{Kerner2015B}, 
	the S$\rightarrow$F instability is  a growing wave of local {\it increase} in speed in synchronized  flow
 (Fig.~\ref{S_F_F_J} (a, b))
that leads to the S$\rightarrow$F transition
  (Fig.~\ref{S_F_F_J} (c))\footnote{The physics of the S$\rightarrow$F instability
and the explanation why this instability governs traffic breakdown has been considered  in details
in the paper~\cite{Kerner2015B}. The explanation of this physics
is out of scope of this mini-review.}. 
	Thus we can make the conclusions:
	\begin{itemize}
	\item Contrary to classical instability of the GM model class and
	string instability of the ACC-vehicles that lead to speed {\it reduction} in 
	free flow (Fig.~\ref{Instability_Com1}),
 the S$\rightarrow$F instability of the three-phase theory
 is  a growing wave of local {\it increase} in speed in synchronized  flow
 (Fig.~\ref{S_F_F_J})~\cite{Kerner2015B}.
	\item Contrary to classical instability of the GM model class and
	string instability of the ACC-vehicles leading to the
	F$\rightarrow$J transition
	(Fig.~\ref{Instability_Com1} (a, b, d, e)),
	the S$\rightarrow$F instability of the three-phase theory
	governs traffic breakdow (F$\rightarrow$S transition)n in free flow~\cite{Kerner2015B}.
	\item In the three-phase theory, no string instability   occurs
	   in a platoon of manual driving vehicles moving at a free flow speed. 
	\end{itemize}

However, it should be noted that the critical conclusion that the classical traffic 
instability of the GM model class
 (see references in  
 reviews

\cite{Brackstone,Gazis,Chowdhury,Helbing,Nagatani,Nagel},

\cite{Maerivoet,Hesham10,TreiberD,Treiber,MiniReview}) failed to explain traffic breakdown in real field traffic data (see
Sec.~\ref{Crit_Cla_S})
is {\it not} related to the string instability of the platoon of the ACC-vehicles.

The reason for this is as follows: In general, the dynamics rules of 
  motion of an ACC vehicle 
are related to a  fixed program written by ACC-developers, {\it not} to some behavior of   manual drivers
in real traffic flow.
Therefore, if the
classical rules 
	(\ref{ACC_dynamics_Eq}) exhibit a string instability
	of the platoon of the ACC-vehicles,
	then this is a feature of the   rules 
	(\ref{ACC_dynamics_Eq}). This feature of the ACC-vehicle should
	{\it not} necessarily be 
 in agreement with real dynamic rules of  motion of
 manual driving vehicles. 

This is  
crucially different for a traffic flow model that should describe
dynamics rules of motion of real manual driving vehicles.
In other words, in contrast with the rules of motion of
the ACC-vehicles, dynamics rules of motion of   manual driving vehicles
in the traffic flow model should be in agreement with
real field traffic data. The real data reflects the behavior of
  real  manual driving vehicles: The real behavior 
	of drivers results in the empirical evidence that
traffic breakdown is the F$\rightarrow$S transition in metastable free flow, {\it not}
the F$\rightarrow$J transition resulting from simulations
 of traffic flow models of the GM model class.

 This explains why the failure of traffic flow models of the GM model class
should not necessarily  be considered as a drawback of the  classical rules 
	(\ref{ACC_dynamics_Eq}) of the dynamics of the ACC-vehicles.
Nevertheless, from the analysis of the effect of the classical ACC-vehicles
	of traffic flow, which   will be made below
	in Sec.~\ref{P2_Aut_Human_S}, we could have an assumption that to improve
	traffic flow through automatic driving vehicles considerably,
	the dynamic behavior of
	  the future ACC-vehicles should learn 
	from some behaviors of   manual driving vehicles in real traffic flow 
	 (see an example in Sec.~\ref{Aut_Hum_S}).

  \subsection{Main objective of analysis of effect of ACC-vehicles on
	traffic flow \label{Crit_Main_S}}
	
	It should be noted that even when  
	condition for string stability (\ref{String_stability})
	is   satisfied, nevertheless, traffic congestion occurs
	in traffic flow consisting of 100$\%$ ACC-vehicles, if the flow rate $q_{\rm sum}$
	exceeds
	the value
	 \begin{equation}
 q_{0}= \frac{3600}{\tau^{\rm (ACC)}_{\rm d}+d/v_{\rm free}}.
 \label{Stable_flow}
\end{equation}
For this reason, in all   simulations of the effect of ACC-vehicles on traffic flow
presented in this mini-review below, we have chosen model parameters
  at which traffic breakdown occurs
in  a mixture traffic flow, only when the flow rate at the bottleneck $q_{\rm sum}$
 is considerably
smaller than $q_{0}$ (\ref{Stable_flow}):
 \begin{equation}
 q_{\rm sum}=q_{\rm on}+q_{\rm in}<q_{0}\approx 2667 \ {\rm [vehicles/h]},
 \label{Stable_flow_q_sum}
\end{equation}
where in formula (\ref{Stable_flow})
we have taken into account that we consider only ACC-vehicles with
 $\tau^{\rm (ACC)}_{\rm d}=1.1$ s,
$v_{\rm free}=$ 30 m/s, and $d=$ 7.5 m.
 Under   condition (\ref{Stable_flow_q_sum}),
 it is often expected  that
if there is  
string stability of a platoon of the ACC-vehicles, then
the ACC-vehicles should   improve traffic flow.
\begin{itemize}
\item
The main objective of our analysis of the effect of the ACC-vehicles on traffic flow
 presented below is to prove that this assumption should not necessarily be valid, even when
  condition
of string stability for the ACC-vehicles (\ref{String_stability}) {\it and}  condition (\ref{Stable_flow_q_sum})
are satisfied.
\item
We will find that  
depending on the coefficients of   ACC adaptation  $K_{1}$ and $K_{2}$ in (\ref{ACC_dynamics_Eq})
(which all satisfy condition
  (\ref{String_stability}) for string 
	stability)
the ACC-vehicles can either improve or  deteriorate the traffic system\footnote{Our simulations show that
 under condition  (\ref{Stable_flow_q_sum})	all results presented in
Secs.~\ref{Crit_Sup_S} and~\ref{Aut_Nature_S}   
can remain qualitatively,  even if condition
  (\ref{String_stability}) for string 
	stability of the ACC-vehicles is not satisfied. However, this is only true for some sets of
	the coefficients    $K_{1}$ and $K_{2}$ in a neighborhood of 
	the threshold of string instability. For example, simulation results shown
	in Figs.~\ref{ACC_FS}--\ref{ACC_stable_r}  remain almost the same, if  we use
	ACC-vehicles with coefficients
	$(K_{1}, \ K_{2})=$(0.1 $s^{-2}$, 0.55 $s^{-1}$) that do not satisfy
	condition of string stability
  (\ref{String_stability}). A  consideration of  special   cases in which ACC-vehicles
	improve traffic flow characteristics, however, the ACC-vehicles
 do not satisfy
	condition of string stability
  (\ref{String_stability}),   
	is out of the scope of this mini-review.  \label{ACC_Inst_Foot}}.
		\end{itemize}
 
Before we consider  cases in which the ACC-vehicles under conditions (\ref{String_stability})  and  
(\ref{Stable_flow_q_sum}) deteriorate the traffic system
(Sec.~\ref{P2_Aut_Human_S}), in
Secs.~\ref{Crit_Sup_S} and~\ref{Aut_Nature_S} we consider 
  expected cases in which the ACC-vehicles   enhance the traffic system, specifically, 
prevent traffic breakdown (Sec.~\ref{Crit_Sup_S}) or decrease the breakdown
 probability (Sec.~\ref{Aut_Nature_S}) .

  \subsection{Suppression of traffic breakdown through 
	automatic driving vehicles \label{Crit_Sup_S}}

  \begin{figure}
\begin{center}
\includegraphics*[width=11 cm]{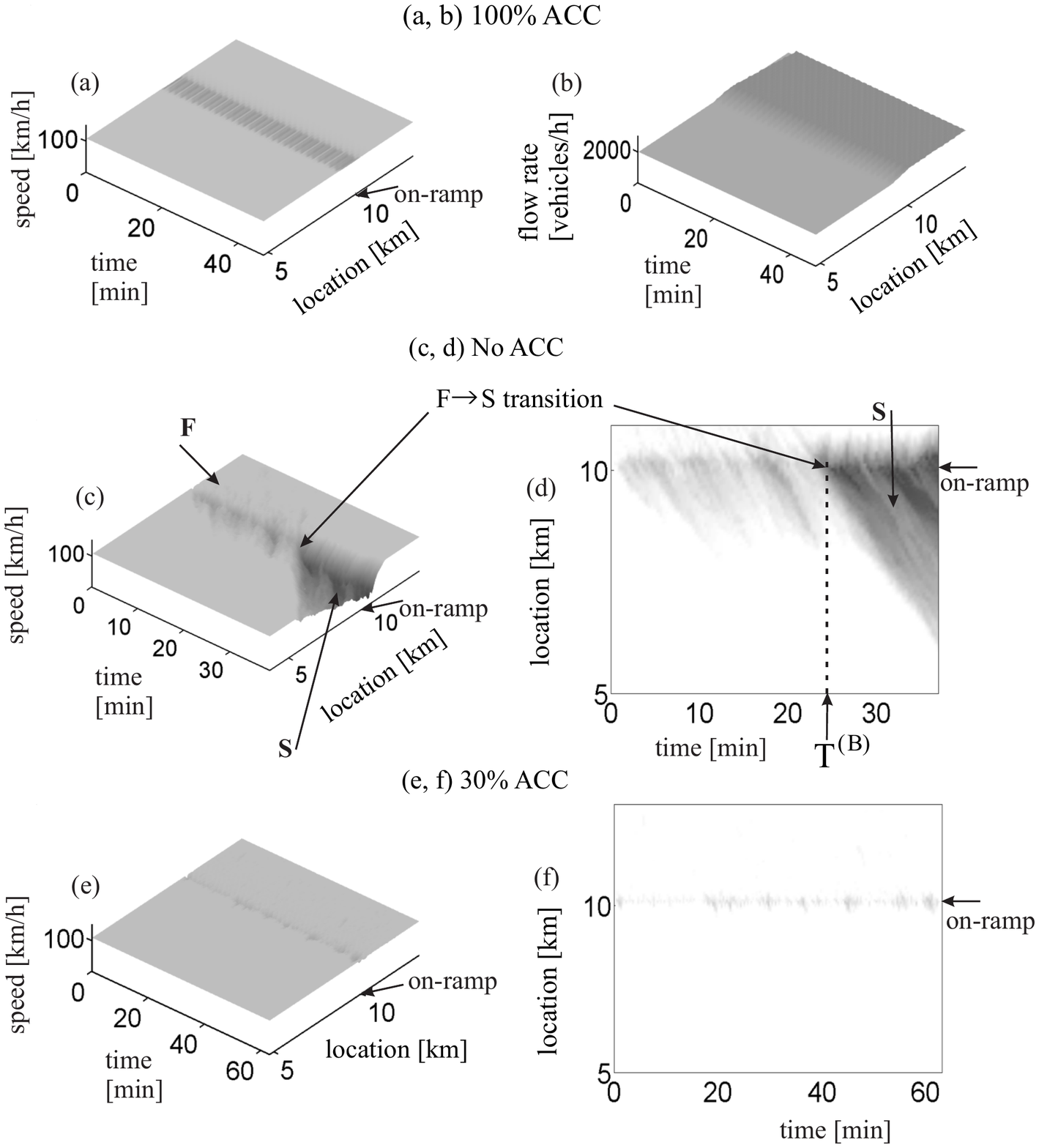}
\end{center}
\caption[]{Suppression of F$\rightarrow$S transition
in traffic flow 
on single-lane road with an on-ramp bottleneck through ACC-vehicles:
(a, b) Vehicle speed   (a) and the flow rate (averaging over 20 vehicles) (b)
 in space and time.
 (c--f)  Vehicle speed in space and time (c, e) and the same speed data presented by regions with variable shades of gray (d, f) 
	(in white regions the speed is equal to 105 km/h, in black regions the speed is equal to zero).
In (a, b), traffic flow with
 $\gamma=100\%$ of ACC vehicles with $\tau^{\rm (ACC)}_{\rm d}=1.1$ s and
  coefficients
  $(K_{1}, \ K_{2})=$(0.14 $s^{-2}$, 0.9 $s^{-1}$)  satisfying 
  (\ref{String_stability}); 
	values $a_{\rm ACC}=b_{\rm ACC}=3 \ {\rm m}/{\rm s}^2$.
In (c, d), no ACC vehicles -- traffic flow  consisting only of human driving vehicles.
In (e, f) $\gamma=30\%$ of   ACC vehicles   with the same parameters
   as those (a, b). 
 Flow rates in all figures are
$q_{\rm on}=320$ vehicles/h, $q_{\rm in}=2000$ vehicles/h ($q_{\rm sum}=q_{\rm on}+q_{\rm in}=2320$ vehicles/h
that satisfies (\ref{Stable_flow_q_sum})).
  F -- free flow, S -- synchronized flow. On-ramp location
	  $x_{\rm on}=$ 10 km.
\label{ACC_FS} }
\end{figure}

 As expected, free flow consisting of $\gamma=100\%$ of   ACC vehicles
that satisfy conditions (\ref{String_stability})  and  
(\ref{Stable_flow_q_sum})
is stable: All speed disturbances occurring due to the merging of ACC-vehicles at the bottleneck
decay over time (Fig.~\ref{ACC_FS} (a, b)). However, the disturbances cause
  a decrease in free flow speed at the bottleneck location
seen in Fig.~\ref{ACC_FS} (a):
Automatic driving vehicles moving initially on the main road
  with the speed $v_{\rm free}$ upstream of the bottleneck
	should decelerate in a neighborhood of the bottleneck
	due to the merging of other automatic driving vehicles  from the on-ramp onto the main road. 

At the same flow rates $q_{\rm on}$ to the on-ramp and on the main road $q_{\rm in}$  
as those in Fig.~\ref{ACC_FS} (a, b), in free flow consisting
only of human driving vehicles (no ACC-vehicles) traffic breakdown occurs
at the bottleneck during the time interval $T_{\rm ob}=$ 30 min
 with the probability $P^{\rm (B)}=$ 0.375 (Fig.~\ref{ACC_FS} (c, d)).
In a simulation realization shown in Fig.~\ref{ACC_FS} (c, d), 
traffic breakdown has occurred after a random time delay $T^{\rm (B)}\approx$ 25 min.

When we consider a mixture traffic flow and  increase the percentage $\gamma$ of ACC vehicles 
in the mixture traffic flow to $\gamma=30 \%$, we get a known result
derived with different traffic flow models in the framework of the three-phase theory
that
no traffic breakdown (F$\rightarrow$S transition) occurs     
(Fig.~\ref{ACC_FS} (e, f))\footnote{The  result of simulations that  $\gamma \sim 30 \%$ of ACC vehicles can suppress traffic congestion
    is also well-known one from many studies of the classical
		traffic flow models (e.g.,~\cite{TreiberD,Treiber,KestingPhil2010,Shrivastava2002A,Zhou2005A,Kesting2008A,Ngoduy2012A,Ngoduy2013A,Shladover2012A,Shladover2002A,Suzuki2003A,Lin2009A,Martinez2007A},
		
		\cite{Papageorgiou2015A,Papageorgiou2015B,Wagner2015A,Friedrich2015A,Papageorgiou2015C,Papageorgiou2015DA,Benz2016A}) that  
		 cannot explain real traffic breakdown
		(F$\rightarrow$S transition) in metastable free flow at a bottleneck (see Sec.~\ref{Crit_Cla_S}). 
		This result  can be explained as follows. Because at chosen flow rates
		and ACC-parameters free flow consisting of 100$\%$
		of ACC vehicles is stable (Figs.~\ref{ACC_FS} (a, b)), we could expect that
		regardless of features of a traffic flow model used for
		simulations of human driving vehicle,
			  there should be a critical percentage of ACC-vehicles when they suppress any instabilities in traffic flow caused by manual driving vehicles in the traffic flow model. Simulations made
				with the classical traffic flow models show that this
		critical percentage of ACC-vehicles is about  $\gamma \sim 30 \%$ 
		of ACC vehicles  (e.g.,~\cite{TreiberD,Treiber,KestingPhil2010,Shrivastava2002A,Zhou2005A,Kesting2008A,Ngoduy2012A,Ngoduy2013A,Shladover2012A,Shladover2002A,Suzuki2003A,Lin2009A,Martinez2007A},
		
		\cite{Papageorgiou2015A,Papageorgiou2015B,Wagner2015A,Friedrich2015A,Papageorgiou2015C,Benz2016A}). 
 However, our main objective is to study the effect of the ACC-vehicles
on the probability of real traffic breakdown (F$\rightarrow$S transition). Classical
traffic flow models used, for example,
		in~\cite{TreiberD,Treiber,KestingPhil2010,Shrivastava2002A,Zhou2005A,Kesting2008A,Ngoduy2012A,Ngoduy2013A,Shladover2012A,Shladover2002A,Suzuki2003A,Lin2009A,Martinez2007A},
		
		\cite{Papageorgiou2015A,Papageorgiou2015B,Wagner2015A,Friedrich2015A,Papageorgiou2015C,Papageorgiou2015DA,Benz2016A,LevinACC2015A}cannot describe an F$\rightarrow$S transition in metastable free flow as observed in real traffic. Therefore, these models cannot be used for an analysis of the effect of ACC-vehicles on the probability of traffic breakdown
		studied in this mini-review.}. The work by Davis~\cite{Davis2004B9} was one of the first  
to obtain this result with another human driver model that had some of the features of
the three-phase theory.
 
 \subsection{Decrease in probability of traffic breakdown through
 automatic driving vehicles \label{Aut_Nature_S}} 
 
\begin{figure}
\begin{center}
\includegraphics*[width=13 cm]{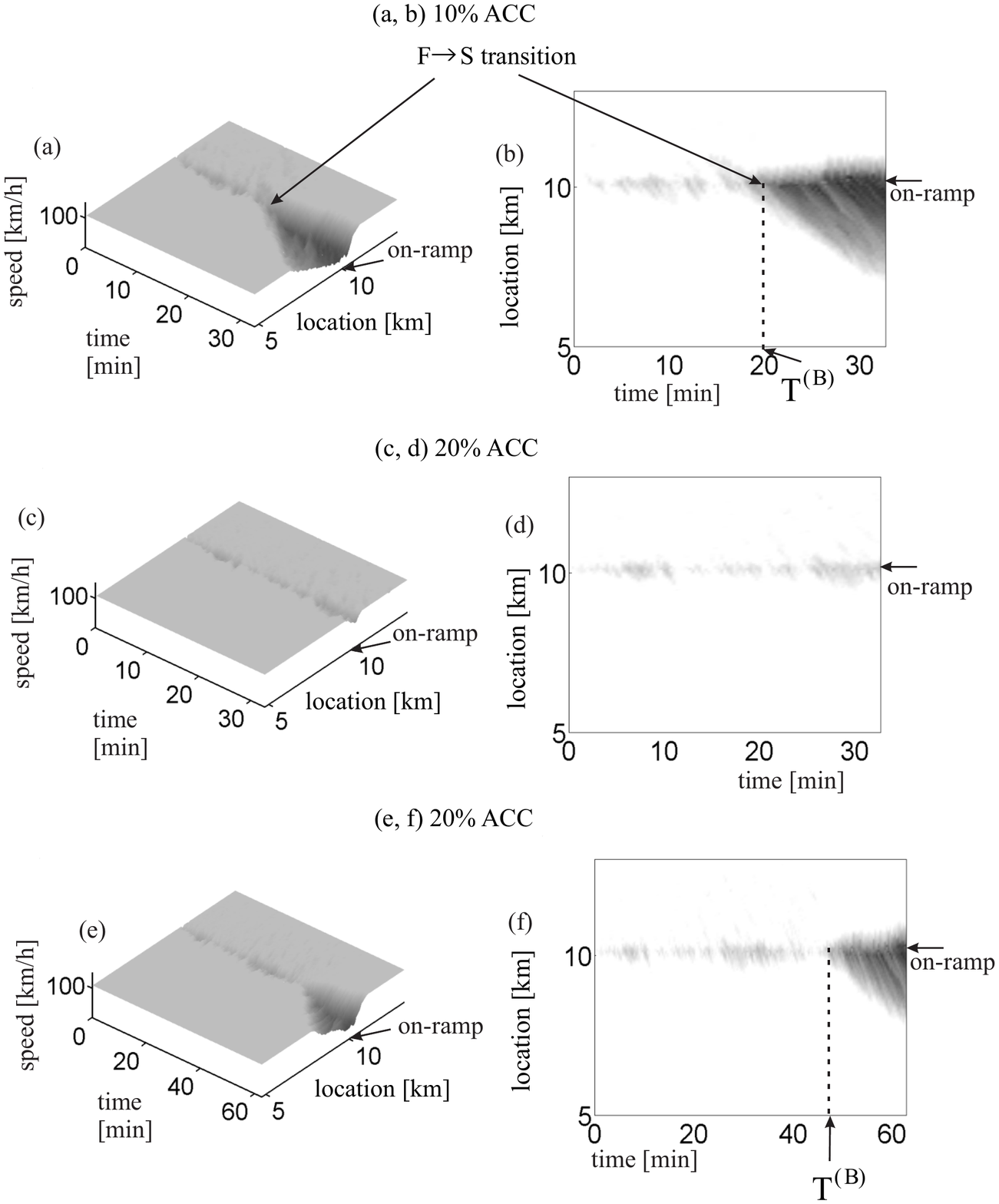}
\end{center}
\caption[]{Simulations of the effect of ACC-vehicles on probabilistic characteristic of
traffic breakdown:
 Vehicle speed in space and time (a, c, e) and the same speed data presented by regions with variable shades of gray (b, d, f) (in white regions the speed is equal to 105 km/h, in black regions the speed is equal to zero).
(a, b)   $\gamma=10 \%$.
(c--f)  $\gamma=20 \%$. ACC-parameters are the same as those in Fig.~\ref{ACC_FS}.
 Arrows F$\rightarrow$S in (a, b) mark the F$\rightarrow$S transition (traffic breakdown) at the location of on-ramp bottleneck.  F -- free flow, S -- synchronized flow. 
\label{ACC_stable} }
\end{figure}

\begin{figure}
\begin{center}
\includegraphics*[width=11 cm]{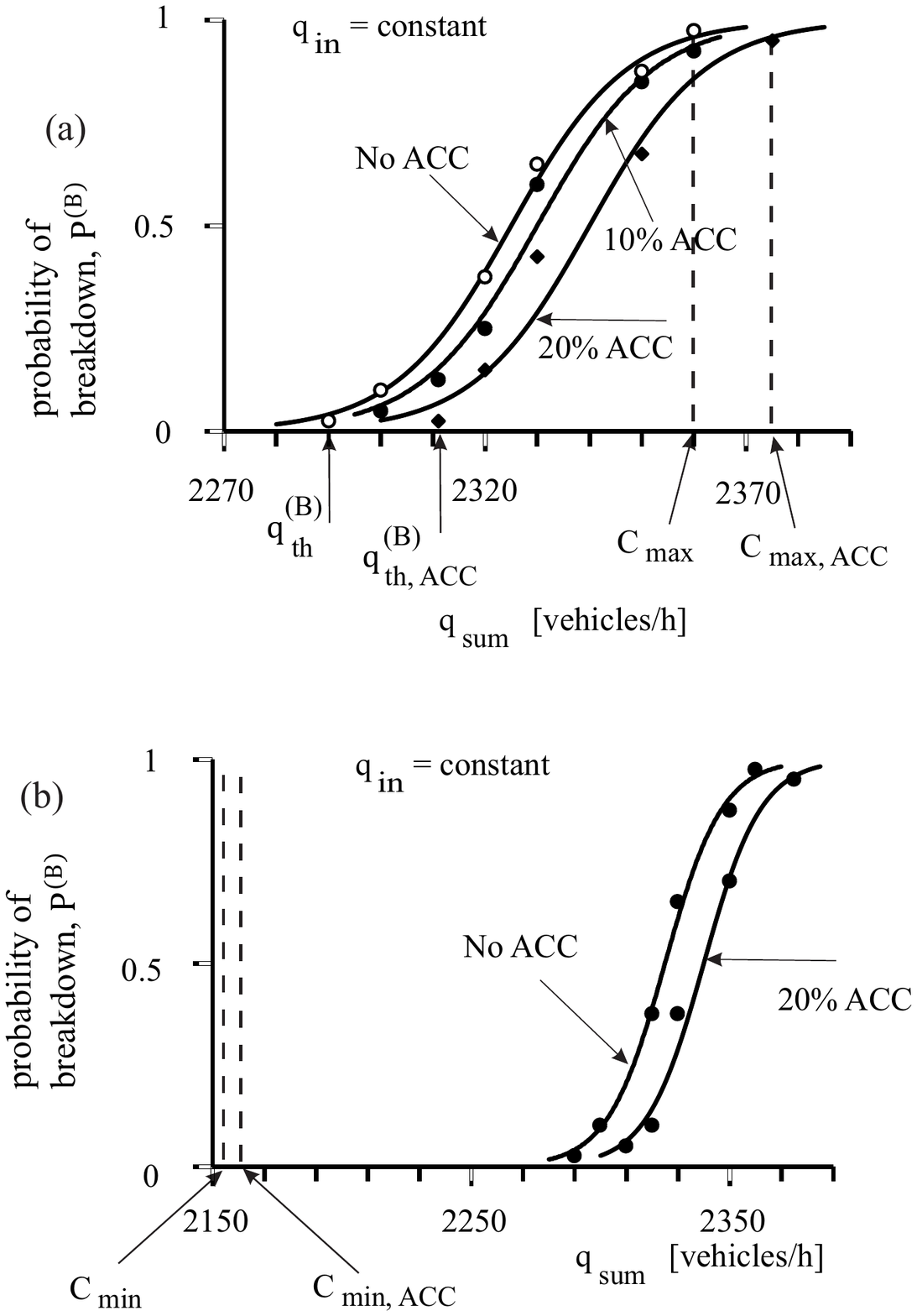}
\caption{Flow-rate functions of probabilities of traffic breakdown
$P^{\rm (B)}(q_{\rm sum})$ in traffic flows without ACC-vehicles
(left curves in (a, b)
labeled by $\lq\lq$No ACC") as well as with 10$\%$  and   20$\%$ of ACC-vehicles
  (right curves 
labeled by $\lq\lq$10$\%$ ACC" 
and $\lq\lq$20$\%$ ACC") as  functions of the flow rate downstream
of the bottleneck $q_{\rm sum}$; curves $\lq\lq$No ACC"
and $\lq\lq$20$\%$ ACC" are shown in different flow-rate scales    in (a) and (b). 
The flow rate $q_{\rm sum}=q_{\rm in}+  q_{\rm on}$ is varied through the change in the
on-ramp inflow rate $q_{\rm on}$ at constant 
$q_{\rm in}=$ 2000 vehicles/h.  To distinguish  
 the cases of traffic flows with the ACC-vehicles and without
 ACC-vehicles, 
 we denote the maximum capacity  $C_{\rm max}$, the minimum capacity $C_{\rm min}$, and the threshold flow rate   $q^{\rm (B)}_{\rm th}$
 for traffic flow with 20$\%$ of ACC-vehicles
  by $C_{\rm max, \ ACC}$, $C_{\rm min, \ ACC}$, and $q^{\rm (B)}_{\rm th, \ ACC}$, respectively.
	Functions $P^{\rm (B)}(q_{\rm sum})$ are described by formula (\ref{Prob_For}).
Other model parameters are the same as those in Fig.~\ref{ACC_FS}.
\label{ACC_Probability1} } 
\end{center}
\end{figure}

    At a smaller  percentage of ACC-vehicles than $\gamma=30 \%$
		in the mixture traffic flow with the same   parameters of the ACC-vehicles and the flow rates
		as those in Fig.~\ref{ACC_FS}, we have found the following results (Fig.~\ref{ACC_stable}).
 As long as the percentage of ACC-vehicles  is appreciably
smaller than 
  $\gamma=10 \%$,   no considerable
change in the probabilistic features of traffic breakdown at the bottleneck is observed. 
Even when the percentage of ACC-vehicles  increases to 
  $\gamma=10 \%$, features of traffic breakdown remains 
	almost the same as those in traffic flow of manual driving vehicles (Fig.~\ref{ACC_stable} (a, b))
	and only a relatively small decrease in the probability
	of the breakdown is observed (Fig.~\ref{ACC_Probability1} (a), curve
	labeled by $\lq\lq$10$\%$  ACC").

\begin{figure}
\begin{center}
\includegraphics*[width=12 cm]{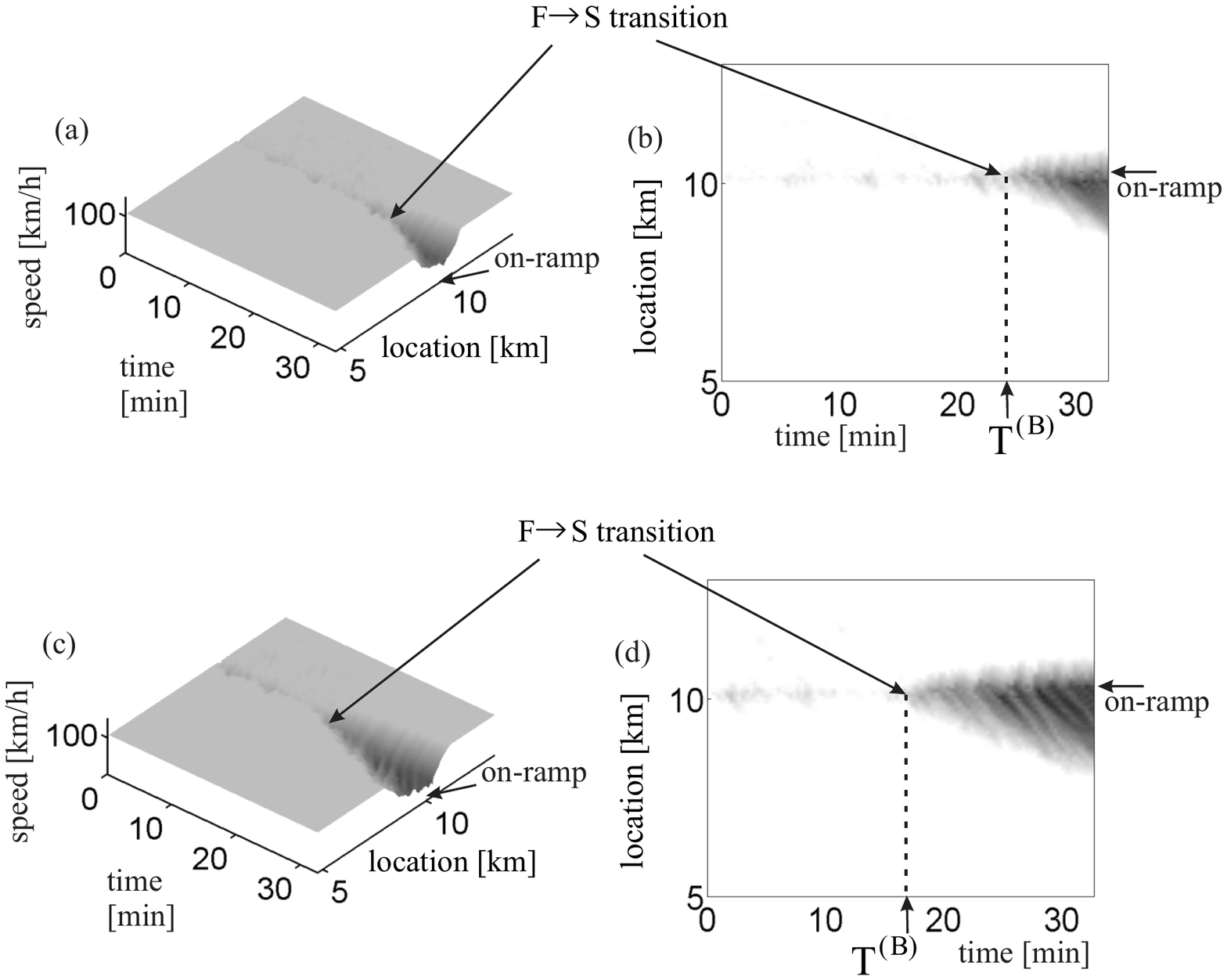}
\end{center}
\caption[]{Two different simulation realizations 2 (a, b) and 3 (c, d)
 of the effect of ACC on probabilistic features of traffic breakdown 
for $\gamma=20\%$ of ACC vehicles with the same time of observation of traffic flow $T_{\rm ob}=$ 30 min as that in simulation realization 1 in which no breakdown occurs  during $T_{\rm ob}=$ 30 min 
(Fig.~\ref{ACC_stable} (c, d)):
 Vehicle speed in space and time (a, c) and the same speed data presented by regions with variable shades of gray (b, d) (in white regions the speed is equal to 105 km/h, 
 in black regions the speed is equal to zero). Other model parameters 
 are the same as those in Fig.~\ref{ACC_FS}.  
\label{ACC_stable_r} }
\end{figure}

Now, in comparison with Fig.~\ref{ACC_stable} (a, b), we increase
the percentage $\gamma$ of automatic driving vehicles  to $\gamma=20 \%$  without any other changes
in simulations. We have found that at $\gamma=20 \%$  automatic driving vehicles
no traffic breakdown occurs during the observation time $T_{\rm ob}=30$ min (Fig.~\ref{ACC_stable} (c, d)).  However, if we continue
  simulations shown Fig.~\ref{ACC_stable} (c, d) during a longer time interval, we have found that after a long time delay $T^{\rm (B)}\approx 47$ min 
traffic breakdown has nevertheless occurred at the bottleneck (Fig.~\ref{ACC_stable} (c, d)).
We have found that the mean time delay  
 of traffic breakdown for $\gamma=20 \%$ of ACC vehicles is considerably longer than
 for traffic flow consisting of 
 only human driving vehicles.    
 
 For   $\gamma=20 \%$ of automatic driving vehicles the probability that traffic breakdown in the mixture traffic flow occurs
 during the observation time $T_{\rm ob}=30$ min is equal to    $P^{\rm (B)}=$ 0.1
for   model parameters used in Fig.~\ref{ACC_stable} (c, d).
Thus we can expect that in comparison with the simulation realization
 shown  in Fig.~\ref{ACC_stable} (c, d), in which no breakdown is observed, there should be other
 simulation realizations made at the same parameters of the mixture traffic flow, in which
 traffic breakdown is observed during the   time interval $T_{\rm ob}=30$ min. Such simulation realizations 
with different random values of time delays $T^{\rm (B)}$ 
to the breakdown  indeed exist (Fig.~\ref{ACC_stable_r}).

 The physics of these results can be explained as follows.
 In traffic flow consisting of human driving vehicles only, traffic breakdown (F$\rightarrow$S transition) occurs (Fig.~\ref{ACC_FS} (c, d)), when a large enough speed disturbance
 (nucleus for the breakdown) occurs in metastable free flow in a neighborhood of the bottleneck. Free flow
 that consists of $100\%$ automatic driving vehicles  
 is stable (Fig.~\ref{ACC_FS} (a, b)). For this reason, we can assume that
  a long enough platoon of ACC-vehicles that propagates through the disturbance can cause the dissolution of the disturbance.
  The larger the percentage $\gamma$ of automatic driving vehicles in the mixture traffic flow, the
  larger the probability of the appearance of the long platoon of ACC-vehicles,
  and, therefore, the larger the probability of the disturbance dissolution and the smaller the breakdown probability $P^{\rm (B)}$.

These qualitative explanations are confirmed by numerical simulations of
  the flow-rate dependence of the probability of 
	traffic breakdown $P^{\rm (B)}(q_{\rm sum})$ presented in Fig.~\ref{ACC_Probability1}.
  Indeed,   we have found that the flow-rate dependence of the probability of traffic breakdown
  for the mixture traffic flow with 20$\%$ automatic driving vehicles
(right curve  $P^{\rm (B)}(q_{\rm sum})$
labeled by $\lq\lq$20$\%$ ACC"  in Fig.~\ref{ACC_Probability1} (a)) is 
shifted to larger flow rates in comparison 
with the function $P^{\rm (B)}(q_{\rm sum})$  for traffic flow consisting of human driving vehicles only  (left curve  
labeled by $\lq\lq$No ACC" in Fig.~\ref{ACC_Probability1} (a)). 

Correspondingly,
 we have found that the maximum capacity $C_{\rm max}$ and the threshold flow rate
$q^{\rm (B)}_{\rm  th}$ for
spontaneous traffic breakdown increase
for the mixture flow. 
Traffic flow consisting of human driving vehicles only
and mixture traffic flow are different traffic flows. We can expect that
the minimum capacities $C_{\rm min}$ can also be different values in these two different traffic flows.
Indeed, we have found that in the mixture traffic flow
the minimum capacity $C_{\rm min, \ ACC}$ is slightly larger
 in comparison with the minimum capacity  $C_{\rm min}$
 in traffic flow consisting of human driving vehicles only (Fig.~\ref{ACC_Probability1} (b)).
 \begin{itemize}
\item Automatic driving vehicles can indeed decrease the probability of traffic breakdown
in the mixture free  flow.
\item Automatic driving vehicles can  increase the threshold flow rate
 for
spontaneous traffic breakdown as well as
  the maximum and minimum capacities of free flow at the bottleneck.
 \end{itemize}

  \section{Deterioration of  performance of traffic  system   through automatic driving vehicles
		\label{P2_Aut_Human_S}}

\begin{figure}
\begin{center}
\includegraphics*[width=11 cm]{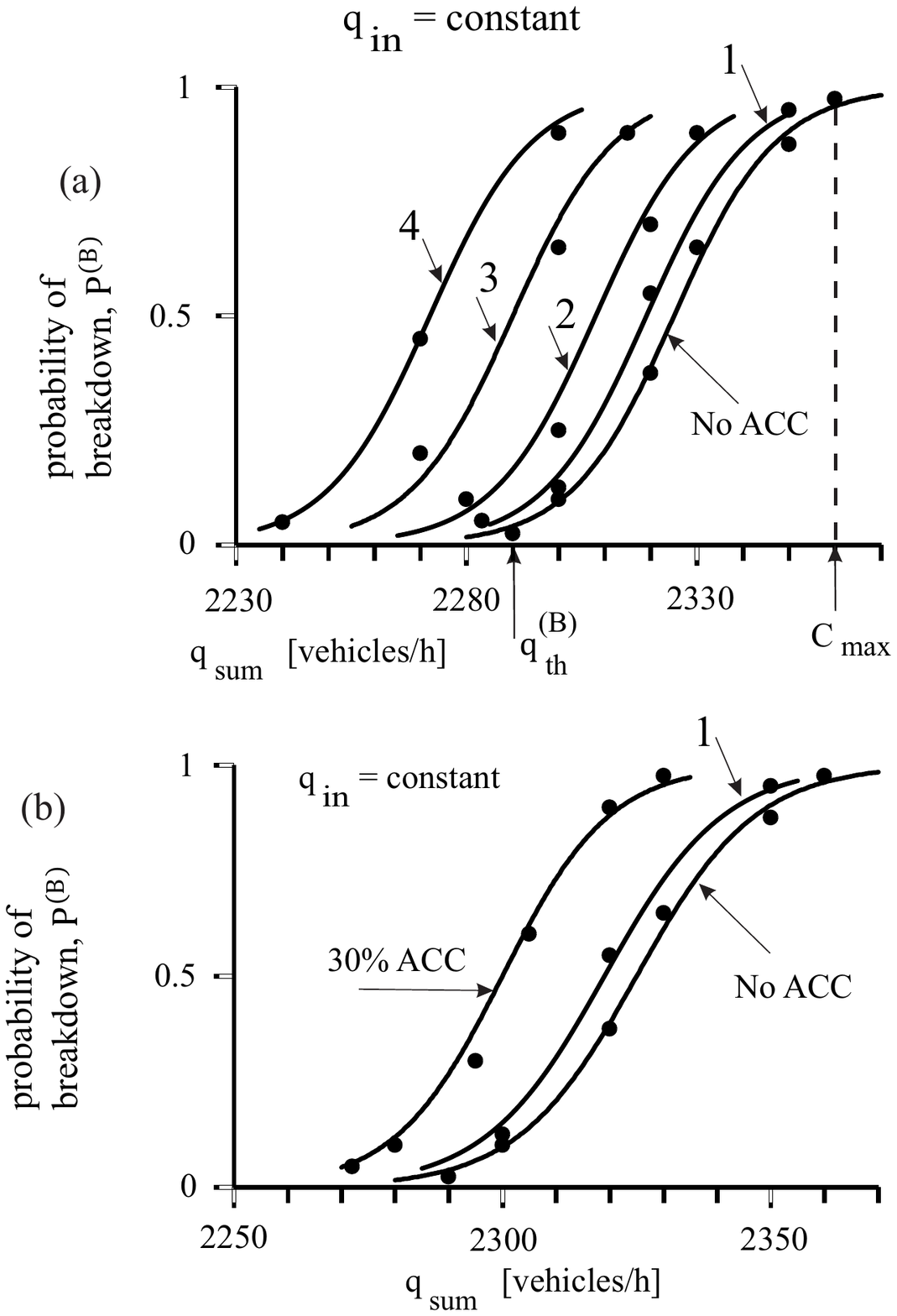}
\caption{Increase in the probability of traffic breakdown
$P^{\rm (B)}$ in mixture free traffic flow through ACC-vehicles
with $\tau^{\rm (ACC)}_{\rm d}=1.1$ s:
In (a, b), the flow-rate dependence of the breakdown probability $P^{\rm (B)}(q_{\rm sum})$
labeled by $\lq\lq$No ACC" is taken from Fig.~\ref{ACC_Probability1} for free flow of manual driving vehicles without ACC-vehicles.
In (a), curves $P^{\rm (B)}(q_{\rm sum})$  labeled by numbers
 1--4 are related to mixture free flows with $\gamma=5 \%$  
 of ACC-vehicles with different sets of coefficients $(K_{1}, \ K_{2})=$
(0.2 $s^{-2}$, 0.82 $s^{-1}$) for curve 1, 
(0.3 $s^{-2}$, 0.77 $s^{-1}$) for curve 2,
(0.5 $s^{-2}$, 0.65 $s^{-1}$) for curve 3,
(0.7 $s^{-2}$, 0.55 $s^{-1}$) for curve 4, which     satisfy 
  (\ref{String_stability}); 
	values $a_{\rm ACC}=b_{\rm ACC}=3 \ {\rm m}/{\rm s}^2$. 
In (b), curve $P^{\rm (B)}(q_{\rm sum})$ labeled by $\lq\lq$30$\%$ ACC" is
related to mixture free flow with $\gamma=30 \%$  
 of ACC-vehicles with   $(K_{1}, \ K_{2})=$
(0.2 $s^{-2}$, 0.82 $s^{-1}$); curve 1 is the same as that in (a).
The flow rate $q_{\rm sum}=q_{\rm in}+  q_{\rm on}$ is varied through the change in the
on-ramp inflow rate $q_{\rm on}$ at constant 
$q_{\rm in}=$ 2000 vehicles/h.  Functions $P^{\rm (B)}(q_{\rm sum})$ are described by formula (\ref{Prob_For}).
\label{ACC_Prob_inc} } 
\end{center}
\end{figure}

\begin{figure} 
\begin{center}
\includegraphics*[width=11 cm]{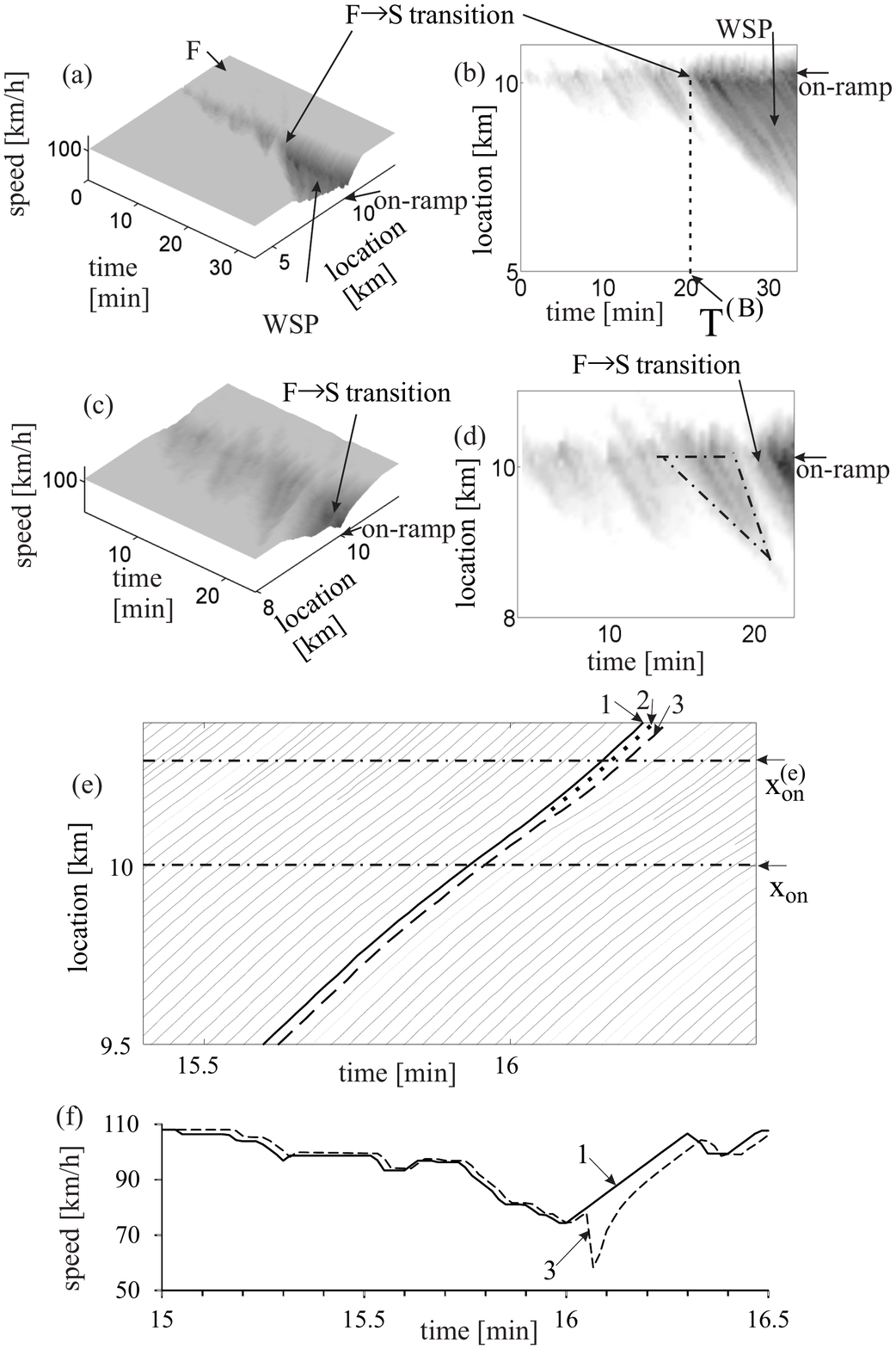}
 \end{center}
\caption{Simulations of dynamics of permanent speed disturbance  
 at on-ramp bottleneck on single-lane road  
 for a mixed traffic flow with
$\gamma=5 \%$  
 of ACC-vehicles with   coefficients $(K_{1}, \ K_{2})=$
(0.5 $s^{-2}$, 0.65 $s^{-1}$): 
(a--d) Speed in space and time  (a, c) and the same data
presented  by regions with variable shades of gray (b, d) (in white regions the speed
is  equal to  or higher than 105  km/h, in black regions the speed is equal to 0 km/h (b) and
smaller than 20 km/h (d));
in (c, d), we show the same data as those in (a, b), however,   for    smaller space and time  intervals. 
(e) Fragment of vehicle trajectories in space and time related to (c, d). (f) Microscopic vehicle speeds along trajectories as time functions
labeled by the same numbers as those in (e). In (d), dashed-dotted lines denote 
  F$\rightarrow$S$\rightarrow$F transitions.
F -- free flow,   WSP -- widening synchronized flow pattern.
$q_{\rm on}=$ 320   vehicles/h,
$q_{\rm in}=$  2000   vehicles/h.  
}
\label{FSF_KKl}
\end{figure}

\begin{figure} 
\begin{center}
\includegraphics*[width=11 cm]{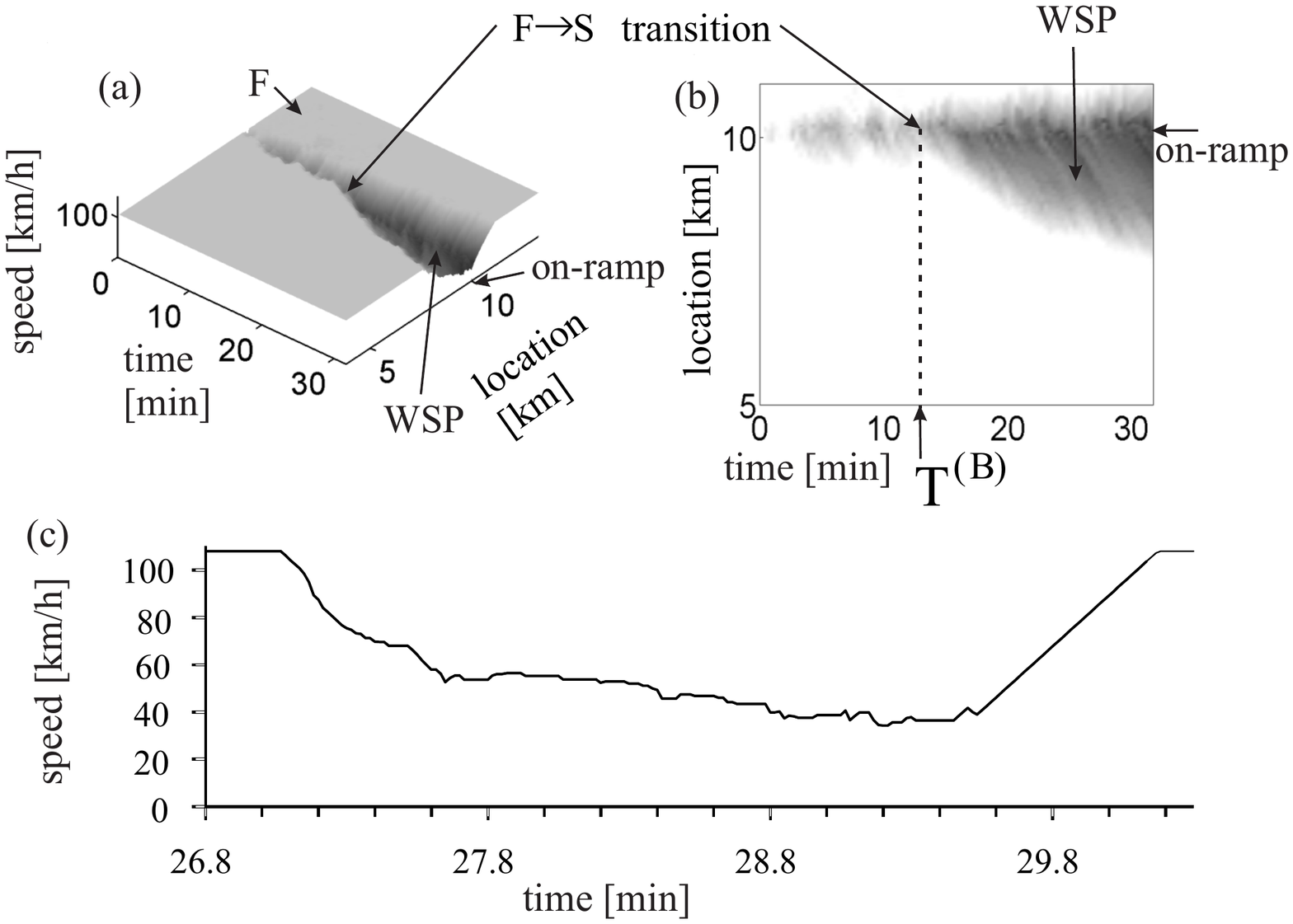}
 \end{center}
\caption{Simulations of F$\rightarrow$S  transition (traffic breakdown)  
 at on-ramp bottleneck on single-lane road for  mixture traffic flow with $\gamma=30 \%$  
 of ACC-vehicles with   $(K_{1}, \ K_{2})=$
(0.2 $s^{-2}$, 0.82 $s^{-1}$) related to curve labeled by $\lq\lq$30$\%$ ACC" in
 Fig.~\ref{ACC_Prob_inc} (b): 
(a, b) Speed in space and time  (a) and the same data
presented  by regions with variable shades of gray (b) (in white regions the speed
is  equal to  or higher than 105  km/h, in black regions the speed is equal to 0 km/h). 
(c) One of vehicle trajectories as   time-function  that propagates through WSP
in (a, b). 
F -- free flow,   WSP -- widening synchronized flow pattern.
$q_{\rm on}=$ 305   vehicles/h,
$q_{\rm in}=$  2000   vehicles/h.    
}
\label{FSF_KKl_ACC}
\end{figure}

			  Rather than the enhancement of traffic flow characteristics
				(Secs.~\ref{Crit_Sup_S} and~\ref{Aut_Nature_S}),  
			 automatic driving vehicles    can also result in the deterioration of  performance of traffic  system.
			To show this, we consider a mixture traffic flow under  condition
(\ref{Stable_flow_q_sum}). In this mixture traffic flow, 
ACC-vehicles exhibit the same short desired time headway $\tau^{\rm (ACC)}_{\rm d}=1.1$ s
 as   used above in Secs.~\ref{Crit_Sup_S} and~\ref{Aut_Nature_S}.
Moreover, all sets of dynamics coefficients $K_{1}$ and $K_{2}$ of
ACC-vehicles used below (see Fig.~\ref{ACC_Prob_inc}--\ref{FSF_KKl_ACC})
satisfy  condition of string stability (\ref{String_stability}).     
 Nevertheless,  we will find that the ACC-vehicles
 can lead to a considerable increase in the probability of traffic breakdown
		at road bottlenecks (Fig.~\ref{ACC_Prob_inc})\footnote{In this mini-review, 
		we do   not  discuss another possible
			case of the deterioration of the performance of traffic 
		system that is often assumed to occur  
		when automatic driving vehicles follow strictly  
		all traffic regulation rules, like a given speed limit.}.

 Indeed, we have found that even a relatively small 
percentage $\gamma=5 \%$ of
the ACC-vehicles in the mixture traffic flow can
 {\it increase} considerably 
the probability of traffic breakdown (Fig.~\ref{ACC_Prob_inc} (a), curves 1--4).
The importance of this result is as follows: In Secs.~\ref{Crit_Sup_S} and~\ref{Aut_Nature_S}, we have mentioned
that the positive effect of the ACC-vehicles on traffic flow, in particular,
the decrease in the probability of traffic breakdown is considerable  only
at   large percentages   of ACC-vehicles $\gamma\approx 20 \%$.
In the next future, we could expect only much smaller 
 percentages   of  automatic driving vehicles in the mixture traffic flow, like 
$\gamma\approx 5 \%$. Therefore, the deterioration of the performance of a mixture traffic flow
shown Fig.~\ref{ACC_Prob_inc} (a) (curves 1--4)
can be a subject of the development of automatic driving vehicles in car-development companies
already during next years.

To understand the deterioration of the performance of the traffic system through 
automatic driving vehicles (Fig.~\ref{ACC_Prob_inc}), firstly
we should note that for each set of dynamic coefficients $(K_{1}, \ K_{2})$ of the ACC-vehicles
used in simulations shown Fig.~\ref{ACC_Prob_inc} (a) (curves 1--4),
free  flow consisting of $\gamma= 100 \%$ of the ACC-vehicles is stable:
We have found   the same results for free flow stability as those presented in
Fig.~\ref{ACC_FS} (a, b).
This means that when in the mixture free
  flow the percentage   of the ACC-vehicles increases, then, at least
under condition  $\gamma\rightarrow 100 \%$, 
no traffic breakdown should be observed any more. This  has  indeed
been found in simulations.

We have found that for any of the sets of dynamic coefficients $(K_{1}, \ K_{2})$ 
used in Fig.~\ref{ACC_Prob_inc}, when 
the percentage   of the ACC-vehicles increases from $\gamma= 5\%$ to   larger values,
firstly,  
the flow rate dependence of the breakdown probability $P^{\rm (B)}(q_{\rm sum})$
is subsequently shifted to the left in the flow rate axis as shown in Fig.~\ref{ACC_Prob_inc} (b) (curve
labeled by $\lq\lq$30$\%$ ACC").
 
 However,
there should be a critical percentage of the ACC-vehicles
denoted by  $\gamma^{\rm (increase)}_{\rm cr}$. When $\gamma=\gamma^{\rm (increase)}_{\rm cr}$,
then the shift
of the function $P^{\rm (B)}(q_{\rm sum})$ to the left in the flow rate axis
should reach its maximum.
 When the percentage of ACC-vehicles increases subsequently, i.e,
 $\gamma>\gamma^{\rm (increase)}_{\rm cr}$, then 
the function $P^{\rm (B)}(q_{\rm sum})$ should begin to be shifted to the right
 in the flow rate axis in comparison with the case 
$\gamma=\gamma^{\rm (increase)}_{\rm cr}$. Finally, as above-mentioned, at 
$\gamma\rightarrow 100 \%$, free flow should occur
as long as condition (\ref{Stable_flow_q_sum}) is satisfied.

The above assumption about the behavior of   the function $P^{\rm (B)}(q_{\rm sum})$ 
under increase in the percentage of the ACC-vehicles is indeed confirmed by   
simulation results, which have been
 made for each of the sets of dynamic coefficients $(K_{1}, \ K_{2})$ of the ACC-vehicles
used in  Fig.~\ref{ACC_Prob_inc}.
Moreover, it turns out that already for dynamic coefficients
$(K_{1}, \ K_{2})=$
(0.2 $s^{-2}$, 0.82 $s^{-1}$), which do not  considerably differ
from $(K_{1}, \ K_{2})=$
(0.14 $s^{-2}$, 0.9 $s^{-1}$) used in Figs.~\ref{ACC_FS}--\ref{ACC_Probability1},
we have found that $\gamma^{\rm (increase)}_{\rm cr}\approx 30 \%$
(curve $P^{\rm (B)}(q_{\rm sum})$ labeled by $\lq\lq$30$\%$ ACC" in Fig.~\ref{ACC_Prob_inc} (b)).
This leads to the following result:
\begin{itemize}
\item	The  deterioration  
   of the performance of the traffic system through 
	automatic driving vehicles
		can occur within   broad ranges 
		of the percentage of ACC-vehicles and the set of coefficients  $(K_{1}, \ K_{2})$, which
		satisfy 
  condition  (\ref{String_stability})
		 of string stability of ACC-vehicles.		
\end{itemize}

\begin{figure} 
\begin{center}
\includegraphics*[width=11 cm]{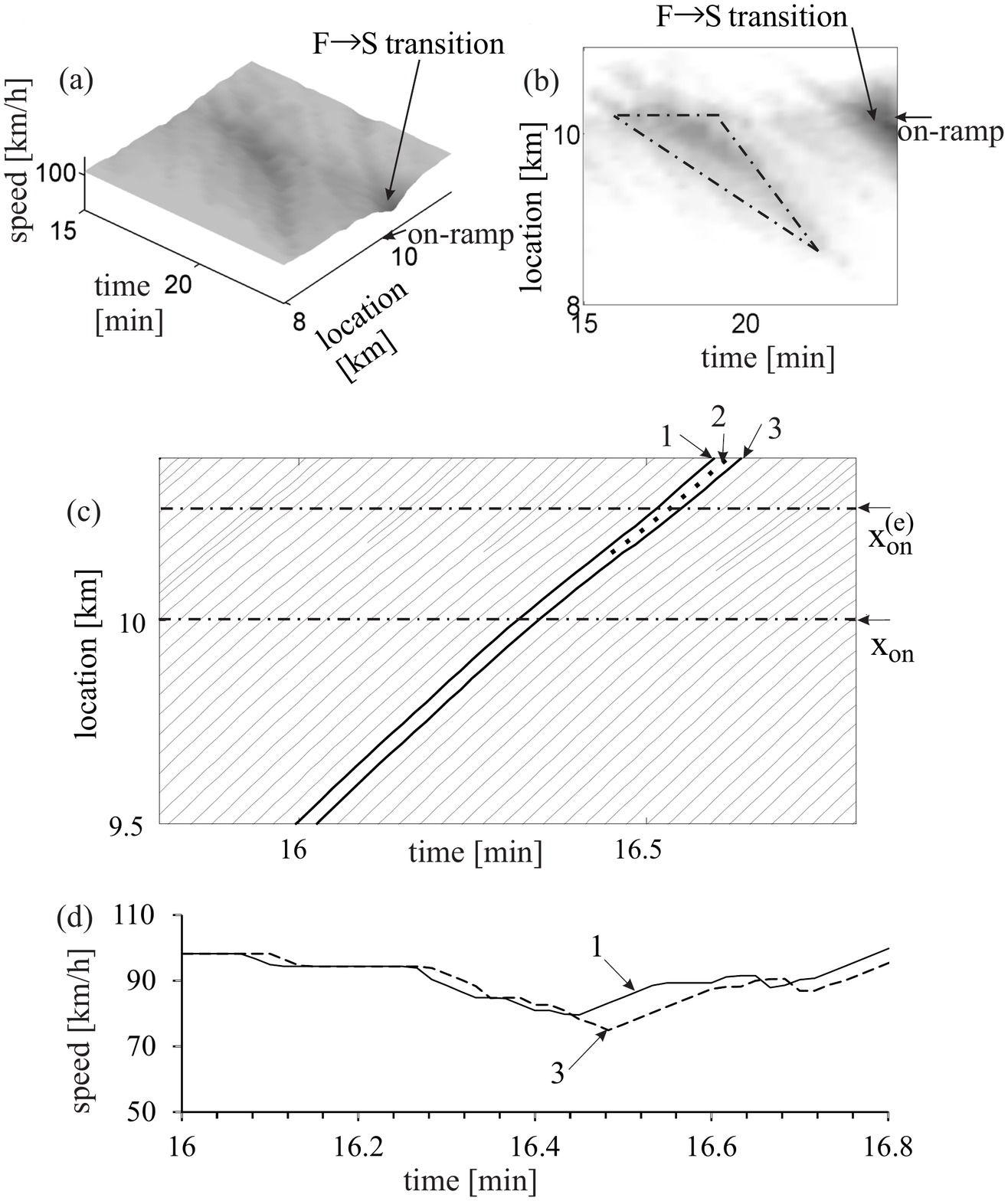}
 \end{center}
\caption{Simulations of dynamics of permanent speed disturbance  
 at on-ramp bottleneck on single-lane road for traffic flow consisting only of
 manual driving vehicles 
 related to Fig.~\ref{ACC_FS} (c, d): 
(a, b) Speed in space and time  (a) and the same data
presented  by regions with variable shades of gray (b) (in white regions the speed
is  equal to  or higher than 105  km/h, in black regions the speed is smaller than   20 km/h);
in (a, b),  we show the same data as those in Fig.~\ref{ACC_FS} (c, d), however,   for  
    smaller space and time  intervals. 
(c) Fragment of vehicle trajectories in space and time related to (a, b). (d) Microscopic vehicle speeds along trajectories as time functions
labeled by the same numbers as those in (c). In (b), dashed-dotted lines denote 
  F$\rightarrow$S$\rightarrow$F transitions.
	$q_{\rm on}=$ 320   vehicles/h,
$q_{\rm in}=$  2000   vehicles/h.  
}
\label{FSF_KKl_NO_ACC}
\end{figure}

To understand this negative effect of the ACC-vehicles on
traffic flow, firstly note that  as long as the percentage of  the ACC-vehicles
in a mixture traffic flow
is not very large,  traffic breakdown
at the bottleneck is qualitatively the same time-delayed  F$\rightarrow$S transition
(Figs.~\ref{FSF_KKl} (a, b)
	and~\ref{FSF_KKl_ACC} (a, b)) as that   in traffic flow consisting of manual driving vehicles only
	(Fig.~\ref{ACC_FS} (c, d)). In both cases,  WSPs result from the breakdown
	(Figs.~\ref{ACC_FS} (c, d),~\ref{FSF_KKl} (a, b), 
	and~\ref{FSF_KKl_ACC} (a, b)).
Moreover, in both cases,  
before traffic breakdown occurs
	(time intervals $0<t<T^{\rm (B)}$ in Figs.~\ref{ACC_FS} (c, d),~\ref{FSF_KKl} (a, b), 
	and~\ref{FSF_KKl_ACC} (a, b)), there are
	many F$\rightarrow$S$\rightarrow$F transitions at the bottlenecks
	(dashed-dotted lines shown in Figs.~\ref{FSF_KKl} (d)
	and~\ref{FSF_KKl_NO_ACC} (b) denote some of the regions of 
  synchronized flow occurring due to a sequence of the F$\rightarrow$S$\rightarrow$F transitions).
	As explained in~\cite{Kerner2015B}, the F$\rightarrow$S$\rightarrow$F transitions determine the dynamics of
a permanent speed disturbance at the bottleneck\footnote{A detailed consideration of the
	  physics of the F$\rightarrow$S$\rightarrow$F transitions 
		and the related dynamics 	of
the permanent speed disturbance at the bottleneck~\cite{Kerner2015B}  is out of scope of this mini-review.}.

		A crucial difference between a mixture traffic flow and traffic flow without ACC-vehicles
		becomes clear, when we consider  
		 the dynamics of a permanent speed disturbance at the bottleneck:
		We have found that  
	 the amplitude of the permanent speed disturbance at the bottleneck occurring in the mixture
		traffic flow can increase considerably in comparison  with that  occurring 
		in free flow consisting only of
		manual driving vehicles (Figs.~\ref{FSF_KKl} (c--f)
		and~\ref{FSF_KKl_NO_ACC}).
			
			When a   vehicle moving at a low speed in the on-ramp lane
			merges from the on-ramp to the main road
			(bold dotted trajectory 2 in Fig.~\ref{FSF_KKl_NO_ACC} (c)) between two vehicles moving on the main
			road (bold   trajectories 1 and 3 in Fig.~\ref{FSF_KKl_NO_ACC} (c)),
		then,
			in comparison  with
			vehicle 1 moving on the main road that is not influenced by the merging vehicle,
		   vehicle 3 (bold   trajectory   3 in Fig.~\ref{FSF_KKl_NO_ACC} (c))
			should decelerate to the speed of this merging vehicle.
			As a result, due to the vehicle merging the speed decreases  within the permanent speed disturbance
			localized at the bottleneck
			(compare speeds of vehicles 1 and   3 in Fig.~\ref{FSF_KKl_NO_ACC} (d)). 
			
			It turns out that the effect of the speed reduction
caused by the merging vehicle can increase considerably, when  
		vehicle 3 is an ACC-vehicle 
		(bold dashed trajectory 3 in Fig.~\ref{FSF_KKl} (e)). Indeed,
		the deceleration of the ACC-vehicle
		(vehicle 3 in Fig.~\ref{FSF_KKl} (f)) due to the merging vehicle 
		(bold dotted trajectory 2 in Fig.~\ref{FSF_KKl} (e)) becomes considerably larger than in the case of
		traffic flow without ACC-vehicles
		(compare  speeds of vehicles 1 and   3 in Fig.~\ref{FSF_KKl} (f) with 
		speeds of vehicles 1 and   3 in Fig.~\ref{FSF_KKl_NO_ACC} (d), respectively).
		Due to a stronger  speed reduction at the bottleneck
		caused by the ACC-vehicles, the probability of traffic breakdown increases
		at the same flow rates as those in the case of traffic flow
		consisting of manual driving vehicles only (Fig.~\ref{ACC_Prob_inc} (a)).
		
		When  the percentage $\gamma$ of the ACC-vehicles increases, the frequency of 
		 large speed disturbances  at the bottleneck
		caused by the ACC-vehicles increases either.
		This explains why at given flow rates the probability of traffic breakdown
		increases when $\gamma$ increases from $\gamma\approx 5\%$ to 
			$\gamma^{\rm (increase)}_{\rm cr}$
(the value $\gamma^{\rm (increase)}_{\rm cr}\approx 30\%$ for parameters of ACC-vehicles used 
in Fig.~\ref{ACC_Prob_inc} (b)). 
Only when the percentage 
of the ACC-vehicles $\gamma>\gamma^{\rm (increase)}_{\rm cr}$,
 long stable platoons of the ACC-vehicle can occur
that lead to a decrease in the breakdown probability.
Indeed, free flow consisting of  $100\%$ ACC-vehicles is stable.
		
		To understand the physics of the deterioration of the performance of
		the traffic system through the ACC-vehicles
		in more details, 
		we should note that in accordance with the hypothesis of
		the three-phase theory about the existence of 2D-states
		of traffic 
		flow~\cite{KernerBook,KernerBook2,Kerner1998E,Kerner1998C,Kerner1999B,Kerner1999D,Kerner2000D,Kerner2000A,Kerner2000B,Kerner2001A,Kerner2001B,Kerner2002A,Kerner2002B,Kerner2002C,Kerner2002D,Kerner2003A,Kerner2004A}, 
		drivers do not control the space gap $g$ to the preceding vehicle
		when condition
		\begin{eqnarray}
 g_{\rm safe}  \leq g \leq G,
\label{q_safe_syn}
\end{eqnarray}
is satisfied, where  
$G$ and $g_{\rm safe}$ are the
synchronization and safe space gaps, respectively.

In contrast to the manual driver behavior (\ref{q_safe_syn}),
in accordance with the classical ACC-model  (\ref{ACC_dynamics_Eq}),
  the ACC-vehicle tries to reach an $\lq\lq$optimal"
space gap  given by formula (\ref{Gap_ACC_steady_For_e}).
This qualitative different dynamic behavior of the ACC-vehicles and manual driving vehicles
could explain the occurrence of large disturbances in free flow at the bottleneck.
When the space gap between the ACC-vehicle and the merging manual vehicle
is smaller than that given by formula (\ref{Gap_ACC_steady_For_e}), the ACC-vehicle decelerates,
 whereas
a manual driving vehicle should not decelerate
 as long as condition (\ref{q_safe_syn}) is satisfied.
The   deceleration of the ACC-vehicle is the stronger, the
larger the coefficient $K_{1}$ in (\ref{ACC_dynamics_Eq}). This can explain the result of simulations that the larger the coefficient $K_{1}$, the more is 
the shift of the function $P^{\rm (B)}(q_{\rm sum})$ to the left in the  flow rate-axis
(curves 1--4 in
Fig.~\ref{ACC_Prob_inc} (a)).
We can make the following conclusion.

\begin{itemize}
\item When dynamics rules of motion of automatic driving vehicles differ considerably from those of human driving vehicles, such automatic driving vehicles can cause
 the deterioration of the performance of
		the traffic system. In particular, speed disturbances
in a mixture traffic flow 
 at road bottlenecks can increase strongly. This can cause
the considerable increase in the probability of traffic breakdown.
\end{itemize}
Naturally, through the use of cooperative driving 
 (Sec.~\ref{Coop_Nature_S}) this negative effect
of automatic driving vehicles on traffic flow   could be 
reduced\footnote{In particular, one  can expect that
cooperative merging could alleviate the problem of large disturbances occurring
during vehicle merging at the bottleneck
in mixture traffic flow illustrated in Fig.~\ref{FSF_KKl}.}.
However, it seems better to develop such dynamics rules of motion of automatic driving vehicles 
that avoid situations at which human drivers can   consider automatic driving vehicles
as  $\lq\lq$obstacles".

\section{Automatic driving   
vehicles learning from driver behavior in real traffic: 
ACC in framework of three-phase theory  \label{Aut_Hum_S}} 

The   deterioration of the performance of
		the traffic system through the ACC-vehicles 
discussed in Sec.~\ref{P2_Aut_Human_S}  could be avoided through
the use of automatic driving systems in vehicles, which learn from  
behaviors of drivers in real traffic as  incorporated in hypotheses of the three-phase  theory.

In particular, in accordance with the hypothesis of three-phase    theory
about 2D-states of traffic flow~\cite{KernerBook,KernerBook2,Kerner1998E,Kerner1998C,Kerner1999B,Kerner1999D,Kerner2000D,Kerner2000A,Kerner2000B,Kerner2001A,Kerner2001B,Kerner2002A,Kerner2002B,Kerner2002C,Kerner2002D,Kerner2003A,Kerner2004A},
 we have introduced  ACC-systems~\cite{KernerPat20039,KernerPat20039A,KernerPat20039B}, in which, in contrast with   the classical model of the ACC-vehicle 
  (\ref{ACC_General}), 
there is {\it no} fixed desired time headway of ACC-vehicle to the preceding vehicle.

\begin{figure}
\begin{center}
\includegraphics*[width=11 cm]{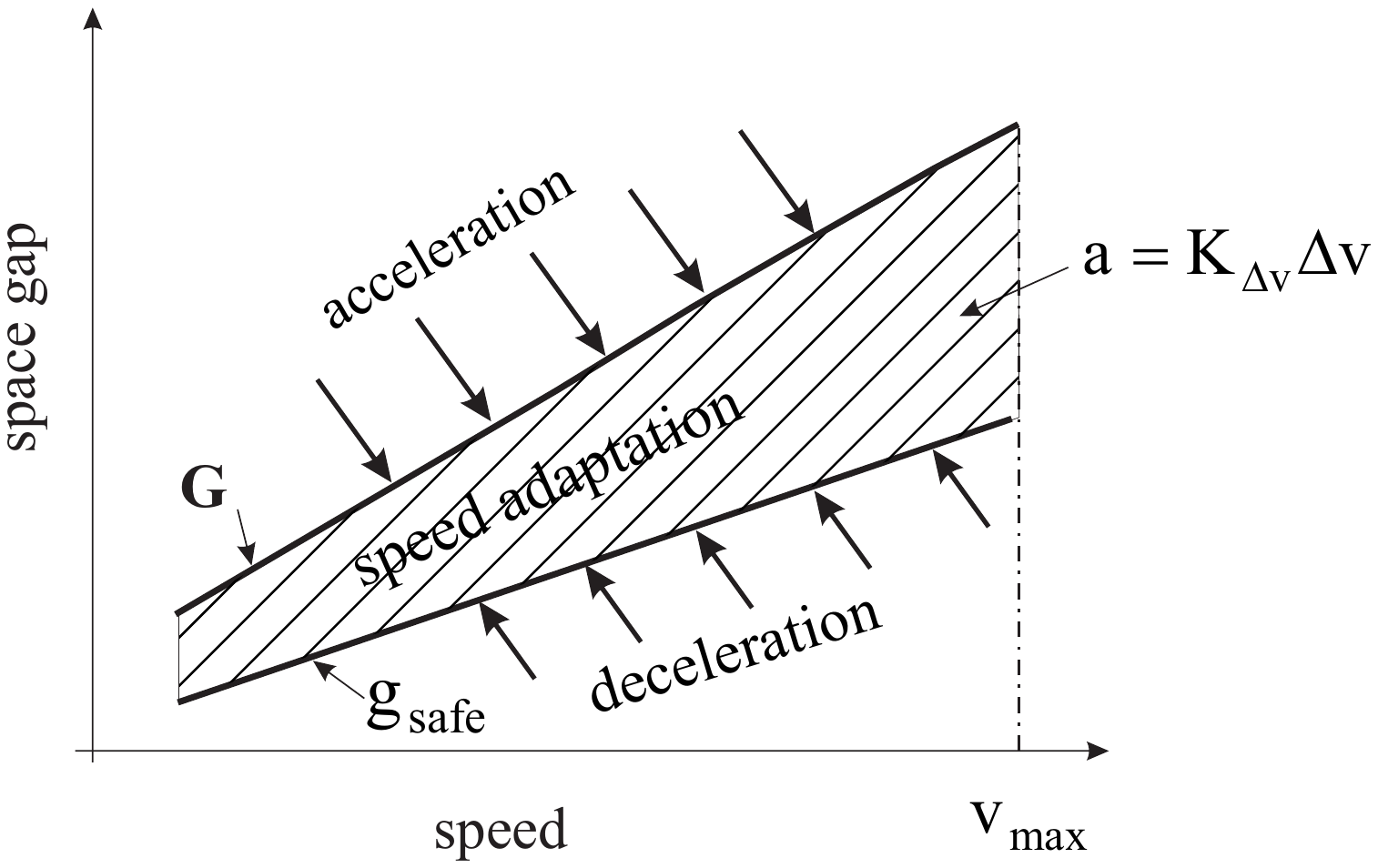}
\end{center}
\caption[]{Explanation of   ACC   in the framework of 
three-phase   theory~\cite{KernerPat20039,KernerPat20039A,KernerPat20039B}: 
 A part of 2D-states of  traffic flow  in
the space-gap--speed plane (dashed region) within which an ACC-vehicle moves in accordance with  
Eq.~(\ref{TPACC_Eq1}).
}
\label{Eco_ACC}
\end{figure}

  In these ACC-systems based on the three-phase   
	theory~\cite{KernerPat20039,KernerPat20039A,KernerPat20039B}, 
 when the space gap   to the preceding vehicle is  within a 2D-region in
the space-gap--speed plane (dashed region in Fig.~\ref{Eco_ACC}), i.e.,  
condition (\ref{q_safe_syn}) is satisfied, then the acceleration (deceleration) of an
 ACC-vehicle is given by formula 
\begin{eqnarray}
a(t)=K_{\rm \Delta v}\Delta v(t).
\label{TPACC_Eq1}
\end{eqnarray}
This means that the ACC-vehicle  
adapts its speed to the speed of the preceding vehicle
 without caring, what the precise space gap (time headway) is.
In (\ref{TPACC_Eq1}),
$K_{\rm \Delta v}$ is a dynamic coefficient ($K_{\rm \Delta v}>0$).   
At $g>G$ the  ACC-vehicle  
accelerates, whereas at $g<g_{\rm safe}$ the  ACC-vehicle decelerates.
 Outside of the 2D-region in
the space-gap--speed plane (Fig.~\ref{Eco_ACC})
  formula (\ref{TPACC_Eq1})   is not 
	applied~\cite{KernerPat20039B,KernerPat20039D,KernerPat20039E,KernerPat20039H}.

Thus, in some traffic situations 
acceleration (deceleration) of 
  the ACC-vehicle  
does not depend on the space gap, i.e., on the time headway
to the preceding vehicle at all. In other words,  
the ACC mode (\ref{TPACC_Eq1}) does not maintain some desired time headway of  the classical model of the ACC-vehicle 
  (\ref{ACC_General}).
Moreover, the dynamic coefficient  
$K_{\rm \Delta v}$ in    (\ref{TPACC_Eq1})
can be chosen to be considerably smaller than
  the dynamic coefficient $K_{2}$ in the classical model of the ACC-vehicle 
  (\ref{ACC_General}).
This explains the following possible advantages   of the  ACC-system based on three-phase  theory
in comparison with the classical ACC-system (\ref{ACC_General}):
\begin{itemize}   
\item The removing of a conflict between the dynamic and comfortable ACC behavior. In particular,
a   much  comfortable
 vehicle motion    is possible. 
 \item The reduction of fuel consumption and $\rm CO_{2}$ emissions    
 while moving in congested traffic.  
\item Because the  ACC-mode (\ref{TPACC_Eq1})  decreases speed changes in traffic flow,
a sequence of such   ACC-vehicles can prevent   traffic breakdown at a bottleneck.
\end{itemize}
Therefore, a development of automatic driving vehicles based on the three-phase theory
  can be a very interesting task for further 
	investigations.

   \section{Conclusions}
   
   (i)	The empirical metastability of free flow with respect to the F$\rightarrow$S transition (traffic breakdown) at
a highway bottleneck
can be considered the  empirical fundament  of transportation science.
   
(ii)	The theoretical fundament   of transportation science
resulting from the above empirical fundament is as follows: 
{\it At any time instant}, there are the infinite number of highway capacities within a range of the flow rate between the minimum capacity and the maximum capacity; 
within this flow rate range, traffic breakdown can be induced at the bottleneck. 

(iii) Additionally to the minimum capacity and the maximum capacity,
an important characteristic of   traffic breakdown is a threshold flow rate for spontaneous
traffic breakdown at a bottleneck.

(iv)	Classical traffic and transportation theories failed to explain  the empirical evidence that traffic breakdown at the bottleneck is an F$\rightarrow$S transition
   occurring in metastable free flow at the bottleneck. 
   For this reasons, traffic flow models, which are based on these classical traffic theories, cannot be used 
   for a reliable analysis of the impact of automatic driving and/or other ITS-applications on traffic flow.  
   Simulations of the effect of automatic driving and/or other ITS-applications on traffic flow with the use of such traffic simulation models and tools lead to incorrect results and invalid conclusions. 

(v)	To perform reliable analysis of the impact of automatic driving and/or other ITS-applications on traffic flow, traffic flow models in the framework of the three-phase   theory should be  used. 
This is because these models can explain the empirical evidence 
that traffic breakdown at the bottleneck is an F$\rightarrow$S transition
   occurring in metastable free flow at the bottleneck.

   (vi) Based on simulations with a stochastic three-phase traffic flow model we have found that depending on the parameters of
     automatic driving vehicles, they 
	   can   either decrease or increase
   the probability of traffic breakdown   in a mixture
	traffic flow consisting of a random distribution
	of automatic driving and manual driving  vehicles. The increase in
   the probability of traffic breakdown at a bottleneck, i.e.,
	the deterioration of the performance of  the
	traffic system can occur already at a small percentage (about 5$\%$) of 
		 automatic driving   vehicles.
		The negative effect of the automatic driving   vehicles on traffic flow can be realized,
	even if any platoon of
	the automatic driving vehicles satisfies condition for string stability;  this effect  occurs
	even at the same flow rates at the bottleneck at which
	there is no traffic breakdown in
	free flow consisting of 100$\%$ of automatic driving vehicles.

   \appendix

   \section{Kerner-Klenov stochastic microscopic model in framework of three-phase traffic theory \label{App2}}

 In a discrete model version of Kerner-Klenov stochastic microscopic 
three-phase model~\cite{KKl2009A} used for simulations in Secs.~\ref{Aut_Cl_S}   
and~\ref{P2_Aut_Human_S}, 
 rather than the continuum space co-ordinate of~\cite{KKl,KKl2003A},   
a discrete space co-ordinate with a small enough value  
of the discrete cell $\delta x=0.01$ m  is used.
 Consequently,  
  the vehicle speed and acceleration (deceleration) discretization intervals are $\delta v$= $\delta x/\tau$
 and   $\delta a$= $\delta v/\tau$, respectively, where time step $\tau=$ 1 s. Because in 
  the discrete model version     discrete (and dimensionless) values of speed and acceleration 
 are used, which are measured respectively in  values $\delta v$ and  $\delta a$, 
and time is  measured in values of $\tau$,
   value $\tau$ in all formulas below is assumed to be the dimensionless value $\tau=1$.

\subsection{Update rules of vehicle motion in road lane \label{Identical_KKl_Up}}
 
Update rules of vehicle motion  are as follows~\cite{KKl,KKl2003A,KKl2009A}:
\begin{equation}
v_{n+1}=\max(0, \min({v_{\rm free}, \tilde v_{n+1}+\xi_{n}, v_{n}+a \tau, v_{{\rm s},n} })),
\label{final}
\end{equation}
\begin{equation}
\label{next_x}
x_{n+1}= x_{n}+v_{n+1}\tau,
\end{equation}
where the index $n$ corresponds 
to the discrete time $t_{\rm n}=\tau n, \ n=0,1,...$, 
$v_{n}$ is the vehicle speed at time step $n$, $a$ is the maximum acceleration,
$\tilde v_{n}$ is the vehicle speed  without  speed fluctuations $\xi_{n}$:
\begin{equation}
 \tilde v_{n+1}= \min(v_{\rm free},  v_{{\rm s},n}, v_{{\rm c},n}),
 \label{final2}
 \end{equation}
\begin{equation}
v_{{\rm c},n}=\left\{
\begin{array}{ll}
v_{ n}+\Delta_{ n} &  \textrm{at $g_{n} \leq G_{ n}$,} \\
v_{ n}+a_{ n}\tau &  \textrm{at $g_{n}> G_{ n}$}, \\
\end{array} \right.  
\label{delta}
\end{equation}
\begin{equation}
\Delta_{n}=\max(-b_{ n}\tau, \min(a_{ n}\tau, \ v_{ \ell,n}-v_{ n})),
 \label{final3}
 \end{equation} 
\begin{equation}
 g_{n}=x_{\ell, n}-x_{n}-d,
 \label{gap_formula}
\end{equation}
the subscript $\ell$  
denotes variables related to the preceding vehicle,
$v_{{\rm s}, n}$ is a safe speed at time step $n$,
$v_{\rm free}$ is the maximum speed in free flow,
  $\xi_{n}$ describes   speed fluctuations;
 $v_{{\rm c},n}$ is a desired speed;
all vehicles have the same length $d$. The vehicle length $d$ includes
the mean space gap between vehicles within a wide moving jam where the speed is  zero.
Values $a_{n}\geq 0$ and $b_{n}\geq 0$ in (\ref{delta}), (\ref{final3}) restrict changes in speed per time step
when the vehicle accelerates or adjusts the speed to that of the preceding vehicle.
 
 \subsection{Synchronization gap  and hypothetical steady states of synchronized flow \label{Syn_Gap_kkl}}
 
 Equations (\ref{delta}), (\ref{final3}) 
describe the adaptation of the vehicle speed to the speed of the preceding vehicle, i.e.,
the speed adaptation effect in synchronized flow. This
vehicle speed adaptation takes place within  the synchronization gap  $G_{n}$: 
At 
\begin{equation}
g_{n}\leq G_{n}
\label{x_x_G_n}
\end{equation}  
the vehicle tends to
adjust its speed to   the preceding vehicle. This means that the vehicle
decelerates if $v_{n}> v_{\ell,n}$, and accelerates if $v_{n}< v_{\ell,n}$~\cite{KKl}.

In the general rules (\ref{final})--(\ref{final3}),
the synchronization gap $G_{n}$ depends on the
vehicle speed $v_{n}$ and on the speed of the preceding vehicle $v_{\ell, n}$:
 \begin{equation}
 G_{n}=G(v_{n}, v_{\ell,n}),
  \label{Syn_Gap}
\end{equation}
 \begin{equation}
G(u, w)=\max(0,  \lfloor k\tau u+  a^{-1}u(u-w) \rfloor),
  \label{Syn_Gap2}
\end{equation}
  $k>1$ is constant.

 \begin{figure}
\begin{center}
\includegraphics[width=13 cm]{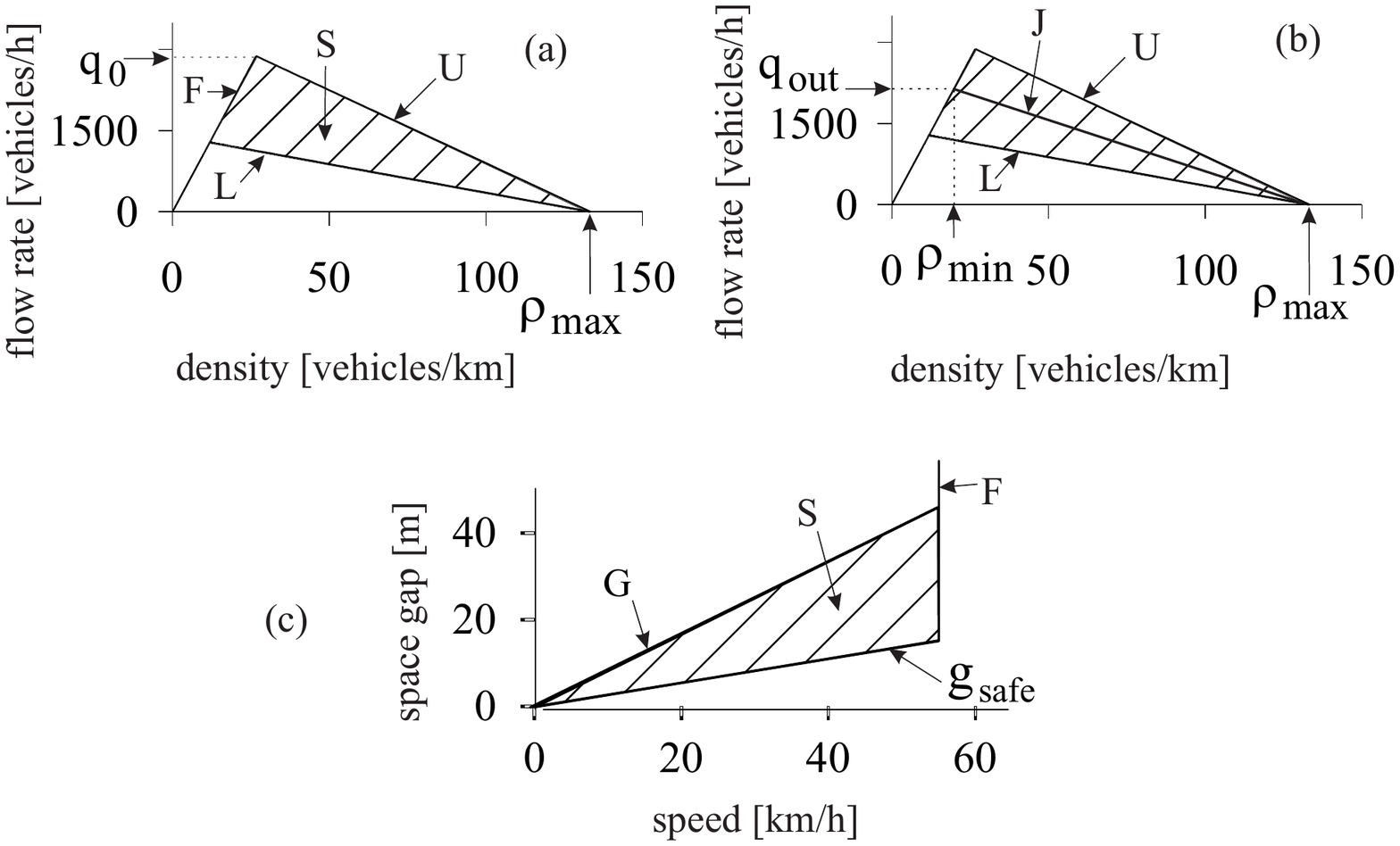}
\end{center}
\caption[]{Steady speed states for the three-phase
traffic flow model in the flow--density   (a, b) and in the space-gap--speed planes (c).
   $L$ and $U$ are, respectively,   lower and upper boundaries of 2D-regions of steady states of synchronized flow, $F$ - free flow, $S$ -- synchronized flow,
 $J$ is line $J$ whose slope is equal to the characteristic mean velocity $v_{\rm g}$ of a wide moving jam;  in the flow--density plane,
 the line $J$
  represents  the propagation of the downstream front of the wide moving jam with time-independent   velocity $v_{\rm g}$.
}
\label{KKl_Steady}
\end{figure}

The speed adaptation effect within the synchronization gap
is related to the hypothesis of three-phase    theory
about 2D-states of traffic flow

\cite{KernerBook,KernerBook2,Kerner1998E,Kerner1998C,Kerner1999B,Kerner1999D,Kerner2000D,Kerner2000A,Kerner2000B},

\cite{Kerner2001A,Kerner2001B,Kerner2002A,Kerner2002B,Kerner2002C,Kerner2002D,Kerner2003A,Kerner2004A}.
Respectively, as for the continuum-in-space
Kerner-Klenov model (see Sec.~16.3 of the book~\cite{KernerBook}), for the discrete model considered here hypothetical steady   states of traffic flow 
   cover a 2D-region in the flow--density plane (Fig.~\ref{KKl_Steady} (a)).
  However, because the speed $v$ and space gap $g$
   are integer in the discrete model,
   the steady states do not form a continuum in the flow--density plane
   as they do in the continuum model. The inequalities
   \begin{equation}
g\leq G(v, \ v) \quad  \rm{and} \quad v\leq  \rm{min}(v_{\rm free}, \ v_{\rm s}(g, \ v))
\label{2D_d}
\end{equation}
 define a 2D-region in the flow--density plane in which the hypothetical steady states
of traffic flow exist for the discrete model, when all model fluctuations are neglected.
 In (\ref{2D_d}), we have taken into account that in the hypothetical 2D-steady states of traffic flow
 vehicle speeds are assumed to be time-independent and the speed of each of the vehicles is equal to the speed of the associated preceding vehicle:
 $v=v_{\ell}$.
  
	It should be emphasized that
due to model fluctuations (see Sec.~\ref{Fluc_KKl_S}), 2D-steady   states of traffic flow are destroyed, i.e.,
 they do not exist in simulations.
This explains the term $\lq\lq$hypothetical" 2D-steady states of traffic flow. Rather than  2D-steady   states of traffic flow, in the Kerner-Klenov stochastic model all 2D-states of traffic flow are non-homogeneous in space and time.
However, in accordance with the
three-phase theory (see explanations in
 Secs.~4.3.4 and~6.3.3 of the book~\cite{KernerBook}),
 the non-homogeneous in space and time
2D-states of traffic flow of the Kerner-Klenov stochastic model
exhibit qualitatively the same features with respect to phase transitions  
  (F$\rightarrow$S, S$\rightarrow$F, and S$\rightarrow$J transitions)
as those  postulated in the three-phase theory for 2D-steady states of traffic
 flow
\cite{KernerBook,KernerBook2,Kerner1998E,Kerner1998C,Kerner1999B,Kerner1999D,Kerner2000D,Kerner2000A,Kerner2000B},
\cite{Kerner2001A,Kerner2001B,Kerner2002A,Kerner2002B,Kerner2002C,Kerner2002D,Kerner2003A,Kerner2004A}. 
  
  \subsection{Model speed fluctuations \label{Fluc_KKl_S}}

 In the   model,   random deceleration and acceleration are applied depending on 
whether the vehicle decelerates or accelerates, or else maintains its speed:
 \begin{equation}
  \xi_{ n}=\left\{
\begin{array}{ll}
\xi_{\rm a} &  \textrm{if  $S_{ n+1}=1$}   \\
- \xi_{\rm b} &  \textrm{if $S_{ n+1}=-1$} \\
\xi^{(0)} &  \textrm{if  $S_{ n+1}=0$}.
\end{array} \right.
\label{noise_CA}
\end{equation}
State of vehicle motion $S$ in (\ref{noise_CA})  
 is determined by   formula
 \begin{equation}
 S_{ n+1}=\left\{
\begin{array}{ll}
-1 &  \textrm{if $\tilde v_{ n+1}< v_{ n}$}   \\
1 &  \textrm{if $\tilde v_{ n+1}> v_{ n}$} \\
0 &  \textrm{if $\tilde v_{ n+1}= v_{ n}$}.
\end{array} \right.
\label{state_CA}
\end{equation}

In  (\ref{noise_CA}), $\xi_{\rm b}$, $\xi^{(0)}$, and $\xi_{\rm a}$ are random sources for deceleration and acceleration that are as follows:
 \begin{equation} 
 \xi_{\rm b}=a^{(\rm b)} \tau \Theta (p_{\rm b}-r),
 \label{xi_dec} 
 \end{equation}
\begin{equation}
\xi^{(0)}=a^{(0)}\tau \left\{
\begin{array}{ll}
-1 &  \textrm{if $r<p^{(0)}$} \\
1 &  \textrm{if $p^{(0)} \leq r<2p^{(0)}$} \quad {\rm and}  \  v_{n}>0 \\
0 &  \textrm{otherwise},
\end{array} \right.
\label{noise_CA_}
\end{equation}
 \begin{equation} 
 \xi_{\rm a}=a^{(\rm a)} \tau \Theta (p_{\rm a}-r),
 \label{xi_acc} 
 \end{equation}
  $p_{\rm b}$ is probability of   
random deceleration, $p_{\rm a}$ is probability of   
random acceleration, $p^{(0)}$
and $a^{(0)}\leq a$ are constants,  
$a^{(\rm a)}=a^{(\rm b)}=a$,
$r={\rm rand (0,1)}$,
 $\Theta (z) =0$ at $z<0$ and $\Theta (z) =1$ at $z\geq 0$.

\subsection{Stochastic time delays in vehicle acceleration and
deceleration}

  To simulate    time delays either in  vehicle
acceleration or in  vehicle
deceleration,  $a_{n}$ and  $b_{n}$ in (\ref{final3}) 
are taken as the following stochastic functions
\begin{equation}
a_{n}=a  \Theta (P_{\rm 0}-r_{\rm 1}),
\label{final_a}
 \end{equation}
 \begin{equation}  
b_{n}=a  \Theta (P_{\rm 1}-r_{\rm 1}),
\label{final_b}
 \end{equation}
\begin{equation}
P_{\rm 0}=\left\{
\begin{array}{ll}
p_{\rm 0} & \textrm{if $S_{ n} \neq 1$} \\
1 &  \textrm{if $S_{ n}= 1$},
\end{array} \right.
\label{final_P_0}
 \end{equation}
 \begin{equation}
P_{\rm 1}=\left\{
\begin{array}{ll}
p_{\rm 1} & \textrm{if $S_{ n}\neq -1$} \\
p_{\rm 2} &  \textrm{if $S_{ n}= -1$},
\end{array} \right.
\label{final_P_1}
 \end{equation}
$r_{1}={\rm rand}(0,1)$, $p_{\rm 1}$ is constant,
$p_{\rm 0}=p_{\rm 0}(v_{n})$ and $p_{\rm 2}=p_{\rm 2}(v_{n})$ are speed functions.

 The physical sense of the functions
$P_{0}$ and $P_{1}$ in (\ref{final_P_0}) and (\ref{final_P_1}) is as follows.  
The function $P_{0}$ in (\ref{final_P_0})  determines
the probability  $\psi_{\rm a}$ of a random time delay  in  vehicle acceleration at  time step $n+1$
corresponding to  
\begin{equation}
 \psi_{\rm a}= 1-P_{0}.
\label{Prob_Delay_Acc_For}
\end{equation}
The function $P_{1}$ (\ref{final_P_1})
determines the probability $\psi_{\rm b}$ of a random time delay  in  vehicle
deceleration at  time step $n+1$ corresponding to  
\begin{equation}
\psi_{\rm b}=1-P_{1}.
\label{Prob_Delay_Dec_For}
\end{equation}

 \subsection{Safe speed  \label{Safe_speed_kkl}}

In the model, the safe speed $v_{{\rm s},n}$ in (\ref{final}) is chosen in the form 
\begin{equation}
v_{{\rm s},n}=
\min{(v^{\rm (safe)}_{ n},  g_{ n}/ \tau+ v^{\rm (a)}_{ \ell})},
\label{safe_kkl}
\end{equation}
$v^{\rm (a)}_{ \ell}$  is an $\lq\lq$anticipation" speed of the preceding vehicle
that will be considered below,
the function    
\begin{equation}
v^{\rm (safe)}_{ n}=\lfloor v^{\rm (safe)} (g_{n}, \ v_{ \ell,n})  \rfloor
  \label{Safe_Speed3} 
\end{equation}
 in (\ref{safe_kkl}) is related to the safe speed $v^{\rm (safe)} (g_{n}, \ v_{ \ell,n})$ in the model by Krau{\ss} et al.~\cite{Kra10,Kra_PhD10}, 
 which is a solution of   the
 Gipps's equation~\cite{Gipps10}  
   \begin{equation}
v^{\rm (safe)} \tau_{\rm safe} + X_{\rm d}(v^{\rm (safe)}) = g_{n}+X_{\rm d}(v_{\ell, n}),
  \label{Gipps_Safe_Speed}
\end{equation}
  where $\tau_{\rm safe}$ is a safe time gap that can be individual for  drivers, 
$X_{\rm d} (u)$ is the braking distance that should be passed by the vehicle
 moving first with the speed $u$ before the vehicle can come to a stop:
    \begin{equation}
X_{\rm d} (u)=b \tau^{2} \bigg(\alpha \beta+\frac{\alpha(\alpha-1)}{2}\bigg),
  \label{Gipps_Safe_Speed2}
\end{equation}
$\alpha=\lfloor u/b\tau \rfloor$ and $\beta=u/b\tau-\alpha$ 
are the integer and  fractional parts  of $u/b\tau$,  
respectively, 
$b$ is constant.

The safe speed $v^{\rm (safe)}$ as a solution of   equation (\ref{Gipps_Safe_Speed}) at the distance $X_{\rm d} (u)$
 given by (\ref{Gipps_Safe_Speed2}) 
and at  $\tau_{\rm safe}=\tau$
has
been found in~\cite{Kra10,Kra_PhD10}
\begin{equation}
v^{\rm (safe)}(g_{n}, v_{\ell, n})=b \tau (\alpha_{\rm safe}+\beta_{\rm safe}),
\label{SafeKr}
\end{equation}
where
\begin{equation}
\alpha_{\rm safe}=\lfloor{\sqrt {2\frac{X_{\rm d}(v_{\ell, n}) + g_{ n}}{b \tau^{2}}+\frac{1}{4}} -\frac{1}{2}}\rfloor,
\label{alpha}
\end{equation}
\begin{equation}
\beta_{\rm safe}=\frac{X_{\rm d}(v_{\ell, n}) +g_{ n}}{(\alpha_{\rm safe}+1)b \tau^{2}}-\frac{\alpha_{\rm safe}}{2}.
\label{beta}
\end{equation}

The safe speed in the model by Krau{\ss} {\it et al.}~\cite{Kra10,Kra_PhD10} provides   collision-less motion of  vehicles
if the time gap $g_{ n}/v_{ n}$ between two vehicles is greater than or equal to the time step $\tau$,
 i.e., if $g_{n} \geq v_{n}\tau$~\cite{Kra_PhD10}. In the  model, it is assumed that
in some cases, mainly due to
lane changing or merging of  vehicles onto the main road within the  merging region of bottlenecks, 
the space gap $g_{n}$ can become less  than  $v_{n}\tau$. In these critical situations,
the collision-less motion of  vehicles in the model is a result of the 
 second term in (\ref{safe_kkl}) in which some prediction ($v^{\rm (a)}_{\ell}$)
 of the speed of the preceding vehicle 
 at the next time step is used. The related $\lq\lq$anticipation" speed $v^{\rm (a)}_{\ell}$  at the 
next time step that is given by formula
    \begin{eqnarray}
v^{\rm (a)}_{\ell}=   
\max(0, \min(v^{\rm (safe)}_{ \ell, n}, v_{ \ell,n}, g_{ \ell, n}/\tau)-a\tau),
  \label{Safe_Speed2}
\end{eqnarray}
where $v^{\rm (safe)}_{ \ell, n}$ is the safe speed (\ref{Safe_Speed3}), (\ref{SafeKr})--(\ref{beta}) for the preceding vehicle,
$g_{ \ell, n}$ is the space gap in front of the preceding vehicle.
Simulations have shown that formulas (\ref{safe_kkl}), (\ref{Safe_Speed3}),  (\ref{SafeKr})--(\ref{Safe_Speed2}) lead to   
collision-less vehicle motion over a wide range of parameters of  the merging region 
of the bottleneck.

\subsection{Boundary  and initial conditions \label{Boundary_Ini_Con_S}}

In the model,  open boundary conditions are applied.
At the beginning of the road new vehicles are generated one after another in each of the lanes
of the road at time moments
\begin{equation}
t^{(k)}=\tau \lceil k \tau_{\rm in}/\tau \rceil, \ k=1,2,....
\label{t_new_KKl}
\end{equation}
In (\ref{t_new_KKl}), $\tau_{\rm in}=  1/q_{\rm in}$,
 $q_{\rm in}$ is the flow rate
in the incoming boundary flow per lane, $\lceil z \rceil$ denotes
 the nearest integer greater than
or equal to $z$.
A new vehicle appears on the road 
only if the distance from the beginning of the road ($x=x_{\rm b}$)
to the position $x=x_{\ell, n}$ of the farthest upstream vehicle on the road
is not smaller than the   distance 
\begin{equation}
x_{\ell, n}-x_{\rm b} \geq v_{\ell, n} \tau+d,
\label{Coordinate_preceding_KKl}
\end{equation}
where $n=t^{(k)}/\tau$. Otherwise,  condition (\ref{Coordinate_preceding_KKl}) is checked
at  time   $(n+1)\tau$ that is the next
 one
 to time  $t^{(k)}$ (\ref{t_new_KKl}), and so on, until
the condition (\ref{Coordinate_preceding_KKl}) is satisfied.
Then the next vehicle appears on the road. After this occurs, the number  $k$ in (\ref{t_new_KKl})
is increased by 1.

The speed  $v_{ n}$ and coordinate $x_{n}$
of the new vehicle
 are  
\begin{eqnarray}
\begin{array}{ll}
v_{ n}= v_{\ell, n}, \\
x_{ n}={\rm max}(x_{\rm b}, x_{\ell, n}-\lfloor{v_{ n}\tau_{\rm in}}\rfloor).
\end{array}
\label{Coordinate_in_KKl}
\end{eqnarray}
The flow rate $q_{\rm in}$ is chosen to have the value $v_{\rm free}\tau_{\rm in}$ integer.
In the initial state ($n=0$), all  vehicles have the  maximum speed $v_{ n}=v_{\rm free}$
 and they are positioned at  space intervals  $x_{\ell, n}-x_{n}=v_{\rm free}\tau_{\rm in}$.

After a vehicle has reached  the end of the road 
it is removed.
Before this occurs, the farthest downstream vehicle maintains its speed.
For the vehicle following the farthest downstream one, the $\lq\lq$anticipation" speed 
$v^{\rm (a)}_{ \ell}$ in (\ref{Safe_Speed2})  is  equal to the speed of the farthest downstream vehicle.

 \subsection{Model of on-ramp bottleneck \label{On_Model_Bott_Sec}}

   \begin{figure}
\begin{center}
\includegraphics[width=10 cm]{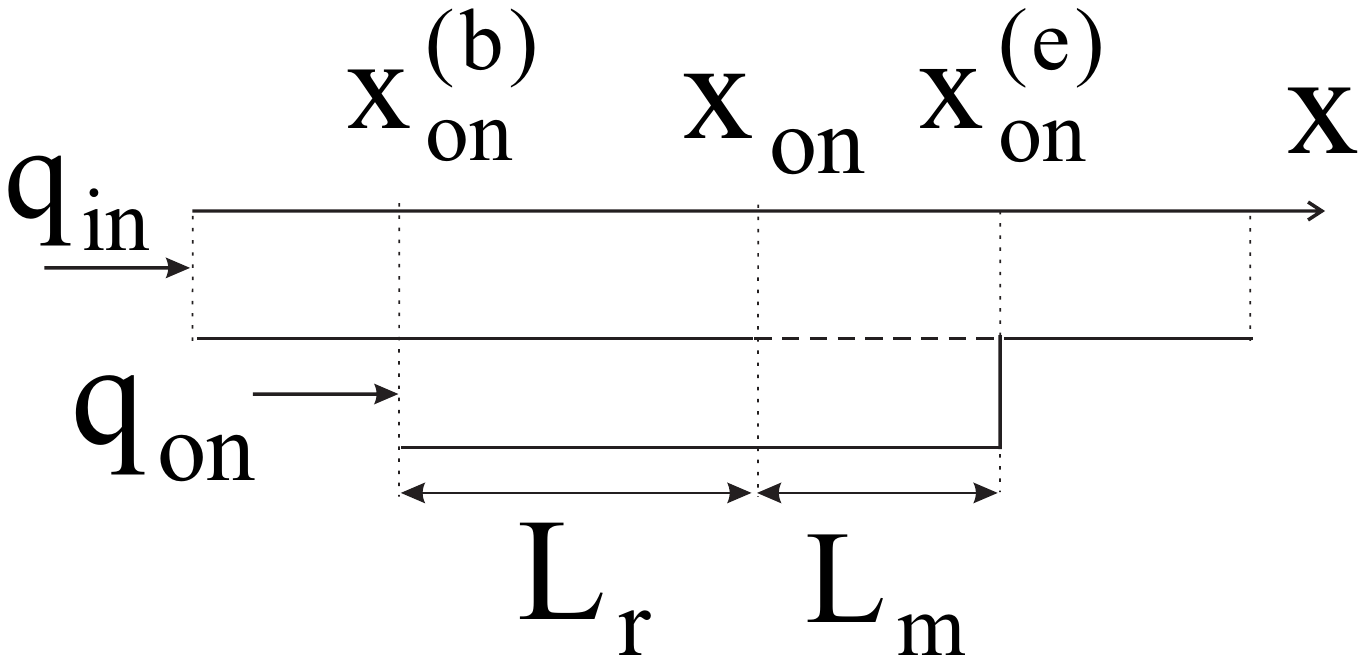}
\end{center}
\caption[]{Model of on-ramp   bottleneck on   single-lane   road.  
}
\label{KKl_onramp_1lane}
\end{figure} 

The on-ramp bottleneck consists of two parts (Fig.~\ref{KKl_onramp_1lane}):
\begin{description}
 \item [(i)] The merging region of  length  $L_{\rm m}$ 
 where vehicle can  
merge onto the main road from the on-ramp lane.
 \item [(ii)] A part of the on-ramp lane of length $L_{\rm r}$
 upstream of the merging region 
where vehicles
move in accordance with  the model
(\ref{final})--(\ref{gap_formula}). The maximal speed of vehicles is
$v_{\rm free}=v_{\rm free \ on}$. 
\end{description}
At the beginning of the 
merging region of the on-ramp lane ($x=x^{\rm (b)}_{\rm on}$) the flow rate to the on-ramp $q_{\rm on}$ 
is given   as in $q_{\rm in}$.

  \subsection{Manual driving vehicle merging at on-ramp bottleneck \label{On_Merg_Bott_Sec}}
 
 A merging region at an on-ramp bottleneck is a road region of length $L_{\rm m}$ within which a vehicle  moving in the on-ramp lane
  merges to the right lane of the main road (Fig.~\ref{KKl_onramp_1lane}).  
   When a vehicle is within the merging region of the bottleneck, the vehicle takes into account 
 the space gaps to the preceding vehicles and their speed
both in the current and target lanes. Respectively, instead of formula (\ref{delta}), 
  in (\ref{final2}) for the speed $v_{{\rm c},n}$ the following formula is used:
      \begin{equation}
v_{{\rm c},n}=\left\{
\begin{array}{ll}
v_{ n}+\Delta^{+}_{ n} &  \textrm{at $g^{+}_{n} \leq G(v_{n}, \hat
v^{+}_{n})$,} \\
v_{ n}+a_{ n}\tau &  \textrm{at $g^{+}_{n}>G( v_{n}, \hat
v^{+}_{n})$}, \\
\end{array}\right.
  \label{Lane_Change2_before}
  \end{equation}
       \begin{equation}
\Delta^{+}_{ n}=\max(-b_{ n}\tau, \min(a_{ n}\tau, \ \hat v^{+}_{n}-v_{
n})),
  \label{Lane_Change2_before2}
  \end{equation}
         \begin{equation}
\hat v^{+}_{n}=\max(0, \min(v_{\rm free}, \  v^{+}_{n}+\Delta
v^{(2)}_{r})),
  \label{Lane_Change2_before3}
  \end{equation}
$\Delta v^{(2)}_{r}$ is  constant. 
Superscripts    $+$   and  $-$  in variables, parameters, and functions 
denote the preceding vehicle and the trailing vehicle 
in the $\lq\lq$target" (neighboring) lane, respectively
(the target lane is the 
lane into which the vehicle wants to change).

The safe speed $v_{{\rm s},n}$ in (\ref{final}), (\ref{final2}) for the vehicle that is the closest one
to the end of merging region  
is chosen in the form 
\begin{equation}
v_{{\rm s},n}=\lfloor v^{\rm (safe)}(x^{\rm (e)}_{ \rm on}- x_{n}, \ 0) \rfloor.
\label{safe_on}
\end{equation}

  Vehicle merging at the bottleneck
 occurs, when     safety conditions ($\ast$) {\it or}   safety conditions  ($\ast \ast$) are satisfied.
Safety conditions ($\ast$) are as follows:
  \begin{equation}
  \begin{array}{ll}
g^{+}_{n} >\min(\hat  v_{n}\tau , \ G(\hat  v_{n}, v^{+}_{n})), \\
g^{-}_{n} >\min(v^{-}_{n}\tau, \ G(v^{-}_{n},\hat  v_{n})),
\label{merging_a}
\end{array} 
\end{equation}
\begin{equation}
 \hat v_{n}=\min(v^{+}_{n},  \ v_{n}+\Delta v^{(1)}_{r}), 
 \label{Atwo}
\end{equation}
 $\Delta v^{(1)}_{r}>0$ is constant.
 
 Safety conditions  ($\ast \ast$) are as follows:
\begin{equation}
x^{+}_{n}-x^{-}_{n}-d > g^{\rm (min)}_{\rm target},
\label{merging_b}
\end{equation}
where
\begin{equation}
g^{\rm (min)}_{\rm target}=\lfloor \lambda_{\rm b}  v^{+}_{n} +d \rfloor,
\label{merging_b2}
\end{equation}
 $\lambda_{\rm b}$ is constant. In addition to conditions (\ref{merging_b}), the safety condition  ($\ast \ast$) includes the condition that
the vehicle should pass the midpoint 
\begin{equation}
x^{\rm (m)}_{n}=\lfloor (x^{+}_{n}+x^{-}_{n})/2 \rfloor
\label{midpoint_f}
\end{equation}
between two neighboring vehicles in the target lane, i.e.,   conditions 
  \begin{equation}
 \begin{array}{ll}
x_{n-1}< x^{\rm (m)}_{n-1} \  \textrm{and} \
 x_{n} \geq x^{\rm (m)}_{n} \\
\ \textrm{\it or} \\
x_{n-1} \geq x^{\rm (m)}_{n-1} \  \textrm{and} \
 x_{n} < x^{\rm (m)}_{n}.
\end{array} 
\label{mid2}
\end{equation}
should also be satisfied.

 The vehicle speed after vehicle merging is equal to
     \begin{equation}
v_{n}=\hat v_{n}.
  \label{Lane_Change2_after}
  \end{equation}
  
  Under conditions ($\ast $),
the vehicle coordinate $x_{n}$  remains the
same.
 Under conditions ($\ast \ast$), the vehicle coordinate $x_{n}$ is equal to 
     \begin{equation}
x_{n} = x^{\rm (m)}_{n}.
  \label{Lane_Change2_after2}
  \end{equation}
	
	\subsection{ACC-vehicle merging at on-ramp bottleneck \label{On_ACC_Bott_Sec}}

In the on-ramp lane, an ACC vehicle
moves in accordance with  the model
(\ref{ACC_dynamics_Eq}), (\ref{next1_ACC}), (\ref{next2_ACC}). The maximal speed of the
ACC vehicle in the on-ramp lane is
$v_{\rm free}=v_{\rm free \ on}$. 
The safe speed $v_{{\rm s},n}$ in (\ref{next2_ACC}) for the ACC vehicle that is the closest one
to the end of merging region  
is the same as that for manual driving vehicles (\ref{safe_on}).

  ACC-vehicle merges from the on-ramp lane onto the main road,
   when   some  ACC-safety conditions ($\ast$) {\it or}   safety conditions  ($\ast \ast$)
	are satisfied.
Safety conditions ($\ast$) for ACC-vehicles are as follows:
  \begin{equation}
  \begin{array}{ll}
g^{+}_{n} >\hat  v_{n}\tau, \quad
g^{-}_{n} >v^{-}_{n}\tau,
\label{merging_a_ACC}
\end{array} 
\end{equation}
where $\hat v_{n}$ is given by formula (\ref{Atwo}).
Safety conditions  ($\ast \ast$) 
are given by formulas
(\ref{merging_b})--(\ref{mid2}), i.e., they are the same as those for manual driving vehicles. Respectively,
as for manual driving vehicles,
the ACC-vehicle speed and its coordinate after ACC-vehicle merging are determined by formulas
  (\ref{Lane_Change2_after})
	and (\ref{Lane_Change2_after2}).

\begin{table}
\caption{Model parameters used in   simulations (other model parameters are given in figure captions)}
\label{table2}
\begin{center}
\begin{tabular}{|l|}
\hline
\multicolumn{1}{|c|}{Vehicle motion in road lane:
}\\
\hline
$\tau_{\rm safe}   = \tau$, $\tau=$1 s,
 $d = 7.5 \  \rm m/\delta x$, $\delta x=$ 0.01 m, \\
$v_{\rm free} = 30 \ {\rm ms^{-1}}/\delta v$, $b = 1 \ {\rm ms^{-2}}/\delta a$, \\ $\delta v= 0.01 \  {\rm ms^{-1}}$, \ $\delta a= 0.01 \  {\rm ms^{-2}}$,
$k=$ 3, \\ $p_{1}=$ 0.3, 
 $p_{b}=   0.1$, $p_{a}=   0.17$, \\
$p^{(0)}= 0.005$, 
$p_{\rm 2}(v_{n})=0.48+ 0.32\Theta{( v_{n}-v_{21})}$, \\
$p_{\rm 0}(v_{n})=0.575+ 0.125\min{(1, v_{n}/v_{01})}$, \\
  $a^{(0)}= 0.2a$, 
  $a^{(\rm a)}=a^{(\rm b)}=  a$, \\
$v_{01} = 10 \ {\rm ms^{-1}}/\delta v$, $v_{21} = 15 \ {\rm ms^{-1}}/\delta v$, 
$a=$ 0.5 ${\rm ms^{-2}}/\delta a$. \\
   \hline
    \multicolumn{1}{|c|}{Model of on-ramp bottleneck:}
\\
 \hline
$\lambda_{\rm b}=$ 0.75, \ 
   $v_{\rm free \ on}=22.2 \ {\rm ms^{-1}}/\delta v$, \\ 
   $\Delta v^{\rm (2)}_{\rm r}=$ 5    ${\rm ms^{-1}}/\delta v$, \\
   $L_{\rm r}=1 \ {\rm km}/\delta x$,   $\Delta v^{\rm (1)}_{\rm r}=10 \ {\rm ms^{-1}}/\delta v$,  \\
   $L_{\rm m}=$ 0.3   \ ${\rm km}/\delta x$. \\
\hline
\end{tabular}
\end{center}
\end{table}
\vspace{1cm}

{\bf Acknowledgments:}

  I thank our partners for their support in the project $\lq\lq$UR:BAN - Urbaner Raum: Benutzergerechte Assistenzsysteme und Netzmanagement", funded by the German Federal Ministry of Economics and Technology by resolution of the German Federal Parliament. 
I   thank   Sergey Klenov  for very helpful suggestions and help in simulations.

\end{document}